\documentclass[twocolumn,
	aps, prd,
	10pt, notitlepage, %a4paper,
        floats, floatfix,
	amsmath, amssymb, amsfonts, eqsecnum,
	superscriptaddress,
	showpacs, showkeys,
	nofootinbib,
 	longbibliography,
]{revtex4-2}

\usepackage{mathtools}
\usepackage{wrapfig}
\usepackage{graphicx} % include figures
\usepackage[dvipsnames]{xcolor}
\usepackage{xspace} % Sensible space treatment at end of simple macros
\usepackage{bm} % bold math
\usepackage{amsmath,amsfonts,amssymb,amsthm,mathtools}
\usepackage[utf8]{inputenc} % for some references from inspirehep.net
\usepackage{subcaption}
\usepackage{multirow}
\usepackage{array}

\xdefinecolor{mylinkcolor}{rgb}{0,0,0.5}
\usepackage[
	bookmarksnumbered, bookmarksopen, bookmarksopenlevel=2,
	breaklinks=true,
	colorlinks=true, filecolor=mylinkcolor, citecolor=mylinkcolor,
	linkcolor=mylinkcolor, urlcolor=mylinkcolor, menucolor=mylinkcolor,
]{hyperref}

\usepackage{verbatim}
\usepackage{mathrsfs}
\usepackage{color}
\usepackage{enumitem}
\usepackage[dvipsnames]{xcolor}
\usepackage{comment}
\usepackage{tikz}
\usetikzlibrary{decorations.pathmorphing}
\usetikzlibrary{arrows}

\tikzset{snake it/.style={decorate, decoration=snake}}

\usepackage[normalem]{ulem}

\newcommand{\nnm}{\nonumber}

\newcommand{\be}{\begin{equation}}
\newcommand{\ee}{\end{equation}}
\newcommand{\bse}{\begin{subequations}}
\newcommand{\ese}{\end{subequations}}

\newcommand{\mr}{\mathrm}

\newcommand{\mc}{\mathcal}

\newcommand{\f}{\frac}

\newcommand{\bpm}{\begin{pmatrix}}
\newcommand{\epm}{\end{pmatrix}}

\newcommand{\AEI}{\affiliation{Max Planck Institute for Gravitational Physics (Albert Einstein Institute), Am M\"uhlenberg 1, Potsdam 14476, Germany}}
\newcommand{\Maryland}{\affiliation{Department of Physics, University of Maryland, College Park, MD 20742, USA}}

\usepackage{colortbl}
\definecolor{blue2}{cmyk}{1, 0.1, 0.1, 0}

\definecolor{pyBlue}{RGB}{31, 119, 180}
\definecolor{pyRed}{RGB}{214, 39, 40}
\definecolor{pyGreen}{RGB}{44, 160, 44}
\definecolor{pyBlue2}{RGB}{0, 111, 237}
\definecolor{pyRed2}{RGB}{224, 52, 36}

\usepackage{colortbl}
\definecolor{summersky}{cmyk}{0.71,0.33,0,0.5}
\definecolor{flamingo}{cmyk}{0,0.51,0.71,0.5}
\definecolor{rp}{cmyk}{0.2, 1, 0.6, 0}
\definecolor{pacificblue}{cmyk}{0.95,0.3,0, 0.5}
\definecolor{gray60}{cmyk}{0.4,0.4,0,0.8}

\newcommand{\red}[1]{\textcolor{pyRed}{#1}}
\newcommand{\blue}[1]{\textcolor{pyBlue}{#1}}

\begin{document}
%\title{Tidal Heating of neutron stars From Scattering Amplitudes}
\title{Investigating tidal heating in neutron stars via gravitational Raman scattering}
\author{M. V. S. Saketh}
\email{msaketh@aei.mpg.de}
\Maryland
\AEI
\author{Zihan Zhou}
\email{zihanz@princeton.edu}
\affiliation{Department of Physics, Princeton University, Princeton, NJ 08540, USA}%
\author{Suprovo Ghosh}
\email{suprovo@iucaa.in}
\affiliation{%
 Inter-University Centre for Astronomy and Astrophysics,
Pune University Campus,
Pune 411007, India\\
}%
\author{Jan Steinhoff}
\email{jan.steinhoff@aei.mpg.de}
\AEI
\author{Debarati Chatterjee}%
\email{debarati@iucaa.in}
\affiliation{%
 Inter-University Centre for Astronomy and Astrophysics,
Pune University Campus,
Pune 411007, India\\
}%
%\author{Alessandra Buonanno}
%\email{alessandra.buonanno@aei.mpg.de}
%\AEI
%\Maryland
\begin{abstract}
We present a scattering amplitude formalism to study the tidal heating effects of nonspinning neutron stars incorporating both worldline effective field theory and relativistic stellar perturbation theory. In neutron stars, tidal heating arises from fluid viscosity due to various scattering processes in the interior. It also serves as a channel for the exchange of energy and angular momentum between the neutron star and its environment. In the interior of the neutron star, we first derive two master perturbation equations that capture fluid perturbations accurate to linear order in frequency. Remarkably, these equations receive no contribution from bulk viscosity due to a peculiar adiabatic incompressibility  which arises in stellar fluid for non-barotropic perturbations. In the exterior, the metric perturbations reduce to the Regge-Wheeler (RW) equation which we solve using the analytical Mano-Suzuki-Takasugi (MST) method. We compute the amplitude for gravitational waves scattering off a neutron star, also known as gravitational Raman scattering. From the amplitude, we obtain expressions for the  electric quadrupolar static Love number and the leading dissipation number to all orders in compactness. We then compute the leading dissipation number for various realistic equation-of-state(s) and estimate the change in the number of gravitational wave cycles due to tidal heating during inspiral in the LIGO-Virgo-KAGRA (LVK) band. We find that the shear-viscous contribution is small. Even for cold ($T\sim 10^5$K) and weakly compact stars it changes the number of cycles by less than $10^{-2}$, about an order of magnitude below the other conservative 4PN contributions, and it decreases as $T^{-2}$ for hotter stars.
\end{abstract}
\maketitle
\section{Introduction}

With the increasing number of events observed by the LIGO-Virgo-KAGRA (LVK) collaboration~\cite{LIGOScientific:2018mvr,LIGOScientific:2020ibl,LIGOScientific:2021usb,LIGOScientific:2021djp}, gravitational-wave (GW) astrophysics has entered an era of precision physics. As a result, the need for highly accurate waveforms has become increasingly apparent in order to accurately model the nature of self-gravitating compact objects, in particular their tidal effects~\cite{Flanagan:2007ix, Buonanno:2022pgc}.
When the components of the binary are far apart, it is sufficient to model them as orbiting point particles, with their intrinsic parameters characterized by the component masses and spins. However, as they inspiral toward each other, finite-size effects such as tidal effects become important. For neutron stars, the detection of tidal effects would be very useful to distinguish between different equations of state (EoS)~\cite{Flanagan:2007ix,Read2009,Buonanno:2022pgc, LIGOScientific:2017vwq,LIGOScientific:2018hze,LIGOScientific:2018cki}, and thus to probe dense matter physics under extreme conditions~\cite{Coughlin2019,Biswas2021,Dietrich2020,O_Boyle_2020,Ghosh2022,Ghosh:2023vrx}.

Tidal effects describe the deformation of a body under external gravitational perturbations. %This involves both the deformation of the body and its gravitational field. 
In Newtonian gravity, they have been greatly explored in the literature 
%the relationship between the deformation of the object and the gravitational field is well known 
(see \cite{love1909yielding, Will:2014kxa,2014ARA&A..52..171O,Terquem:1998ya, PoissonWill} and the references there in). In general relativity (GR), tidal effects are much more complicated due to gravitational non linearities. However, when the external tidal field is weak, we can study tidal deformation from linear perturbation upon the compact objects. Moreover, with the modern Quantum Field Theory (QFT)-inspired techniques, 
tidal effects can also be studied in the context of worldline effective field theory (EFT) by incorporating multipole moments within the worldline action as additional dynamical degrees of freedom~\cite{Goldberger:2004jt,
Goldberger:2005cd,
Goldberger:2009qd,
Porto:2016pyg,
Levi:2018nxp,
Goldberger:2020fot,Saketh:2022xjb, Saketh:2023bul}. For e,g, the leading post-Newtonian (PN) tidal effects for a spherically symmetric object can be included by a worldline action of the form
\begin{alignat}{3}
    S = \int d\tau\left[-m + \mc{L}_{\mr{int}}(Q_{ij},\dot{Q}_{ij}) - \frac{1}{2} Q_{ij} E^{kl}\right]
\end{alignat}
where dynamical quadrupolar degrees of freedom $Q_{ij}$ have been included in addition to the point mass $m$ in the action. The quadrupole is coupled to the external tidal field $E_{ij}=R_{\rho\mu\sigma\nu}u^\rho u^\sigma e_i{}^\mu e_j{}^\nu$, in the frame of the particle. The dynamics of the quadrupole moment is also encoded in the action, via the ``internal" Lagrangian $\mc{L}_{\mr{int}}$, although in practice it is easier make an ansatz for it. That is, the dynamics of the quadrupole moment can be expressed according to linear response theory as~\cite{Goldberger:2020fot, Saketh:2023bul}
\begin{alignat}{3}
\label{eq:linear response intro}
    \langle Q_{ij}(\tau) \rangle = - \frac{1}{2} \int_{-\infty}^\tau d\tau' G^{\rm ret}_{ij, kl}(\tau-\tau')E^{kl}(\tau')
\end{alignat}
where $G^{\mr{ret}}_{ij,kl}(\tau-\tau')$ is the retarded tidal response function. For slowly varying external tidal fields, Eq.~\eqref{eq:linear response intro} can be systematically expanded in terms of time derivatives (corresponding to an expansion in orbital frequency of a binary $\omega$, in Fourier domain)
\begin{alignat}{3}
    Q_{ij}= -  M (GM)^4 \Bigg[ \Lambda^E - (G M) H_\omega^E \frac{d}{d\tau} + \cdots \Bigg] E_{ij}~.
\end{alignat}
Here the coefficients $\Lambda^E$, $H_\omega^E$, $\dots$ characterize the low-frequency tidal response of the particle. Specifically, $\Lambda^E$ is the famous quadrupolar Love number~\cite{love1909yielding}, which characterizes the leading conservative tidal response, and $H_\omega^E$ is often referred to as the dissipation number~\cite{Goldberger:2004jt, Saketh:2022xjb, Saketh:2023bul}, characterizing the leading dissipative part of the tidal response.

Love numbers are known to vanish for black holes \cite{Chia:2020yla,
Charalambous:2021mea,
Ivanov:2022qqt,
Kol:2011vg,
Hui:2020xxx,
Binnington:2009bb,
Damour:2009vw}. For neutron stars, however, they are nonzero and have been studied extensively in the literature \cite{Hinderer:2007mb,Damour:2009vw,
Binnington:2009bb}. At the level of GW observables, Love numbers first formally appear at the 5PN order in the phase of the GW strain. 

The dissipation number, on the other hand, is responsible for the phenomenon of tidal heating. It is nonzero if there exists non-conservative effects, such as viscosity for stars or the event horizon for black holes~\cite{Goldberger:2020fot, Creci:2021rkz, Saketh:2022xjb, Saketh:2023bul}. Physically, dissipative tidal effects irreversibly transfer energy and angular momentum from the surrounding tidal environment into the body\footnote{Famously illustrated by the Earth-Moon system, the tidal lag time caused by the viscosity of the Moon ultimately results in tidal locking, whereby the Moon's rotational frequency synchronizes with its orbital frequency around Earth, which is why we always see the same side of the Moon.}. In the gravitational waveform, the tidal dissipation number first appears at 4PN order for a binary of spherically symmetric compact objects~\cite{Tagoshi:1997jy,
Chatziioannou:2016kem,
Goldberger:2020fot,
Poisson:2005pi,Saketh:2022xjb,Chia:2024bwc}.

Viscous sources in the interior of the neutron stars are known to play an important role in the different evolutionary stages of neutron stars. They have been extensively studied in the context of damping the unstable oscillations of $r$-modes and $f$-modes of rapidly rotating newborn neutron stars~\cite{Jones2001,Lindblom2002,Passamonti_2012,Alford_2011,hernandez2024dampingdensityoscillationsbulk}. More recently, several studies have suggested that viscous dissipation may affect the post-merger oscillations of binary neutron star merger remnants and their stability~\cite{Alford_2017PRL,Most_2021,Celora_2022,Most_2024,chabanov2023impactbulkviscositypostmerger}. In the context of binary neutron star inspirals, previous studies had suggested that tidal heating could potentially cause mass ejection in a pre-merger radiation-driven outflow~\cite{Rees1992} or spin the stars up to corotation before the merger~\cite{Kochanek1992,Bildsten1992}, but these scenarios are not feasible for canonical neutron star viscosities (shear viscosity from $n-n$ or $e-e$ scattering and bulk viscosity from modified Urca reactions). Previous studies by Lai~\cite{Lai:1993di} and more recently by Arras et al.~\cite{Arras:2018fxj} estimated that neutron stars can be heated to a maximum of $\sim 10^8$K during the inspiral due to tidal heating. However, recent studies by Ghosh et al.~\cite{Ghosh:2023vrx} suggest that the viscosity from nonleptonic weak processes involving hyperons is much higher and could heat the star to $10^9-10^{10}$K,  leaving a detectable imprint on the inspiral waveforms. In Ref.~\cite{Ripley:2023qxo}, it was shown that internal dissipative processes entering at 4PN can be constrained to the same extend as the static Love numbers. Later, in Ref.~\cite{Ripley:2023lsq}, the authors constrained the dissipation numbers for the event GW170817, and predictions were made regarding the improvement of constraints with upcoming next-generation detectors. The analysis was extended to include relative 1PN effects in tidal dissipation recently in Ref.~\cite{HegadeKR:2024slr}.

However, most of the literature on the tidal response of neutron stars beyond the static limit has been Newtonian or treated in the mode-sum approximation. A relativistic analysis in the low-frequency regime, without a priori assuming the validity of the mode-sum approximation, was presented in Ref.~\cite{Pitre:2023xsr} for polytropic stellar models without viscosity. Recently, Ref.~\cite{HegadeKR:2024agt} tackled the daunting problem of the relativistic dynamical tidal response by obtaining a resummed all-orders-in-frequency tidal response at linear order in viscosity. This was used to study both conservative and dissipative effects, including resonances for polytropic stellar models and perturbations. The relativistic dynamical tidal response of nonrotating objects has also been studied in Ref.~\cite{Chakrabarti:2013lua} (and applied to conservative neutron-star and dissipative black-hole tides), based on a gauge-dependent matching of the worldine EFT. But starting from quadratic order in a low-frequency expansion, an ambiguity parameter enters the result; while in the scalar-field case it has been shown how the ambiguity could be removed~\cite{Creci:2021rkz}, an implementation in the gravitational case is still missing, and a manifestly gauge-independent matching procedure~\cite{Ivanov:2024sds} would be desirable.

In this paper, we perform a comprehensive first-principle study of the behaviour of realistic neutron stars with bulk and shear viscosity under tidal perturbations in the small frequency regime. We accomplish this by combining  relativistic stellar perturbation theory (SPT), and worldline EFT, for realistic neutron star EoS(s). We compute the amplitude of GWs scattering off the neutron star (gravitational Raman scattering) in SPT and EFT. In SPT, the amplitude depends on the EoS, the compactness, and other internal physics of the neutron star. In the EFT it depends on the tidal response, parametrized by Love and dissipation numbers. By matching the amplitudes obtained in SPT and EFT, we can relate the properties of the neutron star to the tidal response. This method of fixing the world-line action by matching the Raman amplitudes in real and effective theories has already been used for black holes in several previous works~\cite{Saketh:2022wap, Ivanov:2022qqt, Saketh:2022xjb, Saketh:2023bul,Bautista:2021wfy,
Scheopner:2023rzp,
Bautista:2023sdf,
Bautista:2022wjf,Ivanov:2024sds,
Ivanov:2024sds}. Here, we apply it to neutron stars. We give an overview of this work and briefly summarize the main results in the following subsection.

\subsection{Summary of methodology and results}
\label{sec : exec}
\subsubsection{Master equations for SPT \& adiabatic imcompressibility}
To study the relativistic fluid dynamics, we start from the following energy-momentum tensor for viscous fluids
\begin{equation}
\label{eq: energy_stress with viscosity}
    T_{\mu\nu} = \rho u_{\mu} u_{\nu} + (p - \zeta \nabla_{\alpha} u^\alpha) P_{\mu\nu} - 2 \eta P^\alpha_\mu P^\beta_\nu \sigma_{\alpha\beta}
\end{equation}
where $u^\mu$ is the 4-velocity, $P_{\mu\nu}$ is the spatial projector
\begin{equation}
    P_{\mu\nu} \equiv g_{\mu\nu} + u_\mu u_\nu ~,
\end{equation}
and $\sigma_{\mu\nu}$ is the shear tensor
\begin{equation}
    \sigma_{\mu\nu} \equiv \frac{1}{2} \Bigg(\nabla_\mu u_\nu + \nabla_\nu u_\mu - \frac{2}{3} g_{\mu\nu} \nabla_\alpha u^\alpha \Bigg) ~.
\end{equation}
The bulk $\zeta$ and shear $\eta$ viscosities capture the damping associated with volume and layer frictions respectively. After simplifying the Einstein field equation, we get two master equations in Eqs.~(\ref{eq : firstfin}, \ref{eq : secondfin}) which govern the metric perturbations inside the star at linear order in frequency. A crucial feature of these equations is that, at linear order in frequency, they only receive contributions from shear viscosity $\eta$ without any bulk viscosity contributions. We explicitly demonstrate that this is due to a peculiar fluid incompressibility in the static limit a.k.a adiabatic incompressibility, for non-barotropic perturbations, seen earlier in the Newtonian limit in Ref.~\cite{2014ARA&A..52..171O,Terquem:1998ya}. Physically, it indicates that the perturbed fluid packets preserve volume. This result is valid when the `low'-frequency expansion as done in this work is well-defined. Specifically, we argue in Sec.~\ref{sec : static_limit} that the low-frequency expansion in this paper is well-defined when the orbital frequency\footnote{More generally, the frequency of the tidal field. In a binary, it is roughly the orbital frequency.} $\omega$, is sufficiently smaller than the characteristic Brunt-V\"{a}is\"{a}l\"{a} frequency $N_\mr{ch}$, (the frequency of convective oscillations and characteristic frequency of the gravity ($g$-)modes) but much larger than the equilibrium-rate of m-Urca processes. 
The latter-requirement is almost always true, but the former is valid only during early-mid inspiral as $N_\mr{ch}\sim(150-700)$hz depending on the mass and equation of state~\cite{Lai:1993di,Andersson:2019ahb,Ranea_Sandoval_2018,Tran_2023}. The results in Subsec.~V.C in Ref.~\cite{HegadeKR:2024agt} %likely
probe the complementary regime where the frequency $\omega$ is much larger than the Brunt-V\"{a}is\"{a}l\"{a} frequency as they work with barotropic perturbations where it is identically zero through the star. A more detailed study is required to understand the transition between the regimes during inspiral.

\subsubsection{Love number and dissipation number for arbitrary compactness}

To get the Love number and dissipation numbers for neutron stars, we also need to solve the stellar exterior, where the metric perturbations may be reduced to the Regge-Wheeler (RW) equation. With the master equations for the interior of the star, we first numerically obtain the RW variable $\phi$ and its radial derivative at the stellar surface. The information of the stellar interior is fully captured by the quantity  %logarithmic radial derivative of RW variable at stellar surface
\begin{alignat}{3}
    T &\equiv \frac{r}{\phi} \frac{d\phi}{dr_*} \Bigg|_{r=R} %\frac{r}{r_*} \frac{d \log X}{d \log r_*}\Bigg|_{r= R},
    \label{eq : RWT_1}
     = T_0 - i R \omega T_1 +\mc{O}(\omega^2).
\end{alignat}
$T_0$ and $T_1$ encode the conservative and dissipative tidal response respectively. $r_*$ here is the tortoise coordinate. The on-shell information, i.e. the scattering phase shift can be extracted from the ratio of asymptotic expansion at infinity 
\begin{alignat}{3}
\phi(r)|_{r\rightarrow \infty} = A_{\ell,\omega}^{\mr{in}} e^{-i \omega r_*}  + A_{\ell,\omega}^{\mr{out}} e^{+i \omega r_*}.
\end{alignat}
with fixed $T$ boundary condition. We compute the ratio $A_{\ell,\omega}^{\mr{out}}/A_{\ell,\omega}^{\mr{in}}$ analytically to the desired order using the Mano-Suzuki-Takasugi (MST) method following Ref.~\cite{Casals:2015nja}. Subsequently, matching the SPT amplitude to EFT yields their relation to the static Love number and dissipation number. As a result, we provide in Eq.~\eqref{eq: all_compact_Love_Diss}, expressions for the rescaled Love number $k_2^E$, and  rescaled dissipation number $\nu_2^E$, in terms of RW variable $T$ in Eq.~\eqref{eq : RWT_1} for arbitrarily compact objects in GR. Our expression for $k_2^E$ is consistent with the expression in Ref.~\cite{Hinderer:2007mb} while the expression for $\nu_2^E$ has been obtained for the first time.

As a more quantitative study, we use Eq.~\eqref{eq : RWT_1} to compute the Love and dissipation number for various EoS(s) and compactness(s) in Table~\ref{table : dissLove}.\footnote{In versions 1 to 3 of this paper, the numerical values of $H_\omega^E$, $\nu_2^E$ and $\delta\mathcal{N}_{\rm GW}$ in Tables~\ref{table : dissLove} and \ref{table : waveform} were overestimated by a factor of $10^3$ due to an error in converting the shear viscosity in Eq.~(\ref{eq:SV_ee}) from cgs units. The formalism, the analytic results and the Love numbers are unaffected. The error was identified through a comparison with the independent calculation of Ref.~\cite{Vernizzi2026}, whose results agree with the corrected values presented here to within $0.1\%$ for identical stellar backgrounds and viscosity profiles.} The rescaled (w.r.t compactness) Love number $k_2^E$ and dissipation number $\nu_2^E$ are defined in Eq.~(\ref{eq : rescale}). 

As bulk viscosity does not contribute to dissipation number at 4PN, the dissipation numbers here are entirely due to shear-viscosity, the dominant source of which is $e-e$ scattering~\cite{Cutler_1987,Lai:1993di}. The viscosity scales inversely with square of the temperature, see Eq.~(\ref{eq:SV_ee}) and subsequently also the (rescaled) dissipation number(s) in the table. We thus multiply them with the factor $(T_K/10^5)^2$ to cancel this dependence leaving behind the (rescaled) dissipation number at $T=10^5K$. Here $T_K$ simply means temperature in Kelvin. Note that the dissipation number $H_\omega^E$ falls sharply with compactness (roughly as $\sim C^{-6}$), and temperature $\sim T^{-2}$.

\subsubsection{Effect on waveform due to tidal heating}

We then proceed to roughly quantify the effect on the waveform during inspiral due to tidal heating at leading 4PN order in the stationary phase approximation~\cite{Buonanno:2009zt,
Arun:2008kb,
Datta:2020gem,Saketh:2022xjb}. Specifically, we compute the change in the number of GW cycles, assuming the GW frequency to be twice the orbital frequency for a system of two neutron stars as 
\begin{alignat}{3}
&\delta\mathcal{N}_{\text{GW}}=\frac{\delta \phi[(GM\pi\omega_{f})^{1/3}]-\delta \phi[(GM\omega_i\pi)^{1/3}]}{\pi},\label{eq : dN_1}
\nnm \\& = \sum_{a=1,2}\frac{25 R_a^6 \nu_2^{E,a}}{512 m_a^3 (M-m_a) M^2 }\times GM(\omega_i-\omega_f).
\end{alignat}
Here $m_{a=1,2}$ are the masses of the neutron stars, and $M=m_1+m_2$. $R_{a=1,2}$ are the radii, and $\nu_2^{a=1,2}$ are the rescaled dissipation numbers defined in Eq.~(\ref{eq : rescale}).

As mentioned earlier, the bulk viscosity does not contribute at this order. The dominant contribution to the shear viscosity is from electron-electron scattering, which scales inversely with the square of the temperature of the neutron star. We thus evaluate Eq.~(\ref{eq : dN_1}) considering shear viscosity for identical neutron star binaries with various EoS(s) and compactness within the LIGO band. We also compare this with the contributions to the same quantity from other conservative 4PN contributions recently obtained in Ref.~\cite{Blanchet:2023soy}. This is given in Table~\ref{table : waveform} for various EoS(s) and compactness. Cold viscous neutron stars with relatively low compactness have the strongest imprint. However, even in the most favourable case considered here ($T=10^5$K and $M/R\approx 0.1$), tidal heating from shear viscosity changes the number of GW cycles by less than $10^{-2}$, roughly an order of magnitude below the other conservative contributions at 4PN. As the effect scales as $T^{-2}$, it is even smaller at the higher temperatures expected for inspiralling neutron stars. Shear viscosity therefore makes only a small contribution to tidal heating during inspiral.
\begin{widetext}
\begin{center}
\begin{table}
\begin{tabular}{| c | c| c|c| c |c | c|c |} 
 \hline
 \text{EoS} & M (in $M_{\odot}$) & R (in km) & $M/R$ & $k_2$ & $H_\omega^E\times(T_\mr{K}/10^5)^2$  & $(H_\omega^E)_{\text{NS}}/(H_\omega^E)_{\text{BH}}\times(T_\mr{K}/10^5)^2$ & $\nu^E_2\times (T_\mr{K}/10^5)^2$\\ [0.5ex] 
  \hline
  \multirow{5}{4em}{FSU2~\cite{Chen2014}}
     & 1.01 & 14.00 & 0.107 & 0.133 & 22.04 & 30.99 & $4.86\times10^{-5}$\\
  & 1.34 & 13.95 & 0.141 & 0.121 & 3.456 & 4.860 & $4.12\times10^{-5}$ \\ 
 & 1.71 & 13.95 & 0.180 & 0.099 & 0.617 & 0.868 & $3.18\times10^{-5}$ \\ 
  & 2.34 & 13.8 & 0.25 & 0.056 & 0.0478 & 0.0673 & $1.75\times10^{-5}$ \\
 \hline 
 \multirow{5}{4em}{GM1~\cite{GM}} & 1.01 & 12.85 & 0.115 & 0.140 & 17.97 & 25.27 & $6.40\times10^{-5}$ \\
  & 1.34 & 12.85 & 0.154 & 0.118 & 2.462 & 3.463 & $5.02\times10^{-5}$ \\
  & 1.71 & 12.7 & 0.199 & 0.089 & 0.395 & 0.556 & $3.67\times10^{-5}$ \\
  & 2.3 & 11.55 & 0.294 & 0.027 & 0.0139 & 0.0196 & $1.35\times10^{-5}$ \\
 \hline 
 \multirow{5}{4em}{HZTCS~\cite{HZTCS}} & 1.01 & 12.45 & 0.120 & 0.140 & 15.92 & 22.38 & $7.01\times10^{-5}$ \\
 & 1.34 & 12.65 & 0.156  & 0.122 & 2.512 & 3.532 & $5.47\times10^{-5}$\\
 & 1.71 & 12.66 & 0.200  & 0.100 & 0.440 & 0.619 & $4.17\times10^{-5}$\\
 & 2.34 & 12.60 & 0.274  & 0.044 & 0.0284 & 0.0399 & $1.80\times10^{-5}$\\
[1ex] 
 \hline
\end{tabular}
\caption{Tidal response characteristics for various EoS(s) and compactness for non-spinning neutron stars at $T=10^5 K$. Dissipation numbers are computed using shear-viscosity due to electron-electron scattering, which scales inversely with square of temperature. $T_K$ is the temperature of neutron star in Kelvin. Note that the dissipation number $H_\omega^E$ sharply falls with increasing compactness for each EoS. The rescaled dissipation number $\nu_2$ does not change as much. All dissipation numbers are smaller by a factor of $10^3$ than in versions 1 to 3 of this paper. The Love numbers obtained are consistent with known results.}
\label{table : dissLove}
\end{table}

\begin{table}
\begin{tabular}{| c | c| c|c|c| c |c | c|c|} 
 \hline
 \text{EoS} & M (in $M_{\odot}$) & R (in km) & $M/R$ & $H_\omega^E\times(T_\mr{K}/10^5)^2$   & $\nu^E_2\times(T_\mr{K}/10^5)^2$ & $\delta\mathcal{N}_{\text{GW}}\times(T_\mr{K}/10^5)^2$ & $\mathcal{N}^{\text{0PN}}_{\text{GW}}$ &  $\mathcal{N}^{\text{4PN}}_{\text{GW}}$ \\ [0.5ex] 
  \hline
  \multirow{5}{3em}{FSU2}
  & 1.01 & 14.00 & 0.107& 22.04 & $4.86\times10^{-5}$ & $-7.775\times10^{-3}$ & 4428.93 & -0.099  \\
  & 1.34 & 13.95 & 0.141& 3.456 & $4.12\times10^{-5}$ & $-1.618\times10^{-3}$ & 2764.77 & -0.106\\
   & 1.71 & 13.95 & 0.180& 0.617 & $3.18\times10^{-5}$ & $-3.69\times10^{-4}$ & 1841.51 & -0.108 \\
   & 2.34 & 13.8 & 0.25& 0.0478 & $1.75\times10^{-5}$ & $-3.7\times10^{-5}$ & 1091.46 & -0.101 \\ 
    \hline 
  \multirow{5}{3em}{GM1} & 1.01 & 12.86 & 0.115& 17.97 & $6.40\times10^{-5}$ & $-6.339\times10^{-3}$ & 4428.93& -0.099\\
  & 1.34 & 12.85 & 0.154 & 2.462 & $5.02\times10^{-5}$ & $-1.153\times10^{-3}$ & 2764.77 & -0.106\\
 & 1.72 & 12.70 & 0.199& 0.395 & $3.67\times10^{-5}$ & $-2.37\times10^{-4}$ & 1823.70 & -0.108\\
 & 2.30 & 11.55 & 0.294& 0.0139 & $1.35\times10^{-5}$ & $-1.1\times10^{-5}$ & 1123.38 & -0.102\\
   \hline 
  \multirow{5}{3em}{HZTCS} & 1.01 & 12.45 & 0.120& 15.92 & $7.01\times10^{-5}$ & $-5.620\times10^{-3}$ & 4428.93 & -0.099\\
    & 1.34 & 12.65 & 0.156& 2.512 & $5.47\times10^{-5}$ & $-1.176\times10^{-3}$ & 2764.77 & -0.106\\
     & 1.71 & 12.65 & 0.200& 0.440 & $4.17\times10^{-5}$ & $-2.65\times10^{-4}$ & 1841.51& -0.108\\
     & 2.34 & 12.60 & 0.274 & 0.0284 & $1.80\times10^{-5}$ & $-2.2\times10^{-5}$ & 1091.46& -0.101\\
 [1ex]
 \hline
\end{tabular}
\caption{Table showing correction to number of GW cycles due to tidal heating ($\delta \mathcal{N}_{\text{GW}}$) in a symmetric spinless NS-NS binary at fixed temperature $T=10^5K$. $\mathcal{N}_{\text{GW}}^{\text{0PN}}$ is the number of gravitational cycles due to leading order quadrupolar flux, at 0PN. $\mathcal{N}_{\text{GW}}^{4PN}$ is the contribution at 4PN due to other conservative post-Newtonian contributions~\cite{Blanchet:2023bwj}. The GW frequency evolves from 30hz to min($\omega_{\mr{ISCO}}$,1000hz), corresponding to orbital frequencies from 15hz to at most 500hz, consistent with the LVK band. The values of $H_\omega^E$, $\nu_2^E$ and $\delta\mathcal{N}_{\rm GW}$ are smaller by a factor of $10^3$ than in versions 1 to 3 of this paper.}
\label{table : waveform}
\end{table}
\end{center}
\end{widetext}

\subsection{Outline}
The rest of this paper is organized as follows. In Section~\ref{sec : EFT}, we provide a short review discussing the wave scattering formalism in EFT and SPT and the challenges involved isolating the tidal contribution to the amplitude in the latter. In Section~\ref{sec : bg profile}, we review the neutron star background Tolman–Oppenheimer–Volkoff (TOV) equations and the realistic EoS(s) and viscosity profile(s) we use. In Section~\ref{sec : stellar pert}, we derive the linear perturbation equations for fluid and metric, and reduce them to two master equations for the metric perturbations to linear-order in frequency. We show the absence of bulk viscosity in the resultant equations, and its link to the fluid being incompressible in the static limit. In Section~\ref{sec : matching}, we discuss the matching procedure between the neutron star interior and exterior along with the numerical techniques we use. We switch to the RW equation in the exterior of the star, and show how the interior perturbation equations can be integrated to obtain the boundary condition for the RW function at stellar surface. In Section~\ref{sec : scatter}, we compute the Raman scattering amplitude in terms of MST solutions to the RW equation outside the neutron star, and match with the EFT amplitude to provide expressions for the rescaled quadrupole tidal Love number and dissipation number to all order in compactness. In Section.~\ref{sec : dissLovecomp} We use this to compute the electric quadrupolar Love number and dissipation numbers for various EoS(s) and compactness. In Section.~\ref{Sec : quantqual}, we compute the change in number of GW cycles within the LIGO band due to tidal heating by shear viscosity, for various EoS(s) and compactness. We compare this with other conservative post-Newtonian contributions at various orders up to 4PN. We finally conclude in Section.~\ref{Sec : conc} with a discussion of the caveats in the current work, and potential future extensions and refinements.
\section{Wave Scattering Formalism for tides : short review}
\label{sec : EFT}
In this section, we review the tidal dissipation in the worldline EFT formalism. Most of the material here is well-known in the literature and we refer the reader to \cite{Goldberger:2004jt,Goldberger:2005cd,Porto:2007qi,Porto:2016pyg,Goldberger:2020fot,Goldberger:2022ebt, Goldberger:2022rqf, Saketh:2022xjb,Saketh:2023bul,Chia:2024bwc} for comprehensive reviews and detailed discussions.

A convenient starting point modeling a compact object including tidal effects in worldline EFT is by writing down a worldline action. The world-line is characterized by the position $z^{\mu}(\tau)$, $\tau$ being the proper-time. We also attach an orthonormal tetrad comprising of the 4-velocity $u^\mu=dz^\mu/d\tau$, and three space-like vectors $e_i^\mu$. In addition, tidal effects are incorporated by including multipolar moment degrees of freedom $Q_{ij}(\tau),\dots$. We can write
\begin{alignat}{3}
    S  =  \int d\tau &\Big[-M +\mc{L}_{\mr{int}}(Q_{ij}^{E/B},\dots) \nnm
    \\& - \frac{1}{2} Q_{ij}^E E^{ij} - \frac{1}{2} Q_{ij}^B B^{ij} + \cdots \Big] ~,
\end{alignat}
where the electric and magnetic tidal fields are defined as
\be
E_{ij} = u^\mu e_i^\nu u^\rho e_j^\sigma C_{\mu\nu\rho\sigma}\,~B_{ij} = u^\mu e_i^\nu u^\rho e_j^\sigma {}^* C_{\mu\nu\rho\sigma} ~,
\ee
and where $C_{\mu\nu\rho\sigma}$ is the Weyl tensor and ${}^* C_{\mu\nu\rho\sigma}$ stands for its dual. The generalization to higher order multipoles can be done by acting more derivatives on the tidal fields
\begin{equation}
    E_L = \partial_{\langle i_{L-2}} E_{ij\rangle} ~, \quad B_{L} = \partial_{\langle i_{L-2}} B_{ij\rangle} ~,
\end{equation}
with $i_L= i_1 \cdots i_\ell$ the multipole index. $\langle \cdots \rangle$ here represents that the contained indices should be symmetrized and any traces removed%symmetric-trace-free component of tensors
. In this paper, we will just focus on the quadrupolar electric part of the dynamics which corresponds to the $\ell=2$, polar perturbations upon the star. The complete treatment including magnetic field for black holes can be found in \cite{Saketh:2022xjb,Saketh:2023bul,Chia:2024bwc}. However, tidal response to polar perturbations is typically much more important for stellar bodies. For the purpose of getting the evolution of the quadupole moments $Q_{ij}^{E}$ of the compact objects, one can use the linear response theory 
\begin{equation}
    \langle Q_{ij}^E(\tau) \rangle = - \frac{1}{2} \int d\tau' G^{\rm ret}_{ij,kl} (\tau-\tau') E^{kl}(\tau') ~,
\label{eq:linear_response}
\end{equation}
where the retarded Green's function is given by 
\begin{equation}
    G_{ij,kl}^{\rm ret}(\tau-\tau') =  i \left\langle\left[Q_{ij}(\tau), Q_{kl}\left(\tau^{\prime}\right)\right]\right\rangle \Theta\left(\tau-\tau^{\prime}\right) ~.
\end{equation}
In frequency domain, the retarded Green's function admits a well-defined low-frequency expansion
\begin{equation}
    G_{ij,kl}^{\rm ret}(\omega) = F_2(\omega) \delta_{\langle ij \rangle, \langle kl \rangle} ~,
\label{eq:Green function}
\end{equation}
where
\begin{equation}
    F_2(\omega) = 2 (G M)^4 \Big( \Lambda^E + (i G M \omega) H_\omega^E + \cdots \Big) ~.
    \label{eq:low_exp}
\end{equation}
$\Lambda^E$ is known as the static Love number because it is time-reversal even and therefore corresponds to the conservative tidal deformations. Conversely, $H^E_\omega$ is time-reversal odd and therefore corresponds to nonconservative tidal effects. The tensorial structure of the response function is governed by the delta function
\begin{equation}
    \delta_{\langle ij \rangle,\langle kl \rangle} = \frac{1}{2} \Big(\delta_{ik} \delta_{jl} + \delta_{il} \delta_{jk} - \frac{2}{3} \delta_{ij} \delta_{kl} \Big) ~.
\end{equation}
From dimensional analysis, we can figure out that $\Lambda^E \sim (R/G M)^5$ while $H_\omega^E \sim (R/GM)^6$. Therefore, we find it convenient to parametrize the Love number and the dissipation number as
\begin{equation}
    \Lambda^E = \frac{2}{3} k_2^E \Big(\frac{R}{G M}\Big)^5 ~, \quad H^E_\omega = \frac{2}{3} \nu_2^E \Big(\frac{R}{G M}\Big)^6 ~,
    \label{eq : rescale}
\end{equation}
where $k_2^E$ is the well-known rescaled dimensionless Love number in the literature \cite{Hinderer:2007mb, poisson2014gravity}. Similarly, $\nu_2^E$ is the rescaled dimensionless dissipation number. We refer to $k_2^E$ ($\nu_2^E$) as the `rescaled' Love (dissipation) number when distinguishing them from $\Lambda^E$ ($H^E_\omega$). At the microscopic level, the tidal dissipation is related to the tidal lag time $\tau_d$ caused by the fluid kinematic viscosity $\nu_{\rm vis}$
\begin{equation}
    H_\omega^{E} \sim \Lambda^{E} \times \frac{\tau_d}{G m} ~, \quad \tau_d \sim \frac{\nu_{\rm vis}}{\nu_{\rm vis}^{\rm BH}} R ~,
\end{equation}
where for black holes with the same mass $\nu_{\rm vis}^{\rm BH} = 2 G M$.

In the framework of EFT, from the response function provided in Eq.~\eqref{eq:Green function}, we can further calculate the change in mass-energy  of the body due to tidal-heating (horizon energy flux for black holes) as%flux $dE/dt$ 
\begin{equation}
   \frac{dE_\mr{body}}{dt}=\frac{1}{2}M(G M)^5 H_\omega^E \dot{E}^{i j} \dot{E}_{i j}.
\end{equation}
In the stationary phase approximation, we can use the above formulae to compute the effect on the phase of the gravitational waveform. In the nonspinning case, $H_\omega^E$ affects the waveform-phase starting from 4PN \cite{Saketh:2022xjb,Chia:2024bwc, HegadeKR:2024slr}.

The above EFT description can be universally applied to study any compact objects. However, the specfic value of $\Lambda^E$ and $H_\omega^E$ will depend on the internal structure of the objects. In this paper, we are going to fix these coefficients for nonspinning neutron stars by matching the GW scattering amplitude obtained in the EFT, known as gravitational Raman scattering or gravitational Compton amplitude with the one obtained from stellar perturbation theory \cite{Saketh:2022xjb, Saketh:2023bul}. 
In the worldline theory or EFT, the gravitational Compton amplitude due to induced quadrupolar tides is given by the diagram
\begin{equation}
\begin{gathered}
    \begin{tikzpicture}[line width=1,photon/.style={decorate, decoration={snake, amplitude=1pt, segment length=6pt}}]
    \draw[line width = 1, photon] (0,0.5) -- (1,1.5);
    \draw[line width = 1, photon] (0,-0.5) -- (1,-1.5);
    \draw[line width = 1, dashed, double] (0,-0.5) -- (0,0.5);
    \filldraw[fill=black, line width=1.2](0,0.5) circle (0.15) node[left]{\small$Q\;$};
    \filldraw[fill=black, line width=1.2](0,-0.5) circle (0.15) node[left]{\small$Q\;$};
    \end{tikzpicture}
\end{gathered}
    =  i \frac{\omega^4}{16 M_{\mathrm{pl}}^2} G_{ij,kl}^{\rm ret}(\omega) \epsilon_h^{ij}\left(\boldsymbol{k}_{\mathrm{in}}\right) \epsilon_{h'}^{* k l}\left(\boldsymbol{k}_{\mathrm{out}}\right) 
    \label{diag : scat}
\end{equation}
Here, $\epsilon^{ij},\epsilon^{kl}$ are graviton polarization tensors. The scattering amplitude in Eq.~(\ref{diag : scat}) can be related to the scattering phase shift and the degree of absorption after transforming to partial wave basis \cite{Saketh:2023bul} with $\ell=2$ and even parity $P = +1$, corresponding to polar modes
\begin{alignat}{3}
    1- \eta_{2,+}^{\rm EFT} e^{2 i \delta_{2,+}^{\rm EFT}} =& i\mc{A}_{\mr{EFT}}(\ell=2,\omega,+\rightarrow 2,\omega,+) \nnm \\=& - i \frac{\omega^5}{80M_{\mathrm{pl}}^2 \pi}F_2(\omega)(1+GM\omega\pi)\nnm\\&+\mc{O}(\omega^7)\label{eq : ph2amp}
\end{alignat}
In the stellar perturbation theory, the scattering phase and degree of absorption may be computed as follows: in the non-spinning case, the metric perturbation equations in the vacuum outside the star may be reduced to a single source-free Schrodinger-like equation, the RW equation, for both axial and polar perturbations\footnote{We restrict to studying polar perturbations in this work. Axial perturbations couple weakly to neutron stars \cite{Binnington:2009bb}.}. The RW equation is given by
\begin{alignat}{3}
\label{eq : RW2}
   & \frac{d^2 \phi(r)}{d r_*^2} + \left[ \omega^2  - f(r) V(r)\right] \phi(r) =0, \\& V(r)=\Bigg(\frac{\ell(\ell+1)}{r^2}- \frac{6 M}{r^3} \Bigg),~r_*=r+2M\log(r-2M) \nnm.
\end{alignat}
We can solve the RW equation in the vacuum outside a star if the boundary conditions at the surface of the star are provided. The boundary conditions can be obtained by numerical integration of the perturbation equations inside the star, as we show in Sec.~\ref{sec : matching}. Once the RW equation is solved corresponding to the boundary conditions, we take the limit $r\rightarrow \infty$, where it becomes a simple wave equation and is solved by a linear combination of incoming and outgoing waves. Restricting to monochromatic perturbations, we can write
\begin{alignat}{3}
\phi(r)|_{r\rightarrow \infty} = A_{\ell,\omega}^{\mr{in}} e^{-i \omega r_*}  + A_{\ell,\omega}^{\mr{out}} e^{i \omega r_*}.
\end{alignat}
The scattering phase, and the degree of absorption are then given by 
\begin{alignat}{3}
\eta_{\ell}e^{2i\delta_\ell}= (-1)^{\ell+1} \frac{A_{\ell,\omega}^{\mr{out}}}{A_{\ell,\omega}^{\mr{in}}}.
\end{alignat}
The scattering amplitude in SPT can then be obtained using Eq.~(\ref{eq : ph2amp}), as 
\begin{alignat}{3}
1-(-1)^{\ell+1} \frac{A_{\ell,\omega}^{\mr{out}}}{A_{\ell,\omega}^{\mr{in}}} &= i\mc{A}_{\mr{SPT}}(\ell,\omega,+\rightarrow 2,\omega,+) \label{eq : ph2ampPT}
\end{alignat}

At this stage, one might be tempted to directly compare the amplitude obtained above with that in Eq.~(\ref{diag : scat}). However, this is complicated by the fact that the scattering amplitude in the stellar perturbation theory does not just involve the contribution of tidal effects. The GWs also scatter off static background metric (Scwarzschild metric) due to the star. This is not an issue when it comes to extracting the dissipative tidal response, as shown in Ref.~\cite{Saketh:2022xjb}, because non-tidal effects do not contribute to dissipation. However, it complicates computing the conservative tidal response\footnote{Although the focus of this work is dissipative effects. It is a useful test of principle to see whether the conservative tidal response can be extracted, such as Love number using the approach.}. This is shown diagrammatically in Eq.~(\ref{diag : split}). In this, the nontidal (tidal) contributions to the amplitude, are colored in blue (red) and referred to as far (near) zone-contributions following the notation in Ref.~\cite{Ivanov:2022qqt}.

In this work, to also extract the leading conservative tidal response, we isolate the tidal effects from the amplitude computed in stellar perturbation theory in two ways: \textbf{1)} By subtracting the full amplitude obtained using Eq.~(\ref{eq : ph2ampPT}) for a neutron star, with that of a Schwarzschild black hole of the same mass, thus effectively removing all the common contributions (due to GWs scattering off the background gravitational field) leaving behind only\footnote{The vanishing Love number of the Schwarzschild black hole aids us here.} the difference in their tidal contributions. \textbf{2)} By making use of the near-far factorization valid for general $\ell$, as shown in Ref.~\cite{Ivanov:2022qqt}, to isolate the tidal contribution to scattering amplitudes in the stellar perturbation theory.

Once the tidal contribution to the scattering amplitude is isolated, we can compare it with the EFT amplitude in Eq.~(\ref{eq : ph2amp}) and derive the Love number and dissipation number. Both approaches yield the same formulae for Love number and leading dissipation number as presented in Sec.~\ref{sec : scatter}.

\begin{widetext}
\begin{equation}
\begin{gathered}
    \begin{tikzpicture}[line width=1,photon/.style={decorate, decoration={snake, amplitude=1pt, segment length=6pt}}]
    \draw[line width = 1,  dashed] (0,-1.5) -- (0,1.5);
    \draw[line width = 1, photon] (0.25,0.4) -- (1,1.5);
    \draw[line width = 1, photon] (0.25,-0.4) -- (1,-1.5);
    \draw[line width = 1, dashed, double] (0,-0.5) -- (0,0.5);
    \filldraw[fill=White, line width=0.15](0,0.0) circle (0.5) node[]{NS};
    \end{tikzpicture}
\end{gathered}
\quad = \textcolor{blue}{\overbrace{\quad \begin{gathered}
    \begin{tikzpicture}[line width=1,photon/.style={decorate, decoration={snake, amplitude=1pt, segment length=6pt}}]
    \draw[line width = 1,  dashed] (0,-1.5) -- (0,1.5);
    \draw[line width = 1, photon] (1.0,0) -- (2.0,1.5);
    \draw[line width = 1, photon] (1.0,0) -- (2.0,-1.5);
    \draw[line width = 1, photon] (0,0.0) -- (1.0,0.0);
    \filldraw[fill=black, line width=1.2](0,0.0) circle (0.15) node[left]{\small$m\;$};
    \end{tikzpicture}
\end{gathered}+\quad \cdots\quad}^{\text{Far zone}}}
\\\quad \textcolor{red}{\overbrace{+\quad \begin{gathered}
    \begin{tikzpicture}[line width=1,photon/.style={decorate, decoration={snake, amplitude=1pt, segment length=6pt}}]
    \draw[line width = 1,  dashed] (0,-1.5) -- (0,-0.5);
    \draw[line width = 1,  dashed] (0,0.5) -- (0,1.5);
    \draw[line width = 1, photon] (0,0.5) -- (1,1.5);
    \draw[line width = 1, photon] (0,-0.5) -- (1,-1.5);
    \draw[line width = 1, dashed, double] (0,-0.5) -- (0,0.5);
    \filldraw[fill=black, line width=1.2](0,0.5) circle (0.15) node[left]{\small$Q\;$};
    \filldraw[fill=black, line width=1.2](0,-0.5) circle (0.15) node[left]{\small$Q\;$};
    \end{tikzpicture}
\end{gathered}\quad+\quad\cdots\quad}^{\text{Near zone}}}
\label{diag : split}
\end{equation}
\end{widetext}
\section{Stellar interior and viscosity}
\label{sec : bg profile}
In this work, we consider a nonspinning spherically-symmetry neutron star, characterized by a particular density and pressure profile inside. The bounding surface of the star is where the pressure/density drops to zero. Being spherically symmetric, we can set up a polar-like coordinate system ($t,r,\theta,\phi$) where the bounding surface is at a fixed radial coordinate `$r$' and density and pressure are independent of angular coordinates ($\theta, \phi$). We can then write down the most general spherically symmetric metric in these coordinates as~\cite{Kojima:1992ie} 
\begin{equation}
g_{\mu\nu}^{(0)} = 
\begin{pmatrix}
-e^{\nu(r)} & 0 & 0 &0  \\
0 & e^{\lambda(r)} & 0 & 0 \\
0 & 0 & r^2 & 0 \\
0 & 0 & 0 & r^2 \sin^2(\theta)
\end{pmatrix},
\label{eq : bgmetric}
\end{equation}
in the $(-,+,+,+)$ metric signature. Outside, Birkhoff's theorem ensures that the metric becomes Schwarzschild, and we have $\lambda(r)=-\nu(r)$ and $e^{-\lambda(r)}=(1-2M/r)$ for $r>R$, where $R$ is the radial coordinate of the stellar surface. $M$ is the Schwarzschild mass of the neutron star. 

The unknown functions $\nu(r)$ and $\lambda(r)$ may be related to the stellar density and pressure profiles through the Einstein equation if the stress energy tensor of the stellar matter is known. When the star is static and unperturbed, the stellar matter may be treated as an ideal fluid and we can write
\begin{alignat}{3}
T_{\mu\nu}^{(0)} = \rho u_{\mu}u_{\nu} + p P_{\mu\nu},
\label{eq : ideal}
\end{alignat}
where $P_{\mu\nu}= g_{\mu\nu}+u_{\mu}u_{\nu}$. Here $u^{\mu}$ is the four velocity of the fluid elements comprising the star. When  the (unperturbed and nonspinning) star is static, the four velocity is only along the time-like killing vector, and we can write $u^{\mu}=(1/\sqrt{-g_{tt}},0,0,0)=(e^{-\nu/2},0,0,0)$. The four velocity satisfies $u^2=-1$. Now, we can compute the Einstein tensor $G_{\mu\nu}$ using the metric in Eq~(\ref{eq : bgmetric}) and plug it into the Einstein equation $G_{\mu\nu} = \kappa T_{\mu\nu}$, where $\kappa = 8 \pi$.

This yields the well known Tolman–Oppenheimer–Volkoff (TOV) equations, governing the equilibrium configurations of nonrotating relativistic  neutron stars in hydrostatic equilibrium \citep{Tolman:1939jz,Oppenheimer:1939ne},
\begin{eqnarray}
\frac{dM(r)}{dr} &=& 4 \pi \rho(r) r^2 ~, \nonumber \\
\frac{dp(r)}{dr} &=& - \frac{[p(r) + \rho(r)] [M(r)+4 \pi r^3 p(r)] }{r(r-2 m(r))}~, \nonumber \\
\label{eq:tov}
\end{eqnarray}
in units of $c = G = 1$. Here, we have defined $M(r)=(1-e^{-\lambda(r)})(r/2)$, which may be regarded as the mass (energy) bounded by radius $r$ in the nonrelativistic limit. It can be regarded as a parameter which reduces to the Schwarzschild mass (i.e., mass inferred from the Schwarzschild metric outside) at the surface of the star. As it is, Eq.~(\ref{eq:tov}) is not enough to solve for the density and pressure profile. One more relation relating the density and pressure is required, also known as the equation of state (EoS). This crucial quantity depends on the constituents of the neutron star interior and their interactions. Given an EoS, one can integrate the TOV equations from the centre of the star to the surface with the boundary conditions of  vanishing mass, $m|_{r=0}=0$, at the centre of the star, and~a vanishing pressure, $p|_{r=R}=0$, at the surface. By changing the value of the central pressure, one obtains the mass-radius relation of  neutron stars. \\

As we go from surface towards the core of  neutron stars, the density increases rapidly, and the constituents of the  neutron star matter as well as the strong interaction between them are unknown at such high density. So, one must resort to theoretical models to describe the behaviour of dense matter at such high density and compare the predictions of neutron star observable properties with multi-messenger astrophysical data to put constraints on the models and their parameter space. Different theoretical schemes, ab-initio and phenomenological, have been applied to describe dense neutron star matter, including both non-relativistic and relativistic approaches~\citep{EoSreview}. While ab-initio models, such as Chiral Effective Field Theory (Chiral EFT)~\citep{Drischler2020}, provide a reliable microscopic description at sub-saturation densities, the calculations cannot be extended to supra-saturation densities, given our lack of understanding of baryonic three-body forces as well as the possible degrees of freedom (e.g. appearance of strange baryons or ``hyperons", condensates of mesons or deconfined quarks). Phenomenological models, on the other hand, describe baryon-baryon interactions via meson exchange, whose couplings can be fitted to reproduce nuclear saturation properties. In this work, we consider the broad class of the phenomenological Relativistic Mean Field (RMF) model for our core EoS. %\sout{Such models have been widely applied to reproduce both nuclear matter properties as well as neutron star astrophysical data.}
%In the basic framework of the RMF theory~\citep{Chen2014} for nuclear matter, the basic degrees of freedom include baryons (e.g., protons, neutrons, strange baryons or ``hyperons"), interacting via the exchange of mesons, leptons (e.g., electrons and muons) and the photon. In this model, the baryon-baryon interaction is mediated by the exchange of scalar ($\sigma$), vector ($\omega$), and isovector ($\rho$) mesons. In the mean-field approximation, the meson fields are replaced by their mean value, which is obtained assuming baryons are in the ground state. The isoscalar and isovector nuclear coupling constants between the mesons and the baryons are fixed in terms of the different bulk parameters of infinite nuclear matter. For example, the isoscalar nucleon-meson coupling parameters are determined by fixing the nuclear saturation parameters: nuclear saturation density ($n_0$), binding energy per nucleon  at saturation ($E_{sat}$), incompressibility ($K_{sat}$) and the effective nucleon mass ($m^*$) at saturation. On the other hand the isovector couplings are fixed to the symmetry energy ($E_{sym}$) and slope of symmetry energy ($L_{sym}$) at saturation ~\citep{Hornick,Chen2014}. 
The uncertainty in the behaviour of nuclear empirical quantities at higher densities is reflected in the  uncertainty in the determination of the RMF model parameters. Recent studies~\citep{Ghosh2022,Ghosh2022_b,Traversi_2020} have used Bayesian scheme within the RMF model to constrain the parameter space  by imposing information from Chiral EFT at low densities and recent multi-messenger astrophysical observations of neutron stars at high densities, still leaving room for a large uncertainty in determining the behaviour of dense matter at high densities.
To see the dependence and variance of our results with EoS and neutron star compactness, we  have considered a few  standard parametrisations within the RMF model; namely - `FSU2'~\citep{Chen2014}, `GM1'~\citep{GM} and `HZTCS'~\citep{HZTCS}. For the crust of the neutron star which is mainly composed of aggregates of nuclei, we use density-dependent relativistic mean-field model parametrization `DD2'~\citep{Crust} in beta equilibrium as taken from the Compose online database~\citep{Compose}. These EoS satisfy the current multi-messenger observation data of the neutron stars and also, spans the uncertainty range in the EoS. The astrophysical properties (mass and radius) corresponding to these EoSs are given in first three columns of Tables~\ref{table : dissLove},~\ref{table : waveform}.\\

The various EoSs used to solve the TOV equations in Eq.~(\ref{eq:tov}), for obtaining the background stellar profile assume that the various constituents of stellar matter are in chemical equilibrium. However, when the star is perturbed by (say) an external tidal field, the equilibrium condition may be violated depending on the rates of various reactions and the timescale of external perturbations. This can cause the stress energy tensor to deviate slightly from the form of the ideal fluid-stress-energy tensor in Eq.~(\ref{eq : ideal}). In particular, we need to include shear-viscous and bulk-viscous terms in the stress energy tensor, and write
\begin{equation}
\label{eq: energy_stress with viscosity main}
    T_{\mu\nu} = \rho u_{\mu} u_{\nu} + (p - \zeta \nabla_{\alpha} u^\alpha) P_{\mu\nu} - 2 \eta P^\alpha_\mu P^\beta_\nu \sigma_{\alpha\beta}
\end{equation}
where $\sigma_{\mu\nu}$ is the shear tensor
\begin{equation}
    \sigma_{\mu\nu} \equiv \frac{1}{2} \Bigg(\nabla_\mu u_\nu + \nabla_\nu u_\mu - \frac{2}{3} g_{\mu\nu} \nabla_\alpha u^\alpha \Bigg) ~.
\end{equation}
Here, $\zeta$ and $\eta$ are the bulk and shear viscosity respectively. Note that $\zeta$ appears next to the fluid divergence $\nabla_\alpha u^\alpha$. The fluid-divergence may be related to the rate of volumetric change of the fluid packets comprising the neutron stars, and thus the bulk viscosity may be seen as a friction-force resisting such volumetric changes. Shear-viscosity $\eta$ on the other hand enters alongside the shear tensor, which quantifies the shear-deformation of the fluid in the neutron star, and thus is the friction-force resisting shear deformation. There are several out-of-equilibrium viscous processes inside neutron stars that can lead to bulk and shear viscous contributions in the stress energy tensor. The main source of shear viscosity inside the neutron is the momentum transport due to the scattering of the constituents particles like electron, proton, muons and neutron. The shear viscosity generated by these microscopic processes in neutron stars depends on the local density ($\rho$) and the temperature (T) profile. Although neutron stars are born very hot $\sim 10^{11}$K, they cool down rapidly due to neutrino emission~\cite{Yakovlev_2005}. For neutron stars in a binary about to merge, that are very old, we expect the core temperature to be $10^5 - 10^6$K~\cite{Lai:1993di,Haskell:2012vg}. In these temperature regions, the dominant source of shear viscosity comes from the $e-e$ scattering and an approximate fitting formula for the strength is given as~\citep{Flower_1976,Cutler_1987}
\begin{equation}\label{eq:SV_ee}
    \eta_{ee} = 6\times 10^6\rho^2T^{-2} \text{\hspace*{5mm} gm cm$^{-1}$s$^{-1}$}
\end{equation}
where $\rho$ is the density and $T$ is the temperature. There might be other sources of shear viscosity such as neutron-neutron scattering, neutron-muon scattering but they are very sub-dominant at these low temperatures~\citep{Shternin_2008}. Bulk viscosity may also originate from leptonic weak interactions such as direct-Urca and modified-Urca (m-Urca) reactions at the neutron star interior. The bulk viscosity originating from m-Urca reactions is given by~\citep{Sawyer1989}
\begin{equation}\label{eq:BV_mUrca}
    \zeta_{\rm{m-Urca}} = 6\times 10^{-61}\rho^2T^6/\omega^2 \text{\hspace*{5mm} gm cm$^{-1}$s$^{-1}$},
\end{equation}
where $\omega$ is the perturbation frequency\footnote{The factor $1/\omega^2$ seems to make the low-frequency expansion ill-defined. However, the expression in Eq.~(\ref{eq:BV_mUrca}) is obtained in the limit where the equilibrium time scale for Urca processes is taken to be much larger than the orbital time-period $\sim\omega$. Thus, one cannot naively take the limit $\omega\rightarrow 0$.\label{foot:m-Urca}}. From the relative strength, we can see that the m-Urca bulk viscosity for $\omega = 1$ khz only dominate over the shear viscosity at very high temperature $10^9-10^{10}$K. As also mentioned earlier, nonleptonic weak interactions involving hyperons may also produce stronger bulk viscosity than other sources at low temperatures~\cite{Ghosh:2023vrx}. Given the EoS, the bulk viscosity coefficient ($\zeta$) can be calculated in terms of the relaxation time ($\tau$) of the nonleptonic weak process~\cite{Lindblom2002}
\begin{equation}\label{BV}
    \zeta_{hyp} = n_B\frac{\partial P}{\partial n_n} \frac{dn_n}{d n_{B}}\frac{\tau}{1+(\omega\tau)^2},
\end{equation}
where  $P$ is the  pressure, $n_B$ the total baryon number density, $n_n$ the neutron density. For this particular reaction, $\tau$ is proportional to $T^{-2}$ and for frequency $\sim 100$ hz, $\zeta_{hyp}$ reaches its maximum value around $10^8$K~\cite{Chatterjee2006,Ghosh:2023vrx}. Although we expect the binary neutron star core temperature to be low ($\sim 10^5$K) at the early stages of inspiral, the viscous dissipation will cause heating and increase in the temperature. Lai (1994)~\citep{Lai:1993di} has shown that due to the shear viscous dissipation of the dominant $f$-mode energy during binary inspiral, the temperature can reach up to~$\sim 10^7$K. All viscous sources considered in this work are confined to the core of the neutron stars. For the crust, the composition and the state of matter are different requiring separate physics for viscosity sources~\cite{Ofengeim_2015,Yakolev_2005,Yakovlev:2018jia}. In this work, we fix the viscosity at the crust to be identically zero.

\section{Stellar perturbation theory including viscosity}
\label{sec : stellar pert}
We now subject the stellar fluid to polar metric perturbations, and derive the relevant perturbation equations to linear order in frequency. In this work, we restrict to weak external perturbations and thus restrict to linear order in them.
\subsection{Metric and matter perturbations}
The metric given earlier in Eq.~(\ref{eq : bgmetric}) is now linearly perturbed in a manner that depends on all coordinates. The time-dependence may be simplified by considering monochromatic perturbations, i.e., we thus consider all perturbations to have a separable time-dependence as $\exp(-i\omega t)$. This is equivalent to Fourier transforming a generic linear perturbation with respect to time, and then restricting to a single frequency in the Fourier domain. Similarly, dependence on the angular coordinates ($\theta, \phi$) may be simplified using a spherical harmonic decomposition, and then restricting to a given $\ell,m$ mode. The metric perturbations may be further decomposed into polar and axial modes. However, in this work, we restrict our attention to polar modes at $\ell=2$, which have even parity, and induce the electric quadrupolar tidal response. 

Then, the total metric is given by $g_{\mu\nu} = g_{\mu\nu}^{(0)}+\delta g_{\mu\nu}$, where $g_{\mu\nu}^{(0)}$ is the unperturbed metric given in Eq.~(\ref{eq : bgmetric}), and $\delta g_{\mu\nu}$, the perturbation at a given frequency and $\ell,m$ mode. It is given by \cite{Lindblom:1983ps}\footnote{There are some conventional differences here w.r.t. Ref.~\cite{Lindblom:1983ps}.} 
\begin{alignat}{3}
  & \delta g_{\mu\nu} dx^\mu dx^\nu =-e^{-i\omega t} \bar{r}^\ell
  Y_{\ell m}[ H_0 (r)e^\nu dt^2  \\& - 2 i %\left(\frac{r}{R}\right)
  \omega H_1(r) dt dr + e^{\lambda} H_2(r) dr^2 +K(r) r^2 d\Omega^2],\nonumber 
\end{alignat}
where $\bar{r}=r/R$ inside the star ($r<R$), $R$ being the radial coordinate of the unperturbed stellar surface and $\bar{r}=1$ outside, and $d\Omega^2 = d\theta^2+\sin^2(\theta)d\phi^2$.

In addition to metric perturbations, we have matter perturbations within the star. This involves density and pressure perturbations, as well as the shift in fluid elements leading to its deformation. We first parametrize the latter as follows. We can describe the fluid displacements with a vector field $\xi^{\mu}$, so that the perturbed worldline of the fluid element at $x^{\mu}$ is described by $x^{\mu}+\xi^{\mu}$ in the perturbed configuration. We set $\xi_\mu u^\mu=0$. The spatial components may then be parametrized  as follows for polar perturbations. \cite{Lindblom:1983ps}
\begin{alignat}{3}
\label{eq:fluid displacement}
\xi^t& =0, ~ \xi^r = e^{-i\omega t}%\Big(\frac{r}{R}\Big)^\ell
\frac{\bar{r}^\ell}{r}e^{-\frac{\lambda}{2}}W~Y_{\ell m},\\
\{\xi^\theta,\xi^\phi\} &= - e^{-i\omega t} %\Big(\frac{r}{R}\Big)^\ell
\frac{\bar{r}^\ell}{r^2}V\{\partial_\theta Y_{\ell m}, \sin^{-2}\theta \partial_\phi Y_{\ell m}\} \nnm ~.
\end{alignat}
The fluid displacement is related to the perturbation of the four-velocity of the fluid through the relation
\begin{alignat}{3}
\delta u^{\mu} &= \frac{D\xi^\mu}{D\tau}= (\delta u^0, -i\omega \xi^r, -i \omega \xi^\theta,- i \omega \xi^\phi), 
\end{alignat}
with the gravitational redshift factor 
\begin{equation}
    \delta u^0 = -e^{- i \omega t} \frac{e^{-\nu/2}\bar{r}^\ell}{2} H_0 %\Big(\frac{r}{R}\Big)^\ell
    Y_{\ell m}.
\end{equation}

We mentioned in Sec.~\ref{sec : bg profile} that the form of the stress energy tensor is deformed due to perturbations, leading to shear and bulk viscous terms. Similarly, the perturbations also shift the values of pressure and density. They can be obtained from the laws of thermodynamics, and the equation of state. Notice that for relativistic neutron stars, for each fluid element with a fixed number of baryons ($N_B=N_n+N_p$) obeying local charge neutrality ($N_p+N_e$)\footnote{More generally, we can have processes that change Baryon number and introduce additional charged species in the mix (for e.g., Hyperons~\cite{Ghosh:2016qgn}). In that case, if the additional particles are also in chemical equilibrium we can simply define new quantities that are conserved to replace $n_B$. The charge neutrality condition can be simply redefined to include all charged particle species.}, we have the first law of thermodynamics
\begin{alignat}{3}
\Delta e = T\Delta s - p\Delta v + \sum_i \mu_i \Delta x_i,
\end{alignat}
where $e,s,v$ is the average energy, entropy, and volume per baryon. $T\Delta s=\Delta q$ is the heat absorbed/lost by the fluid packet. $x_i=n_i/n_B$ here is the fractional number of each particle species. As we assume that the unperturbed star is in chemical equilibrium, the last term vanishes under the constraints. Furthermore, the predominant contribution to $\Delta q$, is due to neutrino cooling and black body radiation, both of which have timescales much longer than the inspiral timescale~\cite{Rees1992,Ghosh:2023vrx}. Thus, we can just focus on the isentropic perturbations, i.e. $\Delta s=0$ and write, $\Delta e=-p \Delta v$. Now, the Lagrangian variation of energy density for a given fluid element can be obtained as
\begin{alignat}{3}
\Delta \rho &= \Delta(e/v) = \Delta e/v - \rho (\Delta v/v) \\& = -(p+\rho)(\Delta v/v), \nonumber 
\end{alignat}
Since we are tracking a fluid packet with fixed baryon number, we can relate the fractional change of volume to the fractional change of baryon number density, and subsequently to the fluid-displacement vector $\xi^\mu$ via
\begin{alignat}{3}
    \label{eq :dnbn}\frac{\Delta v}{v} =& - \frac{\Delta n_B}{n_B} = D_\mu \xi^\mu + \frac{1}{2} \delta \Big[{}^{(3)} g\Big] \Big/ {}^{(3)} g ~,\\ =& -Y_{\ell m}%\left(\frac{r}{R}\right)^l
    \frac{\bar{r}^\ell e^{-i\omega t}}{2}\Bigg(H_2+2 K-\frac{2\ell(\ell+1)V}{r^2}\nnm \\&-\frac{2 e^{-\frac{\lambda}{2}}[(\ell+1)W+rW']}{r^2}\Bigg)\nnm
\end{alignat}
where $D_\mu$ is the spatial covariant derivative and ${}^{(3)}g$ is the determinant of the spatial metric. Alternatively, one can also write $\Delta v/v = e^{\nu/2}(-i\omega)^{-1}\nabla_\mu u^\mu$.
\subsubsection{pressure perturbation}
\label{sec : Lpres}
Computing the change in pressure is more subtle, and depends on equilibrium equation of state as well as the rates of various chemical reactions in the star, and how they compare with the orbital-time scale\footnote{The orbital time scale is the inverse of the orbital frequency in a binary. Here, it is $1/\omega$, where $\omega$ is the frequency of the tidal perturbation.}. 
Generally, pressure can be viewed as a function of Temperature, density and particle fraction $p(T,\rho,x_i)$. Simply by changing of variables, we can also write the above function as $p(T,\rho,\mu_i)$. This relation between pressure and other intrinsic thermodynamic quantities is referred to as the equation of state relation. In Neutron stars, the effect of temperature in this relation is only relevant when they are very hot $T\geq 10^{10}K$~\cite{Raithel_2019,Raduta:2021coc}. Thus, the pressure is generally assumed to be a two parameter function as  $p\equiv p(\rho,x_p)$ or $p(\rho,\mu)$~\cite{Counsell:2023pqp,Andersson:2019mxp}. The Lagrangian variation in a fluid packet with fixed baryon number $n_B$ may then be written as 
\begin{alignat}{3}
    \Delta p(\rho,x_i) &= \Big(\frac{\partial p}{\partial \rho}\Big)\Big|_{x_i} \Delta \rho + \sum_i \Big(\frac{\partial p}{\partial x_i}\Big)\Big|_{\rho} \Delta x_i ~,\nnm \\
    \Delta p(\rho,\mu_i) &=\Big(\frac{\partial p}{\partial \rho}\Big)\Big|_{\mu_i} \Delta \rho + \sum_i \Big(\frac{\partial p}{\partial \mu_i}\Big)\Big|_{\rho} \Delta \mu_i.
\end{alignat} 
In the general case, one needs to consider the various chemical reactions in the star and their rates to solve for the change in particle fraction or chemical potential~\cite{Counsell:2023pqp}. However, the problem is simplified in the following two extreme regimes. The first extreme is the fast reaction regime where the reaction-time scale for the relevant nuclear reactions are much smaller than the inspiral timescale, in which case the reactions are able to maintain equilibrium even in the perturbed fluid packets. The fractions of various particle species continuously shift to maintain equilibrium. In this case, we can simply set $\Delta \mu_i = 0$, and therefore the change in pressure w.r.t change in density has the same relation as in the background profile because the unperturbed star was also in chemical equilibrium as well . Thus, we have
\begin{equation}
    \textbf{fast reactions}: ~ \Delta p = \Big(\frac{\partial p}{\partial \rho}\Big)\Big|_{\mu_i} \Delta \rho.
\end{equation}
The other regime is the slow reaction regime where the reaction timescales are too large compared to the orbital time-period, and the chemical composition is essentially frozen in the perturbed fluid packet. We can then neglect the change in particle number fraction, i.e. $\Delta x_i=0$ 
and write
\begin{equation}
    \textbf{slow reactions}: ~ \Delta p = \Big(\frac{\partial p}{\partial \rho}\Big)\Big|_{x_i} \Delta \rho.
\end{equation}
%Based on the above two equations, 
We can generally define an %isentropic
index $\gamma$ to be 
\begin{equation}
    \gamma \equiv  \frac{\Delta p/p}{\Delta \rho/(p+\rho)} = - \frac{\Delta p / p}{\Delta v  / v} ~.
\end{equation}
In the fast reaction regime, this index is fully determined by the background since it is also at chemical equilibrium. i.e., $(dp/dr)/(d\rho/dr)=(\partial p/\partial \rho)|_{\mu_i}$, and so 
\begin{equation}
     \gamma_{\rm eq}  %\Big(\frac{\partial \log p}{\partial \log v}\Big)_{s, \mu_i}
     = \frac{\rho + p}{p} \frac{p'}{\rho'}=c_\mr{eq}^2\frac{p+\rho}{p} ~,
     \label{eq : badgamma}
\end{equation}
where ${}^\prime$ corresponds to the radial derivative. For slow reactions instead, $\gamma$ is given by
\begin{equation}
    \gamma_{\rm ins}   %- \frac{\Delta p / p}{\Delta v / v} = \Big(\frac{\partial \log p}{\partial \log v}\Big)_{s, x_i} 
    = \frac{\rho + p }{p} \Big(\frac{\partial p}{\partial \rho}\Big) \Big|_{x_i}=c_\mr{ins}^2\frac{p+\rho}{p} ~.
    \label{eq : goodgamma}
\end{equation}
We have defined the two sound speeds $c_\mr{ins}$ and $c_{\mr{eq}}$ for later use. We will see further below in Sec.~\ref{sec : static_limit}, that the value of $\gamma$ plays a crucial role in determining the validity of the low-frequency expansion in this work. It also affects the frequency of the $g$-modes as shown in the Newtonian limit in Appendix.~\ref{App A}, also see Refs.~\cite{Counsell:2023pqp}. This is due to its relation with convective stability as discussed in Ref.~\cite{HegadeKR:2024agt}.

We show in Sec.~\ref{sec : static_limit} that when Eq.~(\ref{eq : badgamma}) holds, the low-frequency expansion of the perturbation equations in the manner performed in this work is ill defined. This is also the limit in which all the $g$-mode frequencies all collapse to $\omega=0$~\cite{Cowling:1941nqk}. The star is marginally unstable under convective perturbations in this case~\cite{HegadeKR:2024agt}. This turns out to be crucial in the following discussions regarding the contribution of bulk viscosity to tidal heating in Sec.~\ref{sec : static_limit}. 

Realistically, the various nuclear reactions have vastly different timescales, and $\gamma$ lies in between the two extremes. A detailed discussion of the rate of nuclear reactions and its effect on the Lagrangian change in pressure, and on the $g$-mode frequencies may be found in Ref.~\cite{Counsell:2023pqp}. However in our work, we find that the linear-in-frequency perturbation equations are actually insensitive to the value of $\gamma$ provided the low-frequency expansion is valid, which is true when Eq.~(\ref{eq : badgamma}) does not hold. This is true during inspiral within the LIGO band, as the equilibrium-timescale for the dominant m-Urca processes is very large compared to inspiral time scale(s)~\cite{Sawyer1989}. Instead, Eq.~(\ref{eq : goodgamma}) corresponding to the frozen-composition is a good approximation in this case~\cite{Lindblom:1983ps}. Thus, it is important to keep in mind going forward that the `static limit' $\omega \rightarrow 0$ in this work corresponds to neglecting $\omega$ in the perturbation equations but keeping it much higher than the rate of the dominant m-Urca reaction-rates so that Eq.~(\ref{eq : goodgamma}) holds.

\subsubsection{Eulerian perturbation to pressure and density}
Finally, once we obtain the Lagrangian changes in pressure and density, we need the corresponding Eulerian quantities to get the perturbation in pressure and density at a given coordinate point. Thus we have,
\begin{alignat}{3}
\delta \rho &= \Delta \rho - \xi^\mu \nabla_\mu p, \nnm \\
\delta p &= \Delta p - \xi^\mu \nabla_\mu p.  \label{eq : pd_pert_1}
\end{alignat}
For spherically symmetric configurations, this reduces to 
\begin{alignat}{3}
\delta \rho &= \Delta \rho - \xi^r \rho', \nnm \\
\delta p &= \Delta p - \xi^r p'.  \label{eq : pd_pert}
\end{alignat}
These may be directly plugged into the Einstein equation as perturbations to the pressure and density.

Once we have all the above quantities, we are able to derive the linearized Einstein field equation $\delta G_{\mu\nu} = \kappa \delta T_{\mu\nu}$ along with the associated conservation laws $\delta (\nabla_{\mu}T^{\mu}{}_{\nu})=0$ with the stress energy tensor including the contributions from fluid viscosity given in Eq.~\eqref{eq: energy_stress with viscosity}.

\subsection{Equations governing the perturbations to linear order in frequency}
\label{sec : mastereq}

We will restrict ourselves to the $\ell=2$ mode. Before proceeding further, we introduce another function $X(r)$ defined as \cite{Lindblom:1983ps}
\begin{equation}
\label{eq : defX}
    X(r) Y_{\ell m} e^{-i\omega t} \equiv - \frac{\Delta n_B}{n_B} \bar{r}^{-\ell}  %\Big(\frac{r}{R}\Big)^{-\ell}
    (e^{\nu/2}p \gamma-i\omega\zeta) ~,
\end{equation}
which captures the variation of baryon number density, along with some convenient factors.

We first consider the combination $G_{\theta \phi} = \kappa T_{\theta\phi}$, which yields the relation to eliminate $H_2(r)$,
\begin{alignat}{3}
H_2=H_0 - 4 i \omega \kappa e^{-\nu/2}V\eta.
\label{eq : H2sol}
\end{alignat}
We then consider $G_{12}=\kappa T_{12}=0$, yielding 
\begin{alignat}{3}
H_1&= \frac{r}{6} \Big[-2  H_2 +2  r K' -  K(-7+e^\lambda + e^\lambda r^2 \kappa p) \nnm \\& +2 \kappa e^{\frac{\lambda}{2}}W(p+\rho)\Big],\label{eq : H1sol} 
\end{alignat}
Eqs.~(\ref{eq : H2sol}) and (\ref{eq : H1sol}) are useful for eliminating $H_2(r)$ and $H_1(r)$ in all other equations. 
\begin{widetext}
We then consider the conservation law $\nabla_{\mu}T^{\mu}{}_{\theta}=0$, to get the relation
\begin{alignat}{3}
\nnm \omega^2 V(r) &= - \frac{1}{2}e^\nu H_0 - \f{e^{\nu-\frac{\lambda}{2}}(-1+e^{\lambda}+e^{\lambda} r^2 \kappa p)}{2 r^2} W +\f{e^\f{\nu}{2}}{p+\rho} X - i\omega \Bigg[\frac{e^{-\lambda+\f{\nu}{2}}(e^\f{\lambda}{2}W-r V')\eta'}{r(p+\rho)} \\& + \frac{\eta e^{-\lambda+\f{\nu}{2}}(e^\lambda r^2 H_0+ 8 e^\lambda V + 4 e^\f{\lambda}{2}W- 7 r V' - e^\lambda r V' + e^\lambda  r^3 \kappa \rho V' - 2 r^2 V'')}{2 r^2 (p+\rho)} \nnm \\& +\frac{e^{\f{\nu}{2}}X \eta}{3(e^\f{\nu}{2}p\gamma+i\omega \zeta)(p+\rho)} \Bigg]. 
\label{Eq : Vsol}
\end{alignat}
The above relation will be used to simplify other equations containing $V(r)$. We can now proceed to derive the equations governing the metric perturbations $H_0$ and $K$ to linear order-in-frequency. To that end, we start with $G_{23}-\kappa T_{23}=0$, and eliminate $H_1(r)$ and $H_2(r)$ using Eqs.~(\ref{eq : H2sol}, \ref{eq : H1sol}) and truncate to $\mc{O}(\omega^2)$ to get
\begin{equation}
-\frac{2 K}{r}+ H_0'+\frac{ H_0 \left(\kappa  r^2 p e^{\lambda }+e^{\lambda }+1\right)}{r}-K'\textcolor{orange}{-\frac{2 i \kappa  \omega  \eta e^{-\frac{\nu}{2}}
   \left(\kappa  r^2 p e^{\lambda} V+r V'+e^{\lambda} V+V-e^{\frac{\lambda }{2}} W\right)}{r}} = \mathcal{O}(\omega^2)
   \label{eq : firstfin}
\end{equation}

Then, we consider the combination $r(G_{22}-\kappa T_{22}) - 2 (G_{23}-\kappa T_{23})=0$, and use  Eqs.~(\ref{eq : H2sol}, \ref{eq : H1sol}, \ref{Eq : Vsol}) to eliminate $H_2$, $H_1$ and $V$, and then truncate to $\mc{O}(\omega^2)$ get
\begin{equation}
    \begin{aligned}
        & \quad  H_0'+\frac{H_0 e^{\lambda} \left(\kappa  r^2 (p -\rho)+8\right)}{2r}+\frac{1}{2} K' \left(e^{\lambda (r)} \left(\kappa  r^2
   p+1\right)-3\right)+\frac{K \left(e^{\lambda} \left(\kappa  r^2 p-1\right)-3\right)}{r} \\& \textcolor{orange}{-\frac{i \kappa  \omega  \eta e^{-\frac{\nu}{2}}}{2 r}[8\left(4 e^{\lambda}+1\right) V+V' \left(\kappa  r^3 e^{\lambda} \rho-r e^{\lambda}+r\right)-2 r^2 V''+4 e^{\frac{\lambda}{2}} W + r^2 e^{\lambda} (H_0-4 K)]}\\&\textcolor{orange}{ -\frac{i \kappa  r \omega  \eta e^{\lambda-\nu}}{3 p \gamma } X-i e^{-\frac{\nu}{2}}\kappa \omega \eta '(r) \left(e^{\frac{\lambda}{2}} W-r V'\right)}%\\& -\textcolor{orange}{\frac{i \kappa  \omega  e^{-\frac{\nu}{2}} \left(\eta  e^{\lambda} \left(-2 r^2 (2 K+H_0)+r \left(\kappa  r^2 \rho-1\right) V'+32 V\right)-2 r^2 \eta V''-2 r^2 \eta ' V'+r \eta V'+8 \eta V+2 e^{\frac{\lambda }{2}} W \left(r \eta'+2 \eta\right)\right)}{2 r}} \\
   = \mathcal{O}(\omega^2) ~.\end{aligned}
   \label{eq : secondfin}
\end{equation}
\end{widetext}
Eqs.~(\ref{eq : firstfin}) and (\ref{eq : secondfin}) are still coupled to the fluid perturbations $W$ and $V$, but those coupling terms only arise in terms relevant when there is viscosity which are all at linear order in frequency (colored in \textcolor{orange}{orange}). Thus, when solving perturbatively in frequency, one can replace the fluid perturbations in the terms containing viscosity with the static solutions (when $\omega=0$). We derive the expressions for fluid perturbations in the static limit below in Sec.~\ref{sec : static_limit}.

However, another curious observation from Eqs.~(\ref{eq : firstfin}), (\ref{eq : secondfin}) is the absence of the bulk viscosity $\zeta$. The source terms (in \textcolor{orange}{orange}) contain only shear viscosity and its derivative ($\eta,~\eta'$). We show in  the next Sec.~\ref{sec : scatter} that the linear-in-frequency corrections to the perturbation equations contribute to the leading dissipation number which is relevant at 4PN in the flux (subsequently waveform). Thus, the absence of bulk viscosity tells us that it plays no role in dissipation at leading order. This is due to the vanishing of the fluid divergence $\propto X$ in the static-limit as we  also show below in Sec.~\ref{sec : static_limit}.

\subsection{Static limit, incompressibility and validity of the small frequency expansion}
\label{sec : static_limit}
We can solve for the fluid perturbations $W$ and $V$ in terms of the metric perturbations $H$ and $K$ in the static limit as follows. We first again consider the conservation laws $\nabla_{\mu}T^{\mu}{}_{\nu}=0$, but without plugging in the explicit formula for pressure and density perturbations from Eq.~(\ref{eq : pd_pert}). We also set $\omega=0$ and work strictly in the static limit. The conservation law $\nabla_{\mu}T^{\mu}_{\theta}=0$ yields the formula
\begin{alignat}{3}
\delta p &= - %\Big(\frac{r}{R}\Big)^\ell
\frac{\nu^\ell H_0^{\omega^0}}{2}(p+\rho)  Y_{\ell m} e^{-i\omega t},
\label{eq : pres_pert2}
\end{alignat}
where we are using the superscript $\omega^0$ to denote that the quantity has been evaluated in the static limit, i.e., $\omega=0$. Inserting this into $\nabla_{\mu}T^{\mu}_{r}=0$, we can get an expression for the density perturbation in the static limit as 
\begin{alignat}{3}
\delta \rho  &= - %\Big(\frac{r}{R}\Big)^\ell \times
\frac{\nu^\ell H_0^{\omega^0}}{2}(\rho+p)\frac{\rho'}{p'} Y_{\ell m}e^{-i\omega t}~.
\label{eq : dens_pert2}
\end{alignat}
Note that the often assumed statement in the literature, that $\delta p = \delta \rho \times c_\mr{eq}^2$, where $c_{eq}$ is the equilibrium sound speed in Eq.~(\ref{eq : badgamma}), arises automatically in the static limit, without requiring additional assumptions. Now, we demand consistency between the expressions for density and pressure perturbations obtained in Eqs.~(\ref{eq : pres_pert2}), (\ref{eq : dens_pert2}), with those obtained earlier in terms of the fluid perturbations in Eq.~(\ref{eq : pd_pert}). To that end, we rewrite Eq.~(\ref{eq : pd_pert}) by substituting in Eqs.~(\ref{eq : pres_pert2}), (\ref{eq : dens_pert2}) as 
\begin{alignat}{3}
&\frac{H_0^{\omega^0}}{2}(p+\rho) =  e^{-\f{\nu}{2}} X^{\omega^0} + e^{-\f{\lambda}{2}}W^{\omega^0}\frac{p'}{r},\label{eq : WXstateqn} \\
&\frac{H_0^{\omega^0}\rho'}{2p'}(\rho+p)=  e^{-\f{\nu}{2}} X^{\omega^0} \f{p+\rho}{p\gamma}\nnm + e^{-\f{\lambda}{2}}W^{\omega^0}\frac{\rho'}{r}~.
\end{alignat}
The above equation can only be satisfied either for 
\begin{alignat}{3}
X^{\omega^0} =0, ~ W^{\omega^0} =  \frac{H_0^{\omega^0} (\rho+p) e^{\f{\lambda}{2}}r}{p'} ~,
\label{eq : WXsol}
\end{alignat}
or
\begin{alignat}{3}
\gamma  &= \frac{p'}{\rho'}\frac{p+\rho}{p}\label{eq: barotropicEoS}.
\end{alignat}
For the first case, the definition of $X(r)$ in Eq.~(\ref{eq : defX}) reveals that the fractional volume change $dv/v=-dn/n$ vanishes in the static limit. In other words, the fluid becomes incompressible when $\omega \rightarrow 0$. As the bulk viscosity only contributes when the fluid is compressible, this explains the curious disappearance of bulk viscosity at leading order in frequency in Eqs.~(\ref{eq : firstfin}), (\ref{eq : secondfin}). Additionally, we now also have an expression for $W^{\omega^0}(r)$.  This was shown analogously in the Newtonian limit in Ref.~\cite{2014ARA&A..52..171O}.

For the second case, Eq.~\eqref{eq: barotropicEoS} along with the discussion in Sec.~\ref{sec : Lpres} implies that the fluid perturbation is strictly barotropic, meaning the pressure is effectively the same function $p(\rho)$ of the density in both unperturbed and perturbed configurations \cite{PhysRevD.43.1768,ipser1985eulerian}. Additionally, we are unable to obtain unique solutions for $W^{\omega^0}$, $X^{\omega^0}$ implying a degeneracy in the static limit. This is explained by noting that, from the eigen-mode analysis, this is the limit in which all $g$-mode frequencies condense to 0. This leads to infinite dimensional null space of the linear fluid perturbations. As a result, the static fluid configuration can be any linear combination of the $g$-mode eigen-functions. As famously argued by Cowling in \cite{Cowling:1941nqk}, the star is in neutral equilibrium under indefinitely slow perturbations. In other words, we are unable to find a unique-static limit configuration to perform a Taylor expansion about it. It may be possible to use a different expansion scheme to obtain an analytical approximation to the perturbation equations in this case as well, but we do not tackle that here.

However, as discussed in Sec.~\ref{sec : Lpres}, the perturbations are barotropic only when all chemical reactions are fast enough to always maintain equilibrium. The dominant m-Urca processes however are not rapid enough to do so \cite{Sawyer1989}. Thus, for realistic neutron stars, $\gamma$ is instead given by Eq.~(\ref{eq : goodgamma}) and we can accept the solution in Eq.~(\ref{eq : WXsol}). Nevertheless, it is reasonable to assume that the difference between this $\gamma$ and its value in the barotropic case in Eq.~(\ref{eq: barotropicEoS}) likely sets the limits on the validity of the low frequency expansion in this work. More concretly, the perturbing frequency $\omega$, should be smaller than a characteristic frequency scale associated with the difference between $\gamma$ in Eq.~(\ref{eq : goodgamma}) from its barotropic value in Eqs.~(\ref{eq: barotropicEoS},~\ref{eq : badgamma}).

Such a characteristic
frequency scale may be obtained from is the Brunt-V\"{a}is\"{a}l\"{a} frequency $N$, which in the Newtonian limit is given by~\cite{1992ApJ...395..240R}
\begin{equation}
    N =  g\Bigg(\frac{1}{c_\mr{eq}^2}  - \frac{1}{c_\mr{ins}^2} \Bigg)^{1/2} ~,
    \label{eq : bvf}
\end{equation}
where $g=GM(r)/r^2$ is the local acceleration due to gravity inside the star. $c_{\mr{eq}}$ and $c_{\mr{ins}}$ were defined in Eqs.~(\ref{eq : goodgamma}, \ref{eq : badgamma}). $N$ is the oscillation frequency for local convective oscillations inside the star\footnote{Assuming the star is stable under convective oscillations, otherwise $N$ is imaginary. We do not consider that case in this work.}, and proportional to the difference between $\gamma$ (from Eq.~(\ref{eq : goodgamma}) from its barotropic value. $N$ varies with stellar radius, but a  characteristic value, $N_\mr{ch}$
may be computed inside the neutron star. This can be done by substituting typical values of the parameters entering Eq.~(\ref{eq : bvf}), yielding $N_\mr{ch}\sim 500$hz~\cite{1992ApJ...395..240R}. Alternatively, the $g$-mode with the largest resonant frequency may be used as an estimate for $N_\mr{ch}$, typically yielding $N_\mr{ch}\sim 300$hz~\cite{Lai:1993di,Andersson:2019ahb}. Strictly speaking, the linear-in-frequency expansion in this work
may be taken as valid when $\omega\ll N_\mr{ch}$.\footnote{This is more clearly seen in the Newtonian limit, (see Appendix.~\ref{App A} or Ref.~\cite{Terquem:1998ya}}).
Tidal dissipation in the complementary regime where $\omega\gg N_\mr{ch}$ was probed numerically in Ref.~\cite{HegadeKR:2024agt}, where the barotropic case $N=0$ everywhere inside the star, was considered. Realistically, a binary would start in the former regime $\omega\ll  N_\mr{ch}$, and its orbital frequency would eventually cross and exceed the characteristic Brunt-V\"{a}is\"{a}l\"{a} frequency in middle and late inspiral respectively. A more refined study is needed to understand the transition and its implications. In this work, we accept the solution in Eq.~(\ref{eq : WXsol}) as the static limit and work to linear order in frequency about this, and use the Brunt-V\"{a}is\"{a}l\"{a} frequency to estimate its limits of validity.

Yet another subtlety to consider when taking the static limit is that the presence of viscosity adds another frequency scale to the system, as shown in Ref.~\cite{HegadeKR:2024agt}, given by 
\begin{alignat}{3}
\Omega_{\mr{vis}}=\frac{p R}{\eta \boldsymbol{|\xi|}},~\mr{or}~\frac{p R}{\zeta \boldsymbol{|\xi|}}
\end{alignat}
where $|\boldsymbol{\xi}|\leq R$ is the magnitude of typical displacement vector of the fluid due to perturbations. The results obtained via the small-frequency expansion in the presence of viscosity should then be a suitable approximation when
\begin{alignat}{3}
\frac{\omega}{\Omega_{\mr{vis}}} = \frac{\omega \eta |\boldsymbol{\xi}|}{p R}~\mr{or}~\frac{\omega \zeta |\xi|}{p R}\ll 1
\end{alignat}
We have checked that this condition is comfortably valid for inspiral frequencies (10-1000hz) for all (most) sources of shear (bulk) viscosity throughout the star\footnote{In the crust, this condition can be more difficult to satisfy as $p\rightarrow 0$. In this work, we set viscosity to be identically $0$ in the crust.}. The important exception to this is the bulk viscosity due to non-leptonic weak interactions involving hyperons which were shown in Refs.~\cite{Lindblom2002,Chatterjee2006,Ghosh:2023vrx} to exhibit a `resonance' at certain temperatures and frequencies, at which point $\zeta(r)$ can get so large inside the star that this approximation no longer holds for select ranges of temperatures during inspiral. The static limit itself is likely to be significantly modified due to viscosity in this case. However, in this case, the whole scheme of incorporating viscosity by just adding terms linear in the gradient of 4-velocity may be insufficient~\cite{Giovanni_2023}. Thus, in this work, we do not consider this regime and work with the assumption that the viscosity is small enough to only perturbatively affect the small-frequency tidal response of the neutron star.

Thus, having concluded that the fluid in realistic neutron stars for any EoS is incompressible in the static limit, we can use the vanishing of $X^{\omega^0}(r)$ to get the expression for $V^{\omega^0}(r)$ from Eqs.~(\ref{eq : defX},~\ref{eq :dnbn}) to get
\begin{alignat}{3}
\nnm V^{\omega^0}(r) &= \frac{e^{-\f{\lambda}{2}}}{12}[e^\f{\lambda}{2}r^2 H_0^{\omega^0} + 2 e^\f{\lambda}{2}r^2 K^{\omega^0} \\& - 2(3W^{\omega^0}+rW^{\omega^0}{}')]
\label{eq : Vsol}
\end{alignat}

This completes the discussion of the static limit of the perturbations inside a neutron star. We can now substitute the expressions for $V$, $W$ and $X$ in the viscous terms at linear order in frequency (coloured in \textcolor{orange}{orange}) in Eqs.~(\ref{eq : firstfin}), (\ref{eq : secondfin}), in terms of $H_0$ and $K$ in the static limit, and numerically integrate the two first order differential equations to solve for $H_0$ and $K$ in the stellar interior.
\section{Numerical integration and matching}
\label{sec : matching}
In the master equations in Eqs.~(\ref{eq : firstfin}), (\ref{eq : secondfin}), we first substitute the linear-in-frequency expansions as
\begin{alignat}{3}
\nnm H_0(r) &= H_0^{\omega^0}-i\omega H_0^{\omega^1}+\mc{O}(\omega^2),  \\
K(r) &= K^{\omega^0}-i\omega K^{\omega^1}+\mc{O}(\omega^2), \label{eq : HKWVexp}
\\
W(r) & = W^{\omega^0} + \mc{O}(\omega),~ V(r)= V^{\omega^0}+\mc{O}(\omega).\nnm
\end{alignat}
Then, collecting terms at the same order in $\omega$, the master equations split into sub-equations for the static part and the linear-in-frequency part respectively. The static equations need to be first solved, to obtain the source terms (in \textcolor{orange}{orange}) in the master  Eqs.~(\ref{eq : firstfin}), (\ref{eq : secondfin}). Then, the linear-in-frequency pieces may be obtained subsequently. We discuss them separately below.

\subsection{Static part}
\label{Sec : statsolve}
The metric perturbations in the static limit obey the equations
\begin{alignat}{3}
&H_0^{\omega^0}\Bigg(\frac{1+e^{\lambda}}{r}+e^{\lambda}r\kappa p\Bigg)-\frac{2}{r}K^{\omega^0} \label{eq : first_fin_stat} +H^{\omega^0}{}'-K^{\omega^0}{}' =0,
\\ & -\f{3+e^\lambda(1- r^2 \kappa p)}{r}K^{\omega^0} -\frac{3}{2}K^{\omega^0}{}' \nnm  +H_0^{\omega^0}{}'\\& + \frac{e^\lambda [8+r^2 \kappa (p+\rho)]}{2r}H_0^{\omega^0} +\frac{e^\lambda}{2}(1+ r^2 \kappa p)K^{\omega^0}{}' \nnm\\&  =0. \label{eq : second_fin_stat}
\end{alignat}
The above equations may be further reduced to a single second-order differential equation for $H_0^{\omega^0}$. We have verified that doing so yields the same equation (up to difference in conventions) as in Ref.~\cite{Hinderer:2007mb}. 
\subsubsection{Initial conditions}
Eqs.~(\ref{eq : first_fin_stat}), (\ref{eq : second_fin_stat}) comprise two first order differential equations for the functions $H_0^{\omega^{0}}(r)$ and $K^{\omega^{0}}(r)$. To integrate them in the stellar interior, we also need to know the initial value(s), i.e., two parameters characterizing the boundary condition. We can obtain this near $r=0$, by solving the perturbation equations around the origin. We can choose without loss of generality that $H_0^{\omega^0}(0)=1$. Then, plugging in the Taylor expansions for $H_0^{\omega^{0}}(r)$ and $K^{\omega^{0}}(r)$ around $r=0$, we obtain the expressions
\begin{alignat}{3}
H_0^{\omega^0}(r) &\approx 1+ \frac{r^2}{84}\Bigg[-\kappa (33  p_c + \rho_c)+ \frac{18 \rho''_c}{3 p_c+\rho_c}\Bigg], \nnm \\ 
K^{\omega^0}(r) &\approx 1+ \frac{r^2}{14}\Bigg[-\kappa (2  p_c - \rho_c)+ \frac{3 \rho''_c}{3 p_c+\rho_c}\Bigg],\label{eq : statHKnear0}
\end{alignat}
valid to $\mc{O}(r^3)$. Here $\rho_c=\rho(0)$ ($p_c=p(0)$) are the unperturbed central density (pressure). Similarly $\rho''_c=\rho''(0)$ is its second radial derivative at $r=0$. Note that $\rho'(0)=p'(0)=0$. Equipped with the initial conditions in Eq.~(\ref{eq : statHKnear0}), we can integrate from a point close to $r=0$ to the surface. Thus, solving for $H^{\omega^0}$, $K^{\omega^0}$ throughout the interior of the star.

\subsection{Linear-in-frequency part}
The linear-in-frequency part of the metric variables obeys the same master equations as in Eqs.~(\ref{eq : firstfin}), (\ref{eq : secondfin}), but with the source terms computed using the static quantities. We just outlined the process of solving for the $H^{\omega^0}_0$ and $K^{\omega^0}$ in Sec.~\ref{Sec : statsolve}. The fluid perturbations in the static case, i.e., $W^{\omega^0}$ and $V^{\omega^0}$, can be obtained using Eqs.~(\ref{eq : WXsol}), (\ref{eq : Vsol}). We have already shown that $X(r)$ vanishes in the static limit. Thus, the source terms, colored in orange, in the master equations are fully specified. 

\subsubsection{Initial conditions}
To perform the integration from a point near $r=0$, we once again need the near-origin solutions for $H^{\omega^1}_0$ and $K^{\omega^1}$. Similar to the static case, this can be done substituting the Taylor expansions near $r=0$ for the metric and fluid perturbations. Alongside, we also make use of the relations in Eqs.~(\ref{eq : WXsol}), (\ref{eq : Vsol}). This yields the Taylor series expansions to $\mc{O}(r^3)$. We provide the expressions in Eq.~(\ref{eq : linHKnear0}).
\begin{widetext}
\begin{alignat}{3}
H_0^{\omega^1}(r) &=  \frac{e^{-\f{\nu_c}{2}}r^2}{350 \kappa (3p_c+\rho_c)^4}\Bigg\{2\eta_c[50 \kappa^2 (3p_c+\rho_c)^4+15 \kappa (18 p_c^2+9 p_c\rho_c + \rho_c^2)\rho''_c-351 (\rho''_c)^2+75(3p_c+\rho_c)\rho^{(4)}_c] \nnm \\& - 75(3 p_c+\rho_c)[\kappa(21p_c^2+22 p_c\rho_c+5 p_c^2)+6 \rho''_c]\eta''_c+150(3p_c+\rho_c)^2\eta^{(4)}_c\Bigg\},\label{eq : linHKnear0} \\ 
K^{\omega^1}(r) &\approx \frac{6e^{-\f{\nu_c}{2}}\eta_c}{3p_c+\rho_c}+ \frac{e^{-\f{\nu_c}{2}}r^2}{350 \kappa (3p_c+\rho_c)^4}\Bigg\{150(3p_c+\rho_c)^2\eta^{(4)}_c \nnm +150 (3p_c+\rho_c)\eta''_c[\kappa(21 p_c^2+10 p_c \rho_c +\rho_c^2)- 3 \rho''_c]\\& - \eta_c[50\kappa^2(2p_c-3\rho_c)(3p_c+\rho_c)^3+45 \kappa(51 p_c^2 +32 p_c\rho_c+5p_c^2)\rho''_c-702 (\rho''_c)^2+150 (3 p_c+\rho_c)\rho^{(4)}_c]\Bigg\},\nnm
\end{alignat}
where we have set $H_0^{\omega^1}(0)=0$, without loss of generality. Here $\eta_c=\eta(0)$ is the shear viscosity at the origin. Note that $\eta'(0)=0$, due to Eq.~(\ref{eq:SV_ee}). 
\end{widetext}
Note that, due to the expansions in Eq.~(\ref{eq : HKWVexp}), the linear-in-frequency parts are responsible for making the initial conditions for the metric perturbation functions complex. The equations governing the static fields in Eqs.~(\ref{eq : first_fin_stat}), (\ref{eq : second_fin_stat}) also do not have any complex terms in them. Thus, in the absence of shear viscosity, it is possible to choose the overall normalization such that $H_0$ and $K$ are purely real. The presence of shear viscosity necessarily renders them complex. 
\subsection{Matching at surface with the RW function}

Equipped with the initial conditions for the metric perturbations to linear order in frequency given in Eqs.~(\ref{eq : statHKnear0}), (\ref{eq : linHKnear0}), we can integrate the master equations perturbatively to solve for the functions $H_0$ and $K$ to linear order in frequency up to the stellar surface. It is trivial to establish their continuity at the stellar surface, and thus we obtain their values just outside the star.

Outside the star, as mentioned before, the vacuum perturbation equations governing $H_0$ and $K$ may be reduced to a single Schr\"odinger-like equation, the RW equation, Eq.~(\ref{eq : RW2}). This is accomplished using the relations in Eqs.~(59), (60) in Ref.~\cite{Kojima:1992ie}, rewritten here to $\mathcal{O}(\omega^2)$ :
\begin{alignat}{3}
    \nnm \phi & =\frac{-1}{(n+1) M}[\{-n(n+1) r(r-3 M)\} K  \\& + (n+1) r(n r+3 M) e^{-\lambda} H_{0}] ~, \label{eq : HKtoX}\\ 
    \nnm
    \phi'& = \frac{1}{-(n+1) M}[\{n(n+1)\}\{(n+1) r-3 M\} K .  \\
      & \quad .+\{-n(n+1)^2 r-3(n+1) M e^{-\lambda}\} H_0] ~, \nnm
\end{alignat}
where $n= (\ell-1)(\ell+2)/ 2$.
The above two equations can be combined into a dimensionless one via logarithmic derivative of the RW variable w.r.t the radial coordinate
\begin{alignat}{3}
    T &\equiv \frac{r}{\phi} \frac{d\phi}{dr_*} \Bigg|_{r=R} = \frac{r}{r_*} \frac{d \log \phi}{d \log r_*}\Bigg|_{r= R}.
    \label{eq : RWT}
\end{alignat}
This quantity will serve as the boundary condition for solving the RW equation outside the star. To determine the leading-order dissipative response, it is sufficient for us to restrict our discussion to linear order in frequency
\begin{equation}
\label{eq:star_bc}
    T \equiv T_0 - i R \omega T_1 +\mc{O}(\omega^2)~.
\end{equation}
In Sec.~\ref{sec : scatter}, we will see that the leading Love number and dissipation number are fully determined by $T_0$ and $T_1$ respectively.
\section{RW scattering problem using MST}
\label{sec : scatter}
In this section, we relate this boundary condition of the RW scalar in Eq.~(\ref{eq : RWT}) to the static Love number and the leading dissipation number of the star. As outlined in Sec.~\ref{sec : EFT}, we accomplish this by considering the problem of monochromatic GWs scattering off the neutron star. In this case, sufficiently far from the star, i.e, in the limit $r\rightarrow \infty$, the RW equation takes the form
\begin{alignat}{3}
\label{eq : RW_inf}
   & \frac{d^2 \phi(r)}{d r_*^2} +  \omega^2  \phi(r) =0, 
\end{alignat}
which is solved by a linear combination of incoming and outgoing free-wave solutions, i.e, $\phi(r)= A_{\ell, \omega}^{\rm in} \exp(-i \omega r_*) + A_{\ell, \omega}^{\rm out}\exp(i\omega r_*)$. The quantity $A_{\ell, \omega}^{\rm out}/A_{\ell, \omega}^{\rm in}$ can be related to the static Love and leading dissipation numbers, as outlined in Sec.~\ref{sec : EFT}.

To accomplish this, we employ the analytical MST scheme for the RW equation, which was first presented in Ref.~\cite{Mano:1996mf}. In this work, we  follow the conventions and notations employed in Ref.~\cite{Casals:2015nja}. In the MST scheme, we can write down analytical solutions for the RW equation as a sum of infinite hypergeometric functions, convergent in different domains. 

In the near-zone, 
%i.e., $r\omega\ll 1$, 
we can write a general solution to the RW equation as
\begin{alignat}{3}
    \phi(r) = B_{\bar{\nu}} X_0^{\bar{\nu}}(r) + B_{-\bar{\nu}-1} X_0^{-\bar{\nu}-1} (r),
    \label{eq : nzexp}
\end{alignat}
where $X_0^{\bar{\nu}}$ and $X_0^{-\bar{\nu}-1}$ are near-zone Coulomb-type solutions with convergence radius $r_s = 2 G M \leq r < \infty$. $\bar \nu$ stands for ``renormalized" angular momentum. Moreover, these functions have asymptotic behavior $X_0^{\bar{\nu}} \sim r^{\bar{\nu}},X_0^{-\bar{\nu}-1} \sim r^{-\bar{\nu}-1}$ when $r \rightarrow \infty$. Their explicit expansions is given in Ref.~\cite{Casals:2015nja}. We can fix the ratio of $B_{\bar{\nu}}$ and $B_{-\bar{\nu}-1}$ at the surface of the neutron star using Eq.~(\ref{eq : RWT}). 

Solving the scattering problem requires connecting the near-zone solution to the far-zone, i.e., $ \infty \geq r > r_s$. Here, the solution may be written as a combination of incoming and outgoing waves. To connect this to the far-zone, we first switch to the far zone Coulomb-type solutions denoted by $X_C^{\bar{\nu}}, X_C^{-\bar{\nu}-1}$ in \cite{Casals:2015nja} with convergence radius $r_s < r \leq \infty$. In the overlap region, where $r_s < r < \infty$, we can match these two solutions and get
\begin{equation}
    X_0^{\bar{\nu}} = K_{\bar{\nu}} X_C^{\bar{\nu}} ~,  X_0^{-\bar{\nu}-1} = K_{-\bar{\nu}-1} X_C^{-\bar{\nu}-1} ~,
\end{equation}
where the coefficient $K_{\bar{\nu}}$ is given by Eq.(3.32) in Ref.~\cite{Casals:2015nja}. Now, we switch to the basis of incoming and outgoing waves using Eqs.(3.29, 3.34, 3.35) in Ref.~\cite{Casals:2015nja}. Then, in the limit $r\rightarrow \infty$, Eq.~(\ref{eq : nzexp}) takes the form
\begin{alignat}{3}
\phi(r)|_{r\rightarrow \infty} = A_{\ell,\omega}^{\mr{out}} e^{i\omega r_*}  + A_{\ell,\omega}^{\mr{in}} e^{-i\omega r_*},
\end{alignat}
where
\begin{alignat}{3}
\frac{A_{\ell,\omega}^{\mr{out}} }{A_{\ell,\omega}^{\mr{in}}} = \frac{(1+ie^{i\pi \bar{\nu}}\mc{K}_{\text{NS}})}{(1-ie^{-i\pi \bar{\nu}}\frac{\sin[\pi(\bar{\nu}+i\epsilon)]}{\sin[\pi(\bar{\nu}-i\epsilon)]}\mc{K}_{\text{NS}}}\frac{A_-^{\bar{\nu}}}{A_+^{\bar{\nu}}} e^{2i\epsilon \log(\epsilon)}
\label{eq : outbinf}
\end{alignat}
where $\epsilon=2GM\omega$, $\mc{K}_{\mr{NS}}=(B_{-\bar{\nu}-1}K_{-\bar{\nu}-1}/B_{\bar{\nu}} K_{\bar{\nu}})$ and expressions for $A_{-(+)}^{\bar{\nu}}$ given in Eqs.~(3.19, 3.38) in Ref.~\cite{Casals:2015nja}.

As mentioned in Sec.~\ref{sec : EFT}, we cannot immediately relate the ratio in Eq.~(\ref{eq : outbinf}) to the amplitude computed in EFT in Eq.~(\ref{diag : scat}). This is because we have not yet isolated the contributions of tidal effects to the scattering process. As it is, the ratio in Eq.~(\ref{eq : outbinf}) contains the combined effect of both tidal effects, and non-tidal effects (such as scattering of GWs off the gravitational background). Dissipative tidal effects may be easily extracted from the phase following Ref.~\cite{Saketh:2022xjb}, since non-tidal effects do not contribute to dissipation.

The leading conservative tidal response as well may be distilled in two ways. One way is to subtract the total-scattering amplitude of a neutron star with that of a Schwarzschild black hole with same mass, both computed in the perturbation theory framework. This cancels out all common contributions\footnote{Due to scattering of GWs off the background metric}, leaving behind only the difference in their tidal contributions. Then, using the known amplitude due to the black hole's tidal response, we can compute the corresponding result for neutron star\footnote{In fact, as far as the dissipation number computation is concerned, since only tidal effects contribute to dissipation, and thus no such subtraction is required and we can just take the absolute values on both sides of Eq.~(\ref{eq : ph2ampPT}), and write the expression for $\eta_{2}$, as was done in Ref.~\cite{Saketh:2022xjb}.}. This has been schematically shown in Eq.~(\ref{diag : cancel}).
Alternatively, we can make use of the near-far factorization first presented in Ref.~\cite{Ivanov:2022qqt}. This was used to show the vanishing of black hole Love numbers, as well as computation of the renormalization-group flows of the dynamical tidal response in Refs.~\cite{Ivanov:2022qqt, Saketh:2022xjb}. However, this is justified only for the generic-$\ell$ solution to the Teukolsky equation \cite{Bautista:2023sdf}. The essential idea is that the far-zone scattering phase shift corresponds to the wave scattering against the background metric, while the near-zone phase shift is the scattering against the star surface which contains the tidal response:
\begin{widetext}
\begin{equation}
    \frac{A_{\ell,\omega}^{\rm out}}{A_{\ell,\omega}^{\rm in}} = \red{{\underbrace{\frac{A_-^{\bar\nu}}{A_+^{\bar \nu}} e^{2 i \epsilon \log(\epsilon)}}_{\rm far\, zone}}} \times \blue{\underbrace{\frac{1 + i e^{i \pi \bar \nu} \mathcal{K}_{\rm NS}}{1 - i e^{- i \pi \bar \nu} \frac{\sin(\pi(\bar \nu+ i \epsilon))}{\sin(\pi(\bar \nu - i \epsilon))} \mathcal{K}_{\rm NS}}}_{\rm near\, zone}} ~.
    \label{eq : near far split}
\end{equation}
\end{widetext}
From Eq.~\eqref{eq : near far split}, we can straightforwardly get the near-zone phase shift as
\begin{equation}
    \eta_{\ell}^{\rm NZ} e^{2 i \delta_{\ell}^{\rm NZ}} = \frac{1 + i e^{i\pi \bar \nu } \mathcal{K}_{\rm NS} }{1- ie^{-i \pi \bar \nu} \frac{\sin(\pi(\bar \nu+i \epsilon))}{\sin(\pi(\bar \nu - i \epsilon))} \mathcal{K}_{\rm NS}} ~.
\end{equation}

Regardless of method used, we can write down the tidal contribution to the phase shift and absorption as shown in Eqs.~(\ref{eq:stellar_pert_phase_shift}), (\ref{eq : doa}):

\begin{widetext}
\begin{equation}
\begin{gathered}
    \begin{tikzpicture}[line width=1,photon/.style={decorate, decoration={snake, amplitude=1pt, segment length=6pt}}]
    \draw[line width = 1, photon] (0,0.5) -- (1,1.5);
    \draw[line width = 1, photon] (0,-0.5) -- (1,-1.5);
    \draw[line width = 1, dashed, double] (0,-0.5) -- (0,0.5);
    \filldraw[fill=White, line width=0.15](0,0.0) circle (0.5) node[]{NS};
    \end{tikzpicture}
\end{gathered}
\quad - \quad\begin{gathered}
    \begin{tikzpicture}[line width=1,photon/.style={decorate, decoration={snake, amplitude=1pt, segment length=6pt}}]
    \draw[line width = 1, photon] (0,0.5) -- (1,1.5);
    \draw[line width = 1, photon] (0,-0.5) -- (1,-1.5);
    \draw[line width = 1, dashed, double] (0,-0.5) -- (0,0.5);
    \filldraw[fill=White, line width=0.15](0,0.0) circle (0.5) node[]{BH};
    \end{tikzpicture}
\end{gathered}
\quad = \quad \left[\quad \begin{gathered}
    \begin{tikzpicture}[line width=1,photon/.style={decorate, decoration={snake, amplitude=1pt, segment length=6pt}}]
    \draw[line width = 1, photon] (0,0.5) -- (1,1.5);
    \draw[line width = 1, photon] (0,-0.5) -- (1,-1.5);
    \draw[line width = 1, dashed, double] (0,-0.5) -- (0,0.5);
    \filldraw[fill=White, line width=0.15](0,0.0) circle (0.5) node[]{NS};
    \end{tikzpicture}
\end{gathered}\quad\right]_{\mathrm{tidal}}
-\left[\quad \begin{gathered}
    \begin{tikzpicture}[line width=1,photon/.style={decorate, decoration={snake, amplitude=1pt, segment length=6pt}}]
    \draw[line width = 1, photon] (0,0.5) -- (1,1.5);
    \draw[line width = 1, photon] (0,-0.5) -- (1,-1.5);
    \draw[line width = 1, dashed, double] (0,-0.5) -- (0,0.5);
    \filldraw[fill=White, line width=0.15](0,0.0) circle (0.5) node[]{BH};
    \end{tikzpicture}
\end{gathered}\quad\right]_{\rm{tidal}}
\label{diag : cancel}
\end{equation}
\begin{equation}
\begin{aligned}
\label{eq:stellar_pert_phase_shift}
 \delta_{\ell}^{\rm NZ} & = (2 R \omega)^{2\ell + 1} \times  (-1)^{\ell+1}  \frac{\Gamma (-2 \ell ) \Gamma (\ell +2)}{\Gamma (1-\ell ) \Gamma (2 \ell +2)} \\
    & \quad \times \Bigg[ \ell  \Big(-\frac{T_0}{1-2C}+\ell +1\Big) \, _2F_1(-\ell -2,2-\ell, -2 \ell, 2 C)+C
   \left(\ell ^2-4\right) \, _2F_1(-\ell -1,3-\ell, 1-2 \ell, 2 C)\Bigg] \\
   & \quad \times \Bigg[ (\ell +1) \Big(\frac{T_0}{1-2C}+\ell \Big) \, _2F_1(\ell -1,\ell +3,2 (\ell
   +1), 2 C)+C (\ell -1) (\ell +3) \, _2F_1(\ell ,\ell +4, 2 \ell +3, 2 C) \Bigg]^{-1} ~,
\end{aligned}
\end{equation}
\begin{equation}
\begin{aligned}
   \eta_{\ell}^{\rm NZ} & = 1+ (2 R \omega)^{2\ell+2} \times (-1)^\ell \frac{T_1}{1-2C} \frac{\Gamma (-2 \ell ) \Gamma (\ell +2)}{\Gamma (1-\ell ) \Gamma (2 \ell +2)} \\
   & \quad \times \Bigg[C (\ell -2) (\ell +1) (\ell +2) \, _2F_1(-\ell
   -1,3-\ell ;1-2 \ell ;2 C) \, _2F_1(\ell -1,\ell +3,2 (\ell +1),2 C) \\
   & \quad +\ell(\ell +1) (2 \ell +1) {}_2F_1(-\ell -2,2-\ell,-2 \ell,2 C) {}_2F_1(\ell -1,\ell +3,2 (\ell
   +1),2 C) \\
   & \quad +C \ell(\ell -1) (\ell +3) {}_2F_1(-\ell -2,2-\ell,-2 \ell,2 C) {}_2F_1(\ell ,\ell +4,2 \ell +3,2 C) \Bigg]\\
   & \quad \times \Bigg[(\ell +1) \Big(\frac{T_0}{1-2C}+\ell \Big) \, _2F_1(\ell -1,\ell +3,2 (\ell
   +1),2 C)+C (\ell -1) (\ell +3) \, _2F_1(\ell ,\ell +4,2 \ell +3,2 C)\Bigg]^{-2} ~.
\end{aligned}
\label{eq : doa}
\end{equation}
\end{widetext}

\subsection{Matching with EFT}

Finally, we can get the Love number and dissipation number by matching the EFT phase shifts in Eq.~\eqref{eq : ph2amp} with the ones from stellar perturbation theory given in Eqs.~\eqref{eq:stellar_pert_phase_shift}, \eqref{eq : doa}. The expressions for the rescaled Love and dissipation number are as follows: 
\begin{widetext}
\begin{equation}
\begin{aligned}
\label{eq: all_compact_Love_Diss}
    k_2^E & = -\frac{8 C^5 \left(6 C-3+T_0\right)}{5 \left(-2 C \left(C \left(6 C^2+4 C+3\right)+3\right) T_0+6 (C-1) C \left(4 C^3+2 C^2-3\right)-3 \left(6 C-3+T_0\right) \log (1-2
   C)\right)} ~, \\
    \nu_2^E & = \frac{768 C^{10} T_1}{5 \left(2 C \left(C \left(2 C \left(3 C \left(-2 C+T_0+1\right)+2 T_0+3\right)+3 \left(T_0+3\right)\right)+3 \left(T_0-3\right)\right)+3 \left(6
   C+T_0-3\right) \log (1-2 C)\right){}^2} ~.
\end{aligned}
\end{equation}
\end{widetext}
From the above two expressions, we see that the dissipation number vanishes when the boundary condition at the star surface has no linear-in-frequency dependence, i.e. $T_1=0$. This is consistent with the time-reversal property of the retarded Green's function in Eq.~\eqref{eq:low_exp}. Secondly, there is a relation between the rescaled Love number and dissipation number
\begin{equation}
    \nu_2^E = \frac{60 T_1}{(6C-3 + T_0)^2} (k_2^E)^2 ~.
\end{equation}

Eq.~(\ref{eq: all_compact_Love_Diss}) is also consistent with the BH limit, where we have
\begin{equation}
    C\rightarrow \frac{1}{2} ~, \quad T \rightarrow - i \omega r_s ~,
\end{equation}
corresponding to purely ingoing boundary at the horizon. Thus, $T_0 \rightarrow 0,~ T_1 \rightarrow 1$ and Eq.~\eqref{eq: all_compact_Love_Diss} correctly reduces to 
\begin{alignat}{3}
k_2^E=0,~\nu_2^E=\frac{1}{60}
\end{alignat}
Finally, we have also verified that our formula for the rescaled Love number $k_2^E$ is consistent with the one given by T. Hinderer in Ref.~\cite{Hinderer:2007mb} apart from convention-related differences  after rewriting the boundary condition $T_0$ of the RW variable in terms of the logarithmic derivative of the metric perturbation function $H_0(r)$, i.e., $y \equiv R H_0'(R)/H_0(R)$, as
\begin{equation}
\begin{aligned}
    & T_0  = \Big(-3(1+2(-2 + C)C) + 3 (-1+C)(-1+ 2 C)y \Big) \\
    & \quad \Big((3 - 8 C + 6 C^2) + (-1+(5-6 C) C ) y\Big)^{-1} (1-2C).
\end{aligned}
\end{equation}
\section{Observables}
In the previous sections, we have developed the formalism for computing Love and dissipation numbers of viscous neutron stars in the small-frequency limit for nonbarotropic perturbations. We now use it to numerically compute some of the relevant quantities and relate the dissipation number to the waveform contribution at 4PN. 
\subsection{Love and dissipation numbers for various EoS(s) and compactness}
\label{sec : dissLovecomp}
To compute the Love and dissipation number(s), we first integrate the master Eqs.~(\ref{eq : firstfin}), (\ref{eq : secondfin}) and then switch to the RW function using Eq.~(\ref{eq : HKtoX}), and compute the quantities $T_0$ and $T_1$ in Eq.~(\ref{eq:star_bc}). We can then use Eqs.~(\ref{eq: all_compact_Love_Diss}) to compute the Love and dissipation number(s). The compactness $C=M/R$ is obtained from the stellar profiles generated for various EoS(s).

We present the Love and dissipation numbers thus obtained for various EoS(s) and compactness in Table.~\ref{table : dissLove}. Here, the first column labeled EoS lists the equation(s) of state for which the relevant quantities in each row have been computed. The details regarding the various EoS(s) have been discussed before in Sec.~\ref{sec : bg profile}. The remaining columns sequentially show the rescaled Love number $k_2$, the dissipation number $H_\omega^E$, and the rescaled dissipation number $\nu^E_2$, and finally the ratio of the dissipation number $H_\omega^E$ with that for a black hole with the same mass, where $(H_\omega^E)_{\mr{BH}}=32/45$. This is to get a relative measure of the `absorptivity' of a neutron star compared to a black hole. Even at $T=10^5$K, this ratio is at most $\sim 30$ for the least compact configurations and falls below unity for $M\geq 1.71M_\odot$. Since it scales as $T^{-2}$, for $T\geq 10^6$K all configurations in Table~\ref{table : dissLove} are less dissipative than a black hole of the same mass.

As we showed in Sec.~\ref{sec : mastereq}, the bulk viscosity does not affect the (leading) dissipation number. Thus, the dissipation numbers obtained in the table are entirely due to shear viscosity. Here, we consider the most dominant contribution to the shear viscosity which is due to  $e-e$ scattering in the temperature range relevant during inspiral~\cite{Ghosh:2023vrx,Lai:1993di}. The shear-viscosity due to this process scales inversely with the square of temperature (i.e., $\sim 1/T^2$) as seen from Eq.~(\ref{eq:SV_ee}). Thus, colder neutron stars have a greater contribution to shear viscosity. Since the viscosity linearly affects $T_1$, and subsequently the (rescaled) dissipation number $\nu_E$, i.e., they too scale as $1/T^2$. Although neutron stars are born very hot ($\sim 10^{11}$K), they cool rapidly to lower temperatures ($\sim 10^9K$) due to neutrino emission within  $\sim 10^5$ years~\cite{Yakovlev_2005}. The temperature of cold neutron stars about to merge during inspiral is expected to be within $10^{5}-10^{10}K$, with the likely range being within $10^{7}-10^{9}K$~\cite{Lai:1993di,Arras:2018fxj,Ghosh:2023vrx}. We show the results at the coldest possible temperature $T=10^5K$ when the shear viscosity due to $e-e$ scattering is strongest. The temperature dependence is captured in the first row of the table through the factor $(T_K/10^5 K)^2$ next to the relevant quantities, where $T$ is the actual temperature of the neutron star in Kelvin. Since $H_\omega^E$ and $\nu_2^E$ scale as $\sim T^{-2}$, multiplying the factor $(T_K/10^5 K)^2$ renders them constant and gives their value at $T=10^5K$.

Within each EoS, we have considered stars with different compactness in increasing order. It is clear from the table that within any EoS, the dissipation number decreases with increasing compactness. For instance, for the case of `FSU2', as we go from $1\mr{M}_{\odot}$ to $2.34 \mr{M}_{\odot}$, where the radius barely changes ($14.0$ km to $13.8$ km), the dissipation number $H_\omega^E$ falls sharply with compactness by almost 3 orders of magnitude. This is actually expected from Eq.~(\ref{eq: all_compact_Love_Diss}) if we assume that the $T_0$ and $T_1$ do not change a lot with changing compactness. Then, the rest of the expression explains the sharp decrease with increasing compactness. This is seen easily in the Newtonian limit $C\ll 1$, where we have    
\begin{alignat}{3}
H_\omega^E =  \frac{ 10 T_1}{36(2+T_0)^2 C^6},
\end{alignat}
which shows that the dissipation number falls sharply with the inverse of the compactness at the sixth power.

In reality, $T_{0,1}$ do not actually stay constant with changing compactness as they depend on the density and pressure profile(s) which in turn also affects the shear viscosity. However, we find that the change in them is quite small (relatively) and not adequate to compensate for the sharp-dependence on compactness. This can be seen clearly from the last column in Table.~\ref{table : dissLove}, where the rescaled dissipation number is seen to be in the same order of magnitude regardless of compactness (and even EoS(s)).

The importance of compactness in tuning the dissipation number is also seen in the variation of EoS(s). For example, consider the three stars in the table with a mass $M\approx 1.01 M_\odot$. They all correspond to different equations of state. Despite having the same mass, the `HZTCS' star is more compact than `GM1' which in turn is more compact than `FSU2'. Despite the differences in their EoS(s), and consequently the shear-viscosity profile, we find that the dissipation numbers fall with increasing compactness as expected. The compactness does not however capture the complete variation in the dissipation numbers. For the three stars with $M=1.34 M_\odot$ ($M=1.71 M_\odot$), the compactness still increases as one goes from `FSU2' to `GM1' to `HZTCS'. However, while the $H_\omega^E$ for `GM1' and `HZTCS' is smaller than that of `FSU2' as expected, the relative ordering of $H_\omega^E$ between `GM1' and `HZTCS' is not consistent with the ordering of the compactness. %\js{[JS: Here, we switch to a notation where the EoS name is written in single quotes. We should be uniform ]} 
However this is not surprising as the compactness for both of them are very similar (with in $\sim 1\%$). 
\subsection{Correction to number of GW cycles due to dissipation}
\label{Sec : quantqual}

Dissipation in a binary leads to transfer of orbital energy into the mass (thermal energy) of the compact objects, here the neutron stars. This modifies the energy conservation law for quasi-circular inspiral as
\begin{alignat}{3}
\dot{E}= - \mc{F}_\infty - \dot{m}_1-\dot{m}_2.
\end{alignat}
Here $\dot{E}$ is the rate of change of orbital energy, $\mc{F}_\infty$ is the gravitational energy flux leaving the system and $\dot{m}_{1,2}$ is the rate of heating (due to tidal dissipation). We can use this modified energy conservation law to compute the leading (4PN) correction to the phase of the inspiral waveform due to tidal heating in the stationary-phase approximation~\cite{Buonanno:2009zt,
Arun:2008kb,
Datta:2020gem,Saketh:2022xjb}. We do that below as a way to roughly quantify the relevance of shear viscosity.

Since we are only interested in the leading order correction, it is sufficient to compute the correction to the orbital phase due to one star and then linearly sum their contributions. So we first compute the correction to the orbital phase when only one star gets tidally heated.

The rate of heating (or equally rate of increase in mass) for a neutron star of mass $m_1$ in a quasi-circular binary  (with another mass $m_2$) is given by \cite{Saketh:2022xjb} 
\begin{alignat}{3}
\dot{m_1} &= \frac{1}{2}\nonumber m_1(Gm_1)^5 H_\omega^{E,1}\dot{E}_{\rho\sigma}\dot{E}^{\rho\sigma} \\& \nonumber = H_\omega^{E,1}\frac{9m_1^6m_2^2}{GM^8}\Bigg(\frac{GM}{r}\Bigg)^{9} \\&=
H_\omega^{E,1}\frac{9m_1^6m_2^2}{GM^8}x^{18},
\end{alignat}
where $M=m_1+m_2$, $r$ the radius of the orbit, $\Omega=(GM/r^3)^{1/2}$ is the orbital angular frequency, $V=x=(GM\Omega)^{1/3}=(GM/r)^{1/2}$.
The leading quadrupolar flux of the binary to infinity is given by
\begin{alignat}{3}
\mathcal{F}_{\infty}=\frac{32m_1^2m_2^2}{5G M^4}x^{10}.
\end{alignat}
The relative contribution of the tidal heating to the waveform may then be characterized by the ratio 
\begin{alignat}{3}
\dot{m}_1/\mathcal{F}_{\infty}& \nonumber = (45H_\omega^E/32)(m_1/M)^4x^{8}\\& =(15\nu^{E,1}_2/16)(R/m_1)^{2}(R/r)^4.
\end{alignat}
Note that the net effect is enhanced by lower compactness ($m_1/r$), and closeness ($R/r$) of the binary.

Now, the energy conservation law for the system reads
\begin{alignat}{3}
\dot{E} &= -\mathcal{F}_{\infty} - \dot{m}_1-\dot{m}_2,
\\ \implies \dot{x} &=\frac{1}{M\eta x}(\mathcal{F}_{\infty}+\dot{m}_1+\dot{m}_2)
\end{alignat}
where $E=M\eta v^2/2$, with $\eta=(m_1m_2)/M^2$ being the orbital binding energy, sapped away to infinity and into the heat of the bodies. We can now determine the orbital phase of the binary during inspiral as
\begin{alignat}{3}
\label{eq : phase1}& \phi(x)=\int\frac{x^3}{M}\frac{dt}{dx}dx \\& \nonumber \approx \int  \frac{5 M^2}{32m_1m_2 x^{6}}\Bigg[1-\frac{15\nu^{E,1}_2}{16}\frac{R^6}{m_1^2M^4}x^8 - (1\leftrightarrow 2)\Bigg] dx \nnm 
\\& \nnm = -\frac{M^2}{32m_1m_2 x^5}-\frac{25 M^2}{ m_1m_2}\frac{\nu^{E,1}_2}{512}\frac{R^6}{m_1^2M^4} x^3 - (1\leftrightarrow 2)\\& \nnm = \phi_{0\mr{PN}} + \delta\phi(x) .
\end{alignat}
Eq.~(\ref{eq : phase1}) contains $\delta \phi$, which is the correction to the orbital phase due to one neutron stars getting tidally heated. Then, using the fact that the GW frequency in the most dominant mode is twice the orbital frequency, we can compute the change in number of GW cycles due to tidal dissipation within a frequency range as
\begin{alignat}{3}
&\delta\mathcal{N}_{\text{GW}}=\frac{\delta \phi[(GM\pi\omega_{f})^{1/3}]-\delta \phi[(GM\omega_i\pi)^{1/3}]}{\pi},\label{eq : dN}
\nnm \\& = \sum_{a=1,2}\frac{25 R_a^6 \nu_2^{E,a}}{512 m_a^3 (M-m_a) M^2 }\times GM(\omega_i-\omega_f).
\end{alignat}
Here $m_{a=1,2}$ are the masses of the neutron stars, and $M=m_1+m_2$. $R_{a=1,2}$ are the radii, and $\nu_2^{E,a=1,2}$ are the rescaled dissipation numbers defined in Eq.~(\ref{eq : rescale}). The inspiral enters the detector frequency band at GW frequency $\omega_i$ to $\omega_f$.  Eq.~(\ref{eq : dN}) could be evaluated using the dissipation numbers in Table.~\ref{table : dissLove}, for inspiral with in the LVK band for which $\omega_i\approx30\mr{hz}$ and $\omega_f\approx\mr{min}(1000\mr{hz},\omega_{\mr{ISCO}})$, where $\omega_{\mr{ISCO}}=1/(6^{3/2}\pi GM)$ is the GW frequency at the innermost stable circular orbit. However, it is important to remember here the discussion in Sec.~\ref{sec : static_limit}, where we noted that the small-frequency expansion of the perturbation equations as done in this work is valid when the inspiral frequency is small compared to the characteristic Brunt-V\"{a}is\"{a}l\"{a} frequency $N_\mr{ch}$. The latter may be estimated by some averaging scheme~\cite{1992ApJ...395..240R} or by the highest frequency among the g-modes
~\cite{Lai:1993di,Andersson:2019ahb} yielding $N_\mr{ch}\in(150,700)$hz depending on the mass and EoS of the star, and so the dissipation numbers provided in Table.~\ref{table : dissLove} may not suffice throughout the inspiral. Nevertheless, at least for conservative effects, the contribution of the oscillation modes associated with convection, i.e., $g$-modes  is often negligible due to their weak coupling with external tidal fields~\cite{Andersson:2019ahb}. We proceed with the expectation that the same may be true for shear-viscous dissipation as well. Bulk-viscosity on the other hand was shown to be vanishing in the regime $\omega \ll N_\mr{ch}$ in this work. Thus, the correction to it could be a lot more significant and non-negligible in middle and late inspiral. However, such effects are likely to only enhance the contribution to tidal heating and thus the results in this work may be used to obtain a lower-limit for the contribution of tidal heating during inspiral.

We compute Eq.~\eqref{eq : dN} and present the results for various EoS(s) and compactness in Table~\ref{table : waveform}. Once again the first column lists out the various EoS(s) similar to Table~\ref{table : dissLove}. The subsequent five columns show neutron star mass, radius, compactness, $H_\omega^E$ and $\nu_2^E$.

Then, the seventh column shows $\delta \mc{N}_{\mr{GW}}$, change in number of cycles. The eighth column shows the Newtonian (leading) contribution to the total number of cycles within the detector band, which we take to be from 30hz to min($\omega_{\mr{ISCO}}$,1000hz). Finally, the ninth column shows the correction to the number of cycles at 4PN due to other conservative contributions in the spin-less case~\cite{Blanchet:2023bwj,Blanchet:2023sbv,Blanchet:2023soy}. 

As mentioned previously, the dissipation considered here is entirely due to $e-e$ scattering, with a known dependence on temperature given in Eq.~(\ref{eq:SV_ee}). This temperature dependence is cancelled by the factor $(T_K/10^5)^2$ next to $H_\omega^E$, $\nu_2^E$ and $\delta \mc{N}_\mr{GW}$, so that the results in the table correspond to neutron stars $T=10^5K$. Here $T_K$ is the temperature in Kelvin of the neutron star at the start of the inspiral which we have kept fixed while evaluating $\delta \mc{N}_{\text{GW}}$.

In the symmetric NS case, Eq.~(\ref{eq : dN}) reduces to 
\begin{alignat}{3}
\delta \mc{N}_\mr{GW} = \frac{75 H_\omega^E}{2048}GM(\omega_i-\omega_f). 
\end{alignat}
Thus, for a given initial and final frequency, the product of the dissipation number and the mass decides the magnitude of $\delta \mc{N}_\mr{GW}$. We can see from the table that within an EoS, the compactness generally increases with mass (except for the EoS `HZTCS'). Thus, there is a competition between the falling dissipation number (with compactness) and the increasing mass. However, we know that the dissipation number decreases sharply with compactness, ($\sim C^{-6}$ in the Newtonian limit), and thus $\delta \mc{N}_\mr{GW}$ decreases in any EoS, as visible in Table.~\ref{table : waveform}.

Comparing the last three columns of Table~\ref{table : waveform}, we find that the correction to the number of GW cycles due to tidal heating is small. Even at $T=10^5$K and for the least compact configurations, $|\delta\mathcal{N}_{\rm GW}|< 8\times 10^{-3}$, about an order of magnitude below the other conservative contributions at 4PN, and it is further suppressed for more compact stars. Since $\delta\mathcal{N}_{\rm GW}\propto T^{-2}$, at the more likely temperatures $T\sim 10^7$ to $10^9$K the effect is smaller by an additional factor of $10^4$ to $10^8$. Although waveform accuracy requirements for current and future detectors are stringent~\cite{Owen:2023mid,Purrer:2019jcp,Hu:2022rjq}, shear-viscous tidal heating in the low-frequency regime is therefore unlikely to leave an observable imprint unless the stars are exceptionally cold. This is consistent with earlier Newtonian estimates that tidal heating from canonical viscosities is weak~\cite{Lai:1993di,Arras:2018fxj}, and suggests that a measurable dissipative signature during inspiral would require other mechanisms, such as resonant excitation of $g$-modes or bulk viscosity from hyperons~\cite{Ghosh:2023vrx}, which do not contribute within the low-frequency regime considered here.

A finite-frequency analysis, going beyond the low-frequency expansion of this work, may therefore yield a more accurate and potentially larger estimate of tidal dissipation. As discussed above, the expansion used here is valid only for $\omega\ll N_\mr{ch}$, a condition that fails in the middle and late inspiral, and in this regime neither resonances nor bulk viscosity contribute. Recent work supports this possibility. In Ref.~\cite{Ghosh:2026hyp}, finite reaction rates in hyperonic matter were included in the relativistic perturbation equations through a complex, frequency-dependent sound speed. The resulting bulk viscosity damps both the $f$- and $g$-modes, gives rise to a tidal lag, and is expected to be the dominant source of dissipation for $T\sim 10^6$ to $10^9$K, where the equilibration timescales are comparable to the oscillation periods. Similarly, Ref.~\cite{Ghosh:2025tid} showed that the dissipation from hyperon bulk viscosity varies strongly over the course of the inspiral, so that a constant dissipation number fails to capture the resulting phase, and that for massive stars ($\sim 2M_\odot$) tidal heating can exceed the leading tidal-deformability contribution to the phase. A natural framework for such an analysis is the gauge-invariant definition of the dynamical tidal response through gravitational Raman scattering developed in Ref.~\cite{Saketh:2026dyn}, which retains the full frequency dependence of the response, including its resonances, and can be extended to include viscous dissipation, as also anticipated in Ref.~\cite{Ghosh:2026hyp}. Any such extension should reduce at low frequency to the dissipation numbers obtained here, which have now been confirmed independently in Ref.~\cite{Vernizzi2026}.
\section{Conclusion and future Work}
\label{Sec : conc}
In this work, we have studied in detail the tidal response of nonspinning neutron stars in the low-frequency limit, with the goal of determining the leading dissipation numbers due to shear and bulk viscosity, which correspond to the next-to-leading order (in frequency) tidal effects. We found that the bulk viscosity has no contribution to the dissipation number, and demonstrated that this is physically due to the vanishing divergence of the fluid velocity at linear order in frequency, and this is related to a peculiar imcompressibility in the static limit. We showed that this conclusion holds only for non-barotropic perturbations, when the perturbing frequency (inspiral frequency in a binary) $\omega$, is small compared to characteristic Brunt-V\"{a}is\"{a}l\"{a} frequency $N_{\mr{ch}}$ in the star, but much larger than the rate of m-Urca processes.

On the other hand, the leading shear viscous contribution to the dissipation number was seen to be nonzero in this regime. We derived two master equations governing metric perturbations to linear-order-in frequency inside the neutron star, including viscous contributions.

We then considered the problem of scattering GWs off a neutron star i.e., gravitational Raman scattering, and related the scattering phase and degrees of absorption 
%(or amplitude) 
to the static Love and leading dissipation numbers. 

To achieve this, we first integrated the master equations for the metric perturbations from the center to the surface of the star. Outside the star, the metric perturbation equations could be reduced to the RW equation, and the boundary conditions for the RW function at the stellar surface were obtained earlier by integrating the master equations. We then used the analytical MST approach to compute the scattering amplitude as a function of  the RW boundary conditions at the stellar surface. We isolated the tidal contribution to this amplitude and extracted expressions for the rescaled static Love and dissipation numbers. The formula for the rescaled Love number was shown to be consistent with previous formulas by Hinderer~\cite{Hinderer:2007mb}.

We then used the formulae to calculate the Love and dissipation numbers for various EoS(s) and compactness. We found that compactness was the most important parameter tuning the strength of the tidal dissipation number for a given EoS. 

Finally, we computed the correction to the number of GW cycles due to shear viscosity from $e-e$ scattering at 4PN in the waveform for a binary system of two identical neutron stars within the LIGO band. We found that this effect is small. Even for cold ($T\sim 10^5$K) and weakly compact neutron stars, it changes the number of cycles by less than $10^{-2}$, about an order of magnitude below the other conservative contributions at 4PN, and it decreases as $T^{-2}$ for hotter stars. Cold and less compact stars still exhibit the strongest shear-viscous tidal heating at 4PN, but neglecting resonant and spin-related effects, shear viscosity alone is unlikely to leave an observable imprint during inspiral. Our dissipation numbers agree with the independent calculation of Ref.~\cite{Vernizzi2026}, which uses a different method.

The work can be extended in several directions in the future. An obvious next step would be to extend the validity of this work to arbitrary frequencies. The low-frequency expansion of the perturbation equations inside the star is sufficient in a binary system when the inspiral frequency is smaller than any resonant frequencies. While the often considered $f$-mode frequency is much higher than the orbital frequency during inspiral, but the class of $g$-modes have much lower resonant frequencies that can be reached during inspiral. This is also related to the discussions surrounding the Brunt-V\"{a}is\"{a}l\"{a} frequency as it is the characteristic frequency of the $g$-modes. The effect of resonances during inspiral on the conservative tidal response in polytropic systems was recently been studied in Ref.~\cite{HegadeKR:2024agt}. It may be interesting to extend this to realistic EoS(s) and dissipation. This is particularly important in light of the vanishing contribution of bulk viscosity at 4PN order and the smallness of the shear-viscous contribution that we have shown in this paper, as resonant dissipation due to $g$-modes may be the leading bulk-viscous, and possibly the dominant, contribution to tidal heating. Recent work on hyperonic reaction rates~\cite{Ghosh:2026hyp} and on frequency-dependent tidal dissipation~\cite{Ghosh:2025tid} supports this expectation, and the amplitude-based framework of Ref.~\cite{Saketh:2026dyn} provides a route to incorporate such effects consistently within the worldline EFT. 

Second, in this work we restricted ourselves to spherically symmetric systems for simplicity. In addition, we have neglected axial modes~\cite{Flanagan:2006sb, Poisson:2020mdi, Ma:2020oni, Gupta:2020lnv}, since the polar modes dominantly couple to neutron stars. The presence of spin would lead to a violation of spherical symmetry, and thus also couple axial and polar sectors, leading to a more complicated system of equations, and a potentially interesting and perhaps stronger tidal response. It has recently been shown in the Newtonian limit in Ref.~\cite{Yu:2024uxt} that moderately spinning neutron stars in binaries can reach $f$-mode resonances during the inspiral (see also Ref.~\cite{Ma:2020rak} for earlier work and Ref.~\cite{Steinhoff:2021dsn} for a relativistic treatment). They also showed that such a resonance could make the orbit eccentric. Thus, a general-relativistic study of the tidal response of spinning compact objects could lead to the discovery of interesting potentially observable effects. 

Other interesting directions could be the extension to other (possibly exotic) compact objects or to alternative theories of gravity, or simply add more details relevant for a realistic neutron star, such as an in-depth consideration of the various microscopic processes, their rates, and their impact on the observed macro-physics. 

For now, we leave such considerations to future works. With the advent of future detectors~\cite{Purrer:2019jcp, Hu:2022rjq}), and with improved sensitivities, the rich internal physics of neutron stars may be better understood and constrained. Along with the EoS of the neutron star which relates the equilibrium pressure with the density, a knowledge of the out-of-equilibrium transport properties such as viscosity inferred through tidal heating in binary inspiral will be influential in understanding the dense matter behavior as well as the neutron star interior composition. It is thus important that theoretical investigations now rigorously take into account the various possible micro-physics relevant to the neutron star and their observational imprints.

\section*{ACKNOWLEDGEMENTS}
We thank Alessandra Buonanno, Abhishek Hegade, Misha Ivanov, Ben Leather, Julio Parra-Martinez, Justin Ripley, Hector Silva, Teja Venumadhav, Pratik Waghle, Nicol\'as Yunes,  and Matias Zaldarriaga for useful comments and discussions. We thank Swarnim Shirke for providing a comprehensive review of the dissipation sources inside neutron stars. We are grateful to Filippo Vernizzi and collaborators for an independent comparison of our results, which led to the identification of a unit-conversion error in an earlier version of this paper.
\appendix

\section{Newtonian stellar fluid}
\label{App A}
In this appendix, we briefly review the non-viscous linear stellar perturbation theory for neutron stars in Newtonian gravity following \cite{Thorne:1969rba, smeyers2011linear}. The Newtonian limit makes it easier to understand the approach towards static limit $\omega\rightarrow 0$ and the low-frequency expansion presented in the main text starting from complete all-orders-in-$\omega$ perturbation equations. 

With the fluid displacement vector $\boldsymbol{\xi}$, we can write the continuity equation to solve for the density perturbation as
\begin{equation}
    \delta \rho = - \nabla \cdot (\rho \boldsymbol{\xi}) ~, \quad \Delta \rho = -\rho \nabla \cdot \boldsymbol{\xi} ~.
\end{equation}
The perturbed Euler equation (Newton's second law) is given by
\begin{equation}
    \rho \ddot{\boldsymbol{\xi}} = - \nabla \delta p + \delta \rho\nabla U + \rho \nabla \delta U ~,
\end{equation}
where $U$ is the background Newtonian potential and $\delta U$ is its perturbation. We also have the perturbed Poisson equation
\begin{equation}
    \nabla^2 \delta U = - 4 \pi G \delta \rho ~.
\end{equation}
The above equations are not complete, and one needs to solve for the perturbation to pressure as well. As mentioned in the main-text in Sec.~\ref{sec : Lpres}, we can generally define the adiabatic index $\gamma$ and sound speed $c_s$ as
\begin{alignat}{3}
\nnm
    \frac{\Delta p}{\Delta \rho} 
    =  \gamma \frac{p}{\rho} = c_s^2.
\end{alignat}
Then, in the two extremes of very fast reactions and very slow reactions, we have 
\begin{alignat}{3}
    \textbf{fast reactions}: ~ \frac{\Delta p}{\Delta \rho} &= \Big(\frac{\partial p}{\partial \rho}\Big)\Big|_{\mu_i} =c_\mr{eq}^2~.\\
\textbf{slow reactions}: ~ \frac{\Delta p}{\Delta \rho} &= \Big(\frac{\partial p}{\partial \rho}\Big)\Big|_{x_i} =c_\mr{ins}^2~.
\end{alignat}
On the other hand, the unperturbed background profile is in chemical equilibrium and thus 
\begin{alignat}{3}
    \frac{dp/dr}{d\rho/dr} %&= \Bigg[ \Big(\frac{\partial p}{\partial \rho}\Big)_{s,\mu_i} + \Big(\frac{\partial p}{\partial \mu_i}\Big)_{\rho,s} \Big(\frac{\partial \mu_i}{\partial \rho}\Big)_{s,x_i} \Bigg] \Delta \rho~.
    %\\ &
    = \left(\frac{\partial p}{\partial \rho}\right)_{\mu_i}=c_{\mr{eq}}^2,%\Delta. %\rho_i.
\end{alignat}
Generally for neutron stars, the dominant m-Urca reactions have a very large characteristic time-scale for equilibrium compared to the orbital time period in binaries during inspiral, and thus $c_\mr{eq}\neq c_s\approx c_\mr{inst}$.

Now, decomposing the fluid displacement as 
\begin{equation}
    \boldsymbol{\xi}=\left[\xi_{r}(r) \hat{\boldsymbol{r}} +\frac{\xi_{\rm H}(r)}{r}\left(\hat{\boldsymbol{\theta}} \frac{\partial}{\partial \theta} +  \frac{\hat{\boldsymbol{\phi}}}{\sin \theta} \frac{\partial}{\partial \phi}\right)\right] Y_{\ell m}~,
\end{equation}
and defining the following variables for convenience
\begin{equation}
    w_1 = r \xi_r ~, w_2 = \xi_{\rm H} ~, w_3 = \delta U ~, w_4 = \frac{d \delta U}{d \log r} ~.
\end{equation}
The perturbation equation can be rewritten as %follows
\cite{smeyers2011linear}
\begin{alignat}{3}
\label{eq:Newtonian_fluid_pert}
\frac{d w_1}{d r}= & \nnm \left(\frac{g}{c_s^2}-\frac{1}{r}\right) w_1+\left[\frac{\ell(\ell+1)}{r^2}-\frac{\omega^2}{c_s^2}\right] r w_2\\& \nnm +\frac{r}{c_s^2} w_3, \\\nnm 
\frac{d w_2}{d r}= & \left(1-\frac{N^2}{\omega^2}\right) \frac{1}{r} w_1+\frac{N^2}{g} w_2-\frac{1}{\omega^2} \frac{N^2}{g} w_3, \\ 
\frac{d w_3}{d r}= & \frac{1}{r} w_4, \\\nnm 
\frac{d w_4}{d r}= & 4 \pi G \rho \frac{N^2}{g} w_1+\omega^2 \frac{4 \pi G \rho}{c_s^2} r w_2 \\
& +\left[\frac{\ell(\ell+1)}{r^2}-\frac{4 \pi G \rho}{c_s^2}\right] r w_3-\frac{1}{r} w_4 ~,\nnm
\end{alignat}
where the function $g(r)$ here is the local acceleration due to gravity inside the star  
\begin{equation}
    g \equiv U'(r) = \frac{G m(r)}{r^2}
\end{equation}
and $N$ is the Brunt-V\"{a}is\"{a}l\"{a} frequency 
\begin{alignat}{3}
    N^2 &= - \frac{g}{r} \Bigg(\frac{1}{\gamma} \frac{d \log p}{d \log r} - \frac{d \log \rho}{d \log r}\Bigg) ~.
    \\&=   g^2\Bigg(\frac{1}{c_\mr{eq}^2}  - \frac{1}{c_s^2} \Bigg) \approx g^2\Bigg(\frac{1}{c_\mr{eq}^2}  - \frac{1}{c_\mr{ins}^2} \Bigg)~,
    \nnm
\end{alignat}
which is non-zero. Physically, the Brunt-V\"{a}is\"{a}l\"{a} frequency captures the derivation of the perturbations from chemical equilibrium. \footnote{Note that for main sequence stars, perturbations beyond thermal equilibrium can also lead to non-zero $N$ \cite{smeyers2011linear}.} We will see shortly that this yields the characteristic frequencies of gravity ($g$-)modes in the stellar oscillations. 

Starting from the pioneer work by Cowling \cite{Cowling:1941nqk} and Chandrasekhar \cite{chandrasekhar1960general}, it has been shown that the fluid displacement vector $\boldsymbol{\xi}$ governed by the system of Eqs.~\eqref{eq:Newtonian_fluid_pert} admits a decomposition into an Eigenbasis $\boldsymbol{\xi}_i$, which have real eigenvalues $\omega_i^2$, and form a complete basis of the system. The index $i$ is discrete when $\rho$ has compact support.
A great effort has then been devoted into the analysis of these fluid oscillation modes (see \cite{Cowling:1941nqk} for a review). Basically, one can classify the stellar oscillations via their dispersion relations and the corresponding restoring force. For acoustic waves, pressure serves as the restoring force and leads to the dispersion relation
\begin{equation}
    \omega^2 \sim c_s^2 |\boldsymbol{k}|^2
\end{equation}
where we denote the wavenumber as $\boldsymbol{k} \equiv k_r \hat{r} + \boldsymbol{k}_{\rm H}$. These modes are also known as $p-$modes. The second type of waves are the gravity waves known as $g-$modes where buoyancy force serves as the restoring force. The corresponding dispersion relation can be written as
\begin{equation}
    \omega^2 \sim N^2 \frac{|\boldsymbol{k}_{\rm H}|^2}{|\boldsymbol{k}|^2} ~.
    \label{eq : gmode}
\end{equation}
The detailed analysis in Ref.~\cite{Cowling:1941nqk} also shows that all the $p-$ and $g-$modes have infinite number of overtones $n = 1,2, \cdots$. With the increasing of $n$, the eigen-frequency of $p-$modes will increase while it will decrease for $g-$modes. Moreover, depending on the sign of the Brunt-V\"{a}is\"{a}l\"{a} frequency $N$, the eigen-frequency of $g-$modes can be either positive or negative. In this work, we restrict to the case $N^2>0$. Note that the frequency of the g-modes is bounded from above by $N$\footnote{Since $N$ varies throughout the star, its value in Eq.~(\ref{eq : gmode}) should be taken as its characteristic/typical value $N_\mr{ch}$ inside the star.}.

Now, let us consider the approach towards the small frequency limit in the system of Eqs.~\eqref{eq:Newtonian_fluid_pert}. Examining the second equation in them, we find that there exists a formal mathematical ambiguity in the small frequency limit depending on the relative ordering of the Brunt-V\"{a}is\"{a}l\"{a} frequency $N$ and perturbation $\omega$. When $\omega\ll N_\mr{ch}$, assuming that all variables have a well-defined static limit ($\omega=0$), the second equation shrinks to a constraint equation between the fluid radial displacement and the perturbations of Newtonian potential
\begin{equation}
    w_1 = - \frac{r}{g} w_3 ~.
    \label{eq : newton_constraint}
\end{equation}
Combined with the first equation, this yields in the static limit
\begin{equation}
    (r w_1)' - \ell(\ell+1) w_2 = 0 ~,
    \label{eq : newton_div0}
\end{equation}
which is equivalent to the condition $\nabla \cdot \boldsymbol{\xi}=0$. Thus, in the static limit, the fluid is incompressible as shown in the relativistic case in Sec.~\ref{sec : static_limit}.

However, if the regime of interest has $\omega\geq N$, which is true for instance in the fast reaction case, where the Brunt-V\"{a}is\"{a}l\"{a} frequency $N \approx 0$ and thus $\omega\gg N_\mr{ch}$. In this regime, the orbital-frequency also exceeds all the $g$-mode resonant frequencies. We cannot then take $\omega\rightarrow 0$ disregarding all other parameters in Eq.~(\ref{eq:Newtonian_fluid_pert}) as before. In this case, the arguments leading to Eq.~(\ref{eq : newton_constraint}), \eqref{eq : newton_div0} and thus adiabatic-incompressibility is no longer true. While the specific details may differ in the relativistic case, the general feature that the scheme presented in this work is valid for $\omega\ll N_\mr{ch}$ is not likely to change. Thus, the low-frequency expansion presented in the main-text is likely valid in the regime $\omega \ll N_\mr{ch}$, where $N_\mr{ch}$ is the characteristic value of $N$ in the stellar interior.

The results for tidal heating in the barotropic regime ($N=0\equiv N_\mr{ch}\ll \omega$) were shown for relativistic polytropic neutron stars in Ref.~\cite{HegadeKR:2024agt}, where a small but non-zero contribution to the dissipation number due to bulk-viscosity (which is relevant only when $\nabla\cdot\bold{\xi}\neq 0$) was obtained. In an actual inspiral, the binary usually transitions from $\omega\ll N$ to $\omega \geq N$. A more refined study including the transition could help better understand the tidal characteristics and their influence on the waveform, especially the contribution due to bulk-viscosity.
\bibliography{references}

\begin{thebibliography}{134}%
\makeatletter
\providecommand \@ifxundefined [1]{%
 \@ifx{#1\undefined}
}%
\providecommand \@ifnum [1]{%
 \ifnum #1\expandafter \@firstoftwo
 \else \expandafter \@secondoftwo
 \fi
}%
\providecommand \@ifx [1]{%
 \ifx #1\expandafter \@firstoftwo
 \else \expandafter \@secondoftwo
 \fi
}%
\providecommand \natexlab [1]{#1}%
\providecommand \enquote  [1]{``#1''}%
\providecommand \bibnamefont  [1]{#1}%
\providecommand \bibfnamefont [1]{#1}%
\providecommand \citenamefont [1]{#1}%
\providecommand \href@noop [0]{\@secondoftwo}%
\providecommand \href [0]{\begingroup \@sanitize@url \@href}%
\providecommand \@href[1]{\@@startlink{#1}\@@href}%
\providecommand \@@href[1]{\endgroup#1\@@endlink}%
\providecommand \@sanitize@url [0]{\catcode `\\12\catcode `\$12\catcode `\&12\catcode `\#12\catcode `\^12\catcode `\_12\catcode `\%12\relax}%
\providecommand \@@startlink[1]{}%
\providecommand \@@endlink[0]{}%
\providecommand \url  [0]{\begingroup\@sanitize@url \@url }%
\providecommand \@url [1]{\endgroup\@href {#1}{\urlprefix }}%
\providecommand \urlprefix  [0]{URL }%
\providecommand \Eprint [0]{\href }%
\providecommand \doibase [0]{https://doi.org/}%
\providecommand \selectlanguage [0]{\@gobble}%
\providecommand \bibinfo  [0]{\@secondoftwo}%
\providecommand \bibfield  [0]{\@secondoftwo}%
\providecommand \translation [1]{[#1]}%
\providecommand \BibitemOpen [0]{}%
\providecommand \bibitemStop [0]{}%
\providecommand \bibitemNoStop [0]{.\EOS\space}%
\providecommand \EOS [0]{\spacefactor3000\relax}%
\providecommand \BibitemShut  [1]{\csname bibitem#1\endcsname}%
\let\auto@bib@innerbib\@empty
%</preamble>
\bibitem [{\citenamefont {Abbott}\ \emph {et~al.}(2019{\natexlab{a}})\citenamefont {Abbott} \emph {et~al.}}]{LIGOScientific:2018mvr}%
  \BibitemOpen
  \bibfield  {author} {\bibinfo {author} {\bibfnamefont {B.}~\bibnamefont {Abbott}} \emph {et~al.} (\bibinfo {collaboration} {LIGO Scientific, Virgo}),\ }\bibfield  {title} {\bibinfo {title} {{GWTC-1: A Gravitational-Wave Transient Catalog of Compact Binary Mergers Observed by LIGO and Virgo during the First and Second Observing Runs}},\ }\href {https://doi.org/10.1103/PhysRevX.9.031040} {\bibfield  {journal} {\bibinfo  {journal} {Phys. Rev. X}\ }\textbf {\bibinfo {volume} {9}},\ \bibinfo {pages} {031040} (\bibinfo {year} {2019}{\natexlab{a}})},\ \Eprint {https://arxiv.org/abs/1811.12907} {arXiv:1811.12907 [astro-ph.HE]} \BibitemShut {NoStop}%
\bibitem [{\citenamefont {Abbott}\ \emph {et~al.}(2021{\natexlab{a}})\citenamefont {Abbott} \emph {et~al.}}]{LIGOScientific:2020ibl}%
  \BibitemOpen
  \bibfield  {author} {\bibinfo {author} {\bibfnamefont {R.}~\bibnamefont {Abbott}} \emph {et~al.} (\bibinfo {collaboration} {LIGO Scientific, Virgo}),\ }\bibfield  {title} {\bibinfo {title} {{GWTC-2: Compact Binary Coalescences Observed by LIGO and Virgo During the First Half of the Third Observing Run}},\ }\href {https://doi.org/10.1103/PhysRevX.11.021053} {\bibfield  {journal} {\bibinfo  {journal} {Phys. Rev. X}\ }\textbf {\bibinfo {volume} {11}},\ \bibinfo {pages} {021053} (\bibinfo {year} {2021}{\natexlab{a}})},\ \Eprint {https://arxiv.org/abs/2010.14527} {arXiv:2010.14527 [gr-qc]} \BibitemShut {NoStop}%
\bibitem [{\citenamefont {Abbott}\ \emph {et~al.}(2024)\citenamefont {Abbott} \emph {et~al.}}]{LIGOScientific:2021usb}%
  \BibitemOpen
  \bibfield  {author} {\bibinfo {author} {\bibfnamefont {R.}~\bibnamefont {Abbott}} \emph {et~al.} (\bibinfo {collaboration} {LIGO Scientific, VIRGO}),\ }\bibfield  {title} {\bibinfo {title} {{GWTC-2.1: Deep extended catalog of compact binary coalescences observed by LIGO and Virgo during the first half of the third observing run}},\ }\href {https://doi.org/10.1103/PhysRevD.109.022001} {\bibfield  {journal} {\bibinfo  {journal} {Phys. Rev. D}\ }\textbf {\bibinfo {volume} {109}},\ \bibinfo {pages} {022001} (\bibinfo {year} {2024})},\ \Eprint {https://arxiv.org/abs/2108.01045} {arXiv:2108.01045 [gr-qc]} \BibitemShut {NoStop}%
\bibitem [{\citenamefont {Abbott}\ \emph {et~al.}(2021{\natexlab{b}})\citenamefont {Abbott} \emph {et~al.}}]{LIGOScientific:2021djp}%
  \BibitemOpen
  \bibfield  {author} {\bibinfo {author} {\bibfnamefont {R.}~\bibnamefont {Abbott}} \emph {et~al.} (\bibinfo {collaboration} {LIGO Scientific, VIRGO, KAGRA}),\ }\bibfield  {title} {\bibinfo {title} {{GWTC-3: Compact Binary Coalescences Observed by LIGO and Virgo During the Second Part of the Third Observing Run}},\ }\href@noop {} {\  (\bibinfo {year} {2021}{\natexlab{b}})},\ \Eprint {https://arxiv.org/abs/2111.03606} {arXiv:2111.03606 [gr-qc]} \BibitemShut {NoStop}%
\bibitem [{\citenamefont {Flanagan}\ and\ \citenamefont {Hinderer}(2008)}]{Flanagan:2007ix}%
  \BibitemOpen
  \bibfield  {author} {\bibinfo {author} {\bibfnamefont {E.~E.}\ \bibnamefont {Flanagan}}\ and\ \bibinfo {author} {\bibfnamefont {T.}~\bibnamefont {Hinderer}},\ }\bibfield  {title} {\bibinfo {title} {{Constraining neutron star tidal Love numbers with gravitational wave detectors}},\ }\href {https://doi.org/10.1103/PhysRevD.77.021502} {\bibfield  {journal} {\bibinfo  {journal} {Phys. Rev. D}\ }\textbf {\bibinfo {volume} {77}},\ \bibinfo {pages} {021502} (\bibinfo {year} {2008})},\ \Eprint {https://arxiv.org/abs/0709.1915} {arXiv:0709.1915 [astro-ph]} \BibitemShut {NoStop}%
\bibitem [{\citenamefont {Buonanno}\ \emph {et~al.}(2022)\citenamefont {Buonanno}, \citenamefont {Khalil}, \citenamefont {O'Connell}, \citenamefont {Roiban}, \citenamefont {Solon},\ and\ \citenamefont {Zeng}}]{Buonanno:2022pgc}%
  \BibitemOpen
  \bibfield  {author} {\bibinfo {author} {\bibfnamefont {A.}~\bibnamefont {Buonanno}}, \bibinfo {author} {\bibfnamefont {M.}~\bibnamefont {Khalil}}, \bibinfo {author} {\bibfnamefont {D.}~\bibnamefont {O'Connell}}, \bibinfo {author} {\bibfnamefont {R.}~\bibnamefont {Roiban}}, \bibinfo {author} {\bibfnamefont {M.~P.}\ \bibnamefont {Solon}},\ and\ \bibinfo {author} {\bibfnamefont {M.}~\bibnamefont {Zeng}},\ }\bibfield  {title} {\bibinfo {title} {{Snowmass White Paper: Gravitational Waves and Scattering Amplitudes}},\ }in\ \href@noop {} {\emph {\bibinfo {booktitle} {{Snowmass 2021}}}}\ (\bibinfo {year} {2022})\ \Eprint {https://arxiv.org/abs/2204.05194} {arXiv:2204.05194 [hep-th]} \BibitemShut {NoStop}%
\bibitem [{\citenamefont {Read}\ \emph {et~al.}(2009)\citenamefont {Read}, \citenamefont {Lackey}, \citenamefont {Owen},\ and\ \citenamefont {Friedman}}]{Read2009}%
  \BibitemOpen
  \bibfield  {author} {\bibinfo {author} {\bibfnamefont {J.~S.}\ \bibnamefont {Read}}, \bibinfo {author} {\bibfnamefont {B.~D.}\ \bibnamefont {Lackey}}, \bibinfo {author} {\bibfnamefont {B.~J.}\ \bibnamefont {Owen}},\ and\ \bibinfo {author} {\bibfnamefont {J.~L.}\ \bibnamefont {Friedman}},\ }\bibfield  {title} {\bibinfo {title} {Constraints on a phenomenologically parametrized neutron-star equation of state},\ }\bibfield  {journal} {\bibinfo  {journal} {Physical Review D}\ }\textbf {\bibinfo {volume} {79}},\ \href {https://doi.org/10.1103/physrevd.79.124032} {10.1103/physrevd.79.124032} (\bibinfo {year} {2009})\BibitemShut {NoStop}%
\bibitem [{\citenamefont {Abbott}\ \emph {et~al.}(2017)\citenamefont {Abbott} \emph {et~al.}}]{LIGOScientific:2017vwq}%
  \BibitemOpen
  \bibfield  {author} {\bibinfo {author} {\bibfnamefont {B.~P.}\ \bibnamefont {Abbott}} \emph {et~al.} (\bibinfo {collaboration} {LIGO Scientific, Virgo}),\ }\bibfield  {title} {\bibinfo {title} {{GW170817: Observation of Gravitational Waves from a Binary Neutron Star Inspiral}},\ }\href {https://doi.org/10.1103/PhysRevLett.119.161101} {\bibfield  {journal} {\bibinfo  {journal} {Phys. Rev. Lett.}\ }\textbf {\bibinfo {volume} {119}},\ \bibinfo {pages} {161101} (\bibinfo {year} {2017})},\ \Eprint {https://arxiv.org/abs/1710.05832} {arXiv:1710.05832 [gr-qc]} \BibitemShut {NoStop}%
\bibitem [{\citenamefont {Abbott}\ \emph {et~al.}(2019{\natexlab{b}})\citenamefont {Abbott} \emph {et~al.}}]{LIGOScientific:2018hze}%
  \BibitemOpen
  \bibfield  {author} {\bibinfo {author} {\bibfnamefont {B.~P.}\ \bibnamefont {Abbott}} \emph {et~al.} (\bibinfo {collaboration} {LIGO Scientific, Virgo}),\ }\bibfield  {title} {\bibinfo {title} {{Properties of the binary neutron star merger GW170817}},\ }\href {https://doi.org/10.1103/PhysRevX.9.011001} {\bibfield  {journal} {\bibinfo  {journal} {Phys. Rev. X}\ }\textbf {\bibinfo {volume} {9}},\ \bibinfo {pages} {011001} (\bibinfo {year} {2019}{\natexlab{b}})},\ \Eprint {https://arxiv.org/abs/1805.11579} {arXiv:1805.11579 [gr-qc]} \BibitemShut {NoStop}%
\bibitem [{\citenamefont {Abbott}\ \emph {et~al.}(2018)\citenamefont {Abbott} \emph {et~al.}}]{LIGOScientific:2018cki}%
  \BibitemOpen
  \bibfield  {author} {\bibinfo {author} {\bibfnamefont {B.~P.}\ \bibnamefont {Abbott}} \emph {et~al.} (\bibinfo {collaboration} {LIGO Scientific, Virgo}),\ }\bibfield  {title} {\bibinfo {title} {{GW170817: Measurements of neutron star radii and equation of state}},\ }\href {https://doi.org/10.1103/PhysRevLett.121.161101} {\bibfield  {journal} {\bibinfo  {journal} {Phys. Rev. Lett.}\ }\textbf {\bibinfo {volume} {121}},\ \bibinfo {pages} {161101} (\bibinfo {year} {2018})},\ \Eprint {https://arxiv.org/abs/1805.11581} {arXiv:1805.11581 [gr-qc]} \BibitemShut {NoStop}%
\bibitem [{\citenamefont {Coughlin}\ \emph {et~al.}(2019)\citenamefont {Coughlin}, \citenamefont {Dietrich}, \citenamefont {Margalit},\ and\ \citenamefont {Metzger}}]{Coughlin2019}%
  \BibitemOpen
  \bibfield  {author} {\bibinfo {author} {\bibfnamefont {M.~W.}\ \bibnamefont {Coughlin}}, \bibinfo {author} {\bibfnamefont {T.}~\bibnamefont {Dietrich}}, \bibinfo {author} {\bibfnamefont {B.}~\bibnamefont {Margalit}},\ and\ \bibinfo {author} {\bibfnamefont {B.~D.}\ \bibnamefont {Metzger}},\ }\bibfield  {title} {\bibinfo {title} {Multimessenger bayesian parameter inference of a binary neutron star merger},\ }\href {https://doi.org/10.1093/mnrasl/slz133} {\bibfield  {journal} {\bibinfo  {journal} {Monthly Notices of the Royal Astronomical Society: Letters}\ }\textbf {\bibinfo {volume} {489}},\ \bibinfo {pages} {L91–L96} (\bibinfo {year} {2019})}\BibitemShut {NoStop}%
\bibitem [{\citenamefont {Biswas}\ \emph {et~al.}(2021)\citenamefont {Biswas}, \citenamefont {Char}, \citenamefont {Nandi},\ and\ \citenamefont {Bose}}]{Biswas2021}%
  \BibitemOpen
  \bibfield  {author} {\bibinfo {author} {\bibfnamefont {B.}~\bibnamefont {Biswas}}, \bibinfo {author} {\bibfnamefont {P.}~\bibnamefont {Char}}, \bibinfo {author} {\bibfnamefont {R.}~\bibnamefont {Nandi}},\ and\ \bibinfo {author} {\bibfnamefont {S.}~\bibnamefont {Bose}},\ }\bibfield  {title} {\bibinfo {title} {Towards mitigation of apparent tension between nuclear physics and astrophysical observations by improved modeling of neutron star matter},\ }\bibfield  {journal} {\bibinfo  {journal} {Physical Review D}\ }\textbf {\bibinfo {volume} {103}},\ \href {https://doi.org/10.1103/physrevd.103.103015} {10.1103/physrevd.103.103015} (\bibinfo {year} {2021})\BibitemShut {NoStop}%
\bibitem [{\citenamefont {Dietrich}\ \emph {et~al.}(2020)\citenamefont {Dietrich}, \citenamefont {Coughlin}, \citenamefont {Pang}, \citenamefont {Bulla}, \citenamefont {Heinzel}, \citenamefont {Issa}, \citenamefont {Tews},\ and\ \citenamefont {Antier}}]{Dietrich2020}%
  \BibitemOpen
  \bibfield  {author} {\bibinfo {author} {\bibfnamefont {T.}~\bibnamefont {Dietrich}}, \bibinfo {author} {\bibfnamefont {M.~W.}\ \bibnamefont {Coughlin}}, \bibinfo {author} {\bibfnamefont {P.~T.~H.}\ \bibnamefont {Pang}}, \bibinfo {author} {\bibfnamefont {M.}~\bibnamefont {Bulla}}, \bibinfo {author} {\bibfnamefont {J.}~\bibnamefont {Heinzel}}, \bibinfo {author} {\bibfnamefont {L.}~\bibnamefont {Issa}}, \bibinfo {author} {\bibfnamefont {I.}~\bibnamefont {Tews}},\ and\ \bibinfo {author} {\bibfnamefont {S.}~\bibnamefont {Antier}},\ }\bibfield  {title} {\bibinfo {title} {Multimessenger constraints on the neutron-star equation of state and the hubble constant},\ }\href {https://doi.org/10.1126/science.abb4317} {\bibfield  {journal} {\bibinfo  {journal} {Science}\ }\textbf {\bibinfo {volume} {370}},\ \bibinfo {pages} {1450–1453} (\bibinfo {year} {2020})}\BibitemShut {NoStop}%
\bibitem [{\citenamefont {O’Boyle}\ \emph {et~al.}(2020)\citenamefont {O’Boyle}, \citenamefont {Markakis}, \citenamefont {Stergioulas},\ and\ \citenamefont {Read}}]{O_Boyle_2020}%
  \BibitemOpen
  \bibfield  {author} {\bibinfo {author} {\bibfnamefont {M.~F.}\ \bibnamefont {O’Boyle}}, \bibinfo {author} {\bibfnamefont {C.}~\bibnamefont {Markakis}}, \bibinfo {author} {\bibfnamefont {N.}~\bibnamefont {Stergioulas}},\ and\ \bibinfo {author} {\bibfnamefont {J.~S.}\ \bibnamefont {Read}},\ }\bibfield  {title} {\bibinfo {title} {Parametrized equation of state for neutron star matter with continuous sound speed},\ }\bibfield  {journal} {\bibinfo  {journal} {Physical Review D}\ }\textbf {\bibinfo {volume} {102}},\ \href {https://doi.org/10.1103/physrevd.102.083027} {10.1103/physrevd.102.083027} (\bibinfo {year} {2020})\BibitemShut {NoStop}%
\bibitem [{\citenamefont {Ghosh}\ \emph {et~al.}(2022{\natexlab{a}})\citenamefont {Ghosh}, \citenamefont {Chatterjee},\ and\ \citenamefont {Schaffner-Bielich}}]{Ghosh2022}%
  \BibitemOpen
  \bibfield  {author} {\bibinfo {author} {\bibfnamefont {S.}~\bibnamefont {Ghosh}}, \bibinfo {author} {\bibfnamefont {D.}~\bibnamefont {Chatterjee}},\ and\ \bibinfo {author} {\bibfnamefont {J.}~\bibnamefont {Schaffner-Bielich}},\ }\bibfield  {title} {\bibinfo {title} {Imposing multi-physics constraints at different densities on the neutron star equation of state},\ }\bibfield  {journal} {\bibinfo  {journal} {The European Physical Journal A}\ }\textbf {\bibinfo {volume} {58}},\ \href {https://doi.org/10.1140/epja/s10050-022-00679-w} {10.1140/epja/s10050-022-00679-w} (\bibinfo {year} {2022}{\natexlab{a}})\BibitemShut {NoStop}%
\bibitem [{\citenamefont {Ghosh}\ \emph {et~al.}(2024)\citenamefont {Ghosh}, \citenamefont {Pradhan},\ and\ \citenamefont {Chatterjee}}]{Ghosh:2023vrx}%
  \BibitemOpen
  \bibfield  {author} {\bibinfo {author} {\bibfnamefont {S.}~\bibnamefont {Ghosh}}, \bibinfo {author} {\bibfnamefont {B.~K.}\ \bibnamefont {Pradhan}},\ and\ \bibinfo {author} {\bibfnamefont {D.}~\bibnamefont {Chatterjee}},\ }\bibfield  {title} {\bibinfo {title} {{Tidal heating as a direct probe of strangeness inside neutron stars}},\ }\href {https://doi.org/10.1103/PhysRevD.109.103036} {\bibfield  {journal} {\bibinfo  {journal} {Phys. Rev. D}\ }\textbf {\bibinfo {volume} {109}},\ \bibinfo {pages} {103036} (\bibinfo {year} {2024})},\ \Eprint {https://arxiv.org/abs/2306.14737} {arXiv:2306.14737 [gr-qc]} \BibitemShut {NoStop}%
\bibitem [{\citenamefont {Love}(1909)}]{love1909yielding}%
  \BibitemOpen
  \bibfield  {author} {\bibinfo {author} {\bibfnamefont {A.~E.~H.}\ \bibnamefont {Love}},\ }\bibfield  {title} {\bibinfo {title} {The yielding of the earth to disturbing forces},\ }\href@noop {} {\bibfield  {journal} {\bibinfo  {journal} {Proceedings of the Royal Society of London. Series A, Containing Papers of a Mathematical and Physical Character}\ }\textbf {\bibinfo {volume} {82}},\ \bibinfo {pages} {73} (\bibinfo {year} {1909})}\BibitemShut {NoStop}%
\bibitem [{\citenamefont {Will}(2014)}]{Will:2014kxa}%
  \BibitemOpen
  \bibfield  {author} {\bibinfo {author} {\bibfnamefont {C.~M.}\ \bibnamefont {Will}},\ }\bibfield  {title} {\bibinfo {title} {{The Confrontation between General Relativity and Experiment}},\ }\href {https://doi.org/10.12942/lrr-2014-4} {\bibfield  {journal} {\bibinfo  {journal} {Living Rev. Rel.}\ }\textbf {\bibinfo {volume} {17}},\ \bibinfo {pages} {4} (\bibinfo {year} {2014})},\ \Eprint {https://arxiv.org/abs/1403.7377} {arXiv:1403.7377 [gr-qc]} \BibitemShut {NoStop}%
%%CITATION = ARXIV:1403.7377;%%
\bibitem [{\citenamefont {{Ogilvie}}(2014)}]{2014ARA&A..52..171O}%
  \BibitemOpen
  \bibfield  {author} {\bibinfo {author} {\bibfnamefont {G.~I.}\ \bibnamefont {{Ogilvie}}},\ }\bibfield  {title} {\bibinfo {title} {{Tidal Dissipation in Stars and Giant Planets}},\ }\href {https://doi.org/10.1146/annurev-astro-081913-035941} {\bibfield  {journal} {\bibinfo  {journal} {ARA\&A}\ }\textbf {\bibinfo {volume} {52}},\ \bibinfo {pages} {171} (\bibinfo {year} {2014})},\ \Eprint {https://arxiv.org/abs/1406.2207} {arXiv:1406.2207 [astro-ph.SR]} \BibitemShut {NoStop}%
\bibitem [{\citenamefont {Terquem}\ \emph {et~al.}(1998)\citenamefont {Terquem}, \citenamefont {Papaloizou}, \citenamefont {Nelson},\ and\ \citenamefont {Lin}}]{Terquem:1998ya}%
  \BibitemOpen
  \bibfield  {author} {\bibinfo {author} {\bibfnamefont {C.}~\bibnamefont {Terquem}}, \bibinfo {author} {\bibfnamefont {J.~C.~B.}\ \bibnamefont {Papaloizou}}, \bibinfo {author} {\bibfnamefont {R.~P.}\ \bibnamefont {Nelson}},\ and\ \bibinfo {author} {\bibfnamefont {D.~N.~C.}\ \bibnamefont {Lin}},\ }\bibfield  {title} {\bibinfo {title} {{On the tidal interaction of a solar-type star with an orbiting companion: excitation of g mode oscillation and orbital evolution}},\ }\href {https://doi.org/10.1086/305927} {\bibfield  {journal} {\bibinfo  {journal} {Astrophys. J.}\ }\textbf {\bibinfo {volume} {502}},\ \bibinfo {pages} {788} (\bibinfo {year} {1998})},\ \Eprint {https://arxiv.org/abs/astro-ph/9801280} {arXiv:astro-ph/9801280} \BibitemShut {NoStop}%
\bibitem [{\citenamefont {Poisson}\ and\ \citenamefont {Will}(1953)}]{PoissonWill}%
  \BibitemOpen
  \bibfield  {author} {\bibinfo {author} {\bibfnamefont {E.}~\bibnamefont {Poisson}}\ and\ \bibinfo {author} {\bibfnamefont {C.}~\bibnamefont {Will}},\ }\href@noop {} {\emph {\bibinfo {title} {{Gravity: Newtonian, Post-Newtonian, Relativistic}}}}\ (\bibinfo  {publisher} {Cambridge University Press},\ \bibinfo {address} {Cambridge, UK},\ \bibinfo {year} {1953})\BibitemShut {NoStop}%
\bibitem [{\citenamefont {Goldberger}\ and\ \citenamefont {Rothstein}(2006{\natexlab{a}})}]{Goldberger:2004jt}%
  \BibitemOpen
  \bibfield  {author} {\bibinfo {author} {\bibfnamefont {W.~D.}\ \bibnamefont {Goldberger}}\ and\ \bibinfo {author} {\bibfnamefont {I.~Z.}\ \bibnamefont {Rothstein}},\ }\bibfield  {title} {\bibinfo {title} {{An Effective field theory of gravity for extended objects}},\ }\href {https://doi.org/10.1103/PhysRevD.73.104029} {\bibfield  {journal} {\bibinfo  {journal} {Phys. Rev. D}\ }\textbf {\bibinfo {volume} {73}},\ \bibinfo {pages} {104029} (\bibinfo {year} {2006}{\natexlab{a}})},\ \Eprint {https://arxiv.org/abs/hep-th/0409156} {arXiv:hep-th/0409156} \BibitemShut {NoStop}%
\bibitem [{\citenamefont {Goldberger}\ and\ \citenamefont {Rothstein}(2006{\natexlab{b}})}]{Goldberger:2005cd}%
  \BibitemOpen
  \bibfield  {author} {\bibinfo {author} {\bibfnamefont {W.~D.}\ \bibnamefont {Goldberger}}\ and\ \bibinfo {author} {\bibfnamefont {I.~Z.}\ \bibnamefont {Rothstein}},\ }\bibfield  {title} {\bibinfo {title} {{Dissipative effects in the worldline approach to black hole dynamics}},\ }\href {https://doi.org/10.1103/PhysRevD.73.104030} {\bibfield  {journal} {\bibinfo  {journal} {Phys. Rev. D}\ }\textbf {\bibinfo {volume} {73}},\ \bibinfo {pages} {104030} (\bibinfo {year} {2006}{\natexlab{b}})},\ \Eprint {https://arxiv.org/abs/hep-th/0511133} {arXiv:hep-th/0511133} \BibitemShut {NoStop}%
\bibitem [{\citenamefont {Goldberger}\ and\ \citenamefont {Ross}(2010)}]{Goldberger:2009qd}%
  \BibitemOpen
  \bibfield  {author} {\bibinfo {author} {\bibfnamefont {W.~D.}\ \bibnamefont {Goldberger}}\ and\ \bibinfo {author} {\bibfnamefont {A.}~\bibnamefont {Ross}},\ }\bibfield  {title} {\bibinfo {title} {{Gravitational radiative corrections from effective field theory}},\ }\href {https://doi.org/10.1103/PhysRevD.81.124015} {\bibfield  {journal} {\bibinfo  {journal} {Phys. Rev. D}\ }\textbf {\bibinfo {volume} {81}},\ \bibinfo {pages} {124015} (\bibinfo {year} {2010})},\ \Eprint {https://arxiv.org/abs/0912.4254} {arXiv:0912.4254 [gr-qc]} \BibitemShut {NoStop}%
\bibitem [{\citenamefont {Porto}(2016)}]{Porto:2016pyg}%
  \BibitemOpen
  \bibfield  {author} {\bibinfo {author} {\bibfnamefont {R.~A.}\ \bibnamefont {Porto}},\ }\bibfield  {title} {\bibinfo {title} {{The effective field theorist\textquoteright{}s approach to gravitational dynamics}},\ }\href {https://doi.org/10.1016/j.physrep.2016.04.003} {\bibfield  {journal} {\bibinfo  {journal} {Phys. Rept.}\ }\textbf {\bibinfo {volume} {633}},\ \bibinfo {pages} {1} (\bibinfo {year} {2016})},\ \Eprint {https://arxiv.org/abs/1601.04914} {arXiv:1601.04914 [hep-th]} \BibitemShut {NoStop}%
\bibitem [{\citenamefont {Levi}(2020)}]{Levi:2018nxp}%
  \BibitemOpen
  \bibfield  {author} {\bibinfo {author} {\bibfnamefont {M.}~\bibnamefont {Levi}},\ }\bibfield  {title} {\bibinfo {title} {{Effective Field Theories of Post-Newtonian Gravity: A comprehensive review}},\ }\href {https://doi.org/10.1088/1361-6633/ab12bc} {\bibfield  {journal} {\bibinfo  {journal} {Rept. Prog. Phys.}\ }\textbf {\bibinfo {volume} {83}},\ \bibinfo {pages} {075901} (\bibinfo {year} {2020})},\ \Eprint {https://arxiv.org/abs/1807.01699} {arXiv:1807.01699 [hep-th]} \BibitemShut {NoStop}%
\bibitem [{\citenamefont {Goldberger}\ \emph {et~al.}(2021)\citenamefont {Goldberger}, \citenamefont {Li},\ and\ \citenamefont {Rothstein}}]{Goldberger:2020fot}%
  \BibitemOpen
  \bibfield  {author} {\bibinfo {author} {\bibfnamefont {W.~D.}\ \bibnamefont {Goldberger}}, \bibinfo {author} {\bibfnamefont {J.}~\bibnamefont {Li}},\ and\ \bibinfo {author} {\bibfnamefont {I.~Z.}\ \bibnamefont {Rothstein}},\ }\bibfield  {title} {\bibinfo {title} {{Non-conservative effects on spinning black holes from world-line effective field theory}},\ }\href {https://doi.org/10.1007/JHEP06(2021)053} {\bibfield  {journal} {\bibinfo  {journal} {JHEP}\ }\textbf {\bibinfo {volume} {06}},\ \bibinfo {pages} {053}},\ \Eprint {https://arxiv.org/abs/2012.14869} {arXiv:2012.14869 [hep-th]} \BibitemShut {NoStop}%
\bibitem [{\citenamefont {Saketh}\ \emph {et~al.}(2023)\citenamefont {Saketh}, \citenamefont {Steinhoff}, \citenamefont {Vines},\ and\ \citenamefont {Buonanno}}]{Saketh:2022xjb}%
  \BibitemOpen
  \bibfield  {author} {\bibinfo {author} {\bibfnamefont {M.~V.~S.}\ \bibnamefont {Saketh}}, \bibinfo {author} {\bibfnamefont {J.}~\bibnamefont {Steinhoff}}, \bibinfo {author} {\bibfnamefont {J.}~\bibnamefont {Vines}},\ and\ \bibinfo {author} {\bibfnamefont {A.}~\bibnamefont {Buonanno}},\ }\bibfield  {title} {\bibinfo {title} {{Modeling horizon absorption in spinning binary black holes using effective worldline theory}},\ }\href {https://doi.org/10.1103/PhysRevD.107.084006} {\bibfield  {journal} {\bibinfo  {journal} {Phys. Rev. D}\ }\textbf {\bibinfo {volume} {107}},\ \bibinfo {pages} {084006} (\bibinfo {year} {2023})},\ \Eprint {https://arxiv.org/abs/2212.13095} {arXiv:2212.13095 [gr-qc]} \BibitemShut {NoStop}%
\bibitem [{\citenamefont {Saketh}\ \emph {et~al.}(2024)\citenamefont {Saketh}, \citenamefont {Zhou},\ and\ \citenamefont {Ivanov}}]{Saketh:2023bul}%
  \BibitemOpen
  \bibfield  {author} {\bibinfo {author} {\bibfnamefont {M.~V.~S.}\ \bibnamefont {Saketh}}, \bibinfo {author} {\bibfnamefont {Z.}~\bibnamefont {Zhou}},\ and\ \bibinfo {author} {\bibfnamefont {M.~M.}\ \bibnamefont {Ivanov}},\ }\bibfield  {title} {\bibinfo {title} {{Dynamical tidal response of Kerr black holes from scattering amplitudes}},\ }\href {https://doi.org/10.1103/PhysRevD.109.064058} {\bibfield  {journal} {\bibinfo  {journal} {Phys. Rev. D}\ }\textbf {\bibinfo {volume} {109}},\ \bibinfo {pages} {064058} (\bibinfo {year} {2024})},\ \Eprint {https://arxiv.org/abs/2307.10391} {arXiv:2307.10391 [hep-th]} \BibitemShut {NoStop}%
\bibitem [{\citenamefont {Chia}(2021)}]{Chia:2020yla}%
  \BibitemOpen
  \bibfield  {author} {\bibinfo {author} {\bibfnamefont {H.~S.}\ \bibnamefont {Chia}},\ }\bibfield  {title} {\bibinfo {title} {{Tidal deformation and dissipation of rotating black holes}},\ }\href {https://doi.org/10.1103/PhysRevD.104.024013} {\bibfield  {journal} {\bibinfo  {journal} {Phys. Rev. D}\ }\textbf {\bibinfo {volume} {104}},\ \bibinfo {pages} {024013} (\bibinfo {year} {2021})},\ \Eprint {https://arxiv.org/abs/2010.07300} {arXiv:2010.07300 [gr-qc]} \BibitemShut {NoStop}%
\bibitem [{\citenamefont {Charalambous}\ \emph {et~al.}(2021)\citenamefont {Charalambous}, \citenamefont {Dubovsky},\ and\ \citenamefont {Ivanov}}]{Charalambous:2021mea}%
  \BibitemOpen
  \bibfield  {author} {\bibinfo {author} {\bibfnamefont {P.}~\bibnamefont {Charalambous}}, \bibinfo {author} {\bibfnamefont {S.}~\bibnamefont {Dubovsky}},\ and\ \bibinfo {author} {\bibfnamefont {M.~M.}\ \bibnamefont {Ivanov}},\ }\bibfield  {title} {\bibinfo {title} {{On the Vanishing of Love Numbers for Kerr Black Holes}},\ }\href {https://doi.org/10.1007/JHEP05(2021)038} {\bibfield  {journal} {\bibinfo  {journal} {JHEP}\ }\textbf {\bibinfo {volume} {05}},\ \bibinfo {pages} {038}},\ \Eprint {https://arxiv.org/abs/2102.08917} {arXiv:2102.08917 [hep-th]} \BibitemShut {NoStop}%
\bibitem [{\citenamefont {Ivanov}\ and\ \citenamefont {Zhou}(2023)}]{Ivanov:2022qqt}%
  \BibitemOpen
  \bibfield  {author} {\bibinfo {author} {\bibfnamefont {M.~M.}\ \bibnamefont {Ivanov}}\ and\ \bibinfo {author} {\bibfnamefont {Z.}~\bibnamefont {Zhou}},\ }\bibfield  {title} {\bibinfo {title} {{Vanishing of Black Hole Tidal Love Numbers from Scattering Amplitudes}},\ }\href {https://doi.org/10.1103/PhysRevLett.130.091403} {\bibfield  {journal} {\bibinfo  {journal} {Phys. Rev. Lett.}\ }\textbf {\bibinfo {volume} {130}},\ \bibinfo {pages} {091403} (\bibinfo {year} {2023})},\ \Eprint {https://arxiv.org/abs/2209.14324} {arXiv:2209.14324 [hep-th]} \BibitemShut {NoStop}%
\bibitem [{\citenamefont {Kol}\ and\ \citenamefont {Smolkin}(2012)}]{Kol:2011vg}%
  \BibitemOpen
  \bibfield  {author} {\bibinfo {author} {\bibfnamefont {B.}~\bibnamefont {Kol}}\ and\ \bibinfo {author} {\bibfnamefont {M.}~\bibnamefont {Smolkin}},\ }\bibfield  {title} {\bibinfo {title} {{Black hole stereotyping: Induced gravito-static polarization}},\ }\href {https://doi.org/10.1007/JHEP02(2012)010} {\bibfield  {journal} {\bibinfo  {journal} {JHEP}\ }\textbf {\bibinfo {volume} {02}},\ \bibinfo {pages} {010}},\ \Eprint {https://arxiv.org/abs/1110.3764} {arXiv:1110.3764 [hep-th]} \BibitemShut {NoStop}%
\bibitem [{\citenamefont {Hui}\ \emph {et~al.}(2021)\citenamefont {Hui}, \citenamefont {Joyce}, \citenamefont {Penco}, \citenamefont {Santoni},\ and\ \citenamefont {Solomon}}]{Hui:2020xxx}%
  \BibitemOpen
  \bibfield  {author} {\bibinfo {author} {\bibfnamefont {L.}~\bibnamefont {Hui}}, \bibinfo {author} {\bibfnamefont {A.}~\bibnamefont {Joyce}}, \bibinfo {author} {\bibfnamefont {R.}~\bibnamefont {Penco}}, \bibinfo {author} {\bibfnamefont {L.}~\bibnamefont {Santoni}},\ and\ \bibinfo {author} {\bibfnamefont {A.~R.}\ \bibnamefont {Solomon}},\ }\bibfield  {title} {\bibinfo {title} {{Static response and Love numbers of Schwarzschild black holes}},\ }\href {https://doi.org/10.1088/1475-7516/2021/04/052} {\bibfield  {journal} {\bibinfo  {journal} {JCAP}\ }\textbf {\bibinfo {volume} {04}},\ \bibinfo {pages} {052}},\ \Eprint {https://arxiv.org/abs/2010.00593} {arXiv:2010.00593 [hep-th]} \BibitemShut {NoStop}%
\bibitem [{\citenamefont {Binnington}\ and\ \citenamefont {Poisson}(2009)}]{Binnington:2009bb}%
  \BibitemOpen
  \bibfield  {author} {\bibinfo {author} {\bibfnamefont {T.}~\bibnamefont {Binnington}}\ and\ \bibinfo {author} {\bibfnamefont {E.}~\bibnamefont {Poisson}},\ }\bibfield  {title} {\bibinfo {title} {{Relativistic theory of tidal Love numbers}},\ }\href {https://doi.org/10.1103/PhysRevD.80.084018} {\bibfield  {journal} {\bibinfo  {journal} {Phys. Rev. D}\ }\textbf {\bibinfo {volume} {80}},\ \bibinfo {pages} {084018} (\bibinfo {year} {2009})},\ \Eprint {https://arxiv.org/abs/0906.1366} {arXiv:0906.1366 [gr-qc]} \BibitemShut {NoStop}%
\bibitem [{\citenamefont {Damour}\ and\ \citenamefont {Nagar}(2009)}]{Damour:2009vw}%
  \BibitemOpen
  \bibfield  {author} {\bibinfo {author} {\bibfnamefont {T.}~\bibnamefont {Damour}}\ and\ \bibinfo {author} {\bibfnamefont {A.}~\bibnamefont {Nagar}},\ }\bibfield  {title} {\bibinfo {title} {{Relativistic tidal properties of neutron stars}},\ }\href {https://doi.org/10.1103/PhysRevD.80.084035} {\bibfield  {journal} {\bibinfo  {journal} {Phys. Rev. D}\ }\textbf {\bibinfo {volume} {80}},\ \bibinfo {pages} {084035} (\bibinfo {year} {2009})},\ \Eprint {https://arxiv.org/abs/0906.0096} {arXiv:0906.0096 [gr-qc]} \BibitemShut {NoStop}%
\bibitem [{\citenamefont {Hinderer}(2008)}]{Hinderer:2007mb}%
  \BibitemOpen
  \bibfield  {author} {\bibinfo {author} {\bibfnamefont {T.}~\bibnamefont {Hinderer}},\ }\bibfield  {title} {\bibinfo {title} {{Tidal Love numbers of neutron stars}},\ }\href {https://doi.org/10.1086/533487} {\bibfield  {journal} {\bibinfo  {journal} {Astrophys. J.}\ }\textbf {\bibinfo {volume} {677}},\ \bibinfo {pages} {1216} (\bibinfo {year} {2008})},\ \bibinfo {note} {[Erratum: Astrophys.J. 697, 964 (2009)]},\ \Eprint {https://arxiv.org/abs/0711.2420} {arXiv:0711.2420 [astro-ph]} \BibitemShut {NoStop}%
\bibitem [{\citenamefont {Creci}\ \emph {et~al.}(2021)\citenamefont {Creci}, \citenamefont {Hinderer},\ and\ \citenamefont {Steinhoff}}]{Creci:2021rkz}%
  \BibitemOpen
  \bibfield  {author} {\bibinfo {author} {\bibfnamefont {G.}~\bibnamefont {Creci}}, \bibinfo {author} {\bibfnamefont {T.}~\bibnamefont {Hinderer}},\ and\ \bibinfo {author} {\bibfnamefont {J.}~\bibnamefont {Steinhoff}},\ }\bibfield  {title} {\bibinfo {title} {{Tidal response from scattering and the role of analytic continuation}},\ }\href {https://doi.org/10.1103/PhysRevD.104.124061} {\bibfield  {journal} {\bibinfo  {journal} {Phys. Rev. D}\ }\textbf {\bibinfo {volume} {104}},\ \bibinfo {pages} {124061} (\bibinfo {year} {2021})},\ \bibinfo {note} {[Erratum: Phys.Rev.D 105, 109902 (2022)]},\ \Eprint {https://arxiv.org/abs/2108.03385} {arXiv:2108.03385 [gr-qc]} \BibitemShut {NoStop}%
\bibitem [{\citenamefont {Tagoshi}\ \emph {et~al.}(1997)\citenamefont {Tagoshi}, \citenamefont {Mano},\ and\ \citenamefont {Takasugi}}]{Tagoshi:1997jy}%
  \BibitemOpen
  \bibfield  {author} {\bibinfo {author} {\bibfnamefont {H.}~\bibnamefont {Tagoshi}}, \bibinfo {author} {\bibfnamefont {S.}~\bibnamefont {Mano}},\ and\ \bibinfo {author} {\bibfnamefont {E.}~\bibnamefont {Takasugi}},\ }\bibfield  {title} {\bibinfo {title} {{PostNewtonian expansion of gravitational waves from a particle in circular orbits around a rotating black hole: Effects of black hole absorption}},\ }\href {https://doi.org/10.1143/PTP.98.829} {\bibfield  {journal} {\bibinfo  {journal} {Prog. Theor. Phys.}\ }\textbf {\bibinfo {volume} {98}},\ \bibinfo {pages} {829} (\bibinfo {year} {1997})},\ \Eprint {https://arxiv.org/abs/gr-qc/9711072} {arXiv:gr-qc/9711072} \BibitemShut {NoStop}%
\bibitem [{\citenamefont {Chatziioannou}\ \emph {et~al.}(2016)\citenamefont {Chatziioannou}, \citenamefont {Poisson},\ and\ \citenamefont {Yunes}}]{Chatziioannou:2016kem}%
  \BibitemOpen
  \bibfield  {author} {\bibinfo {author} {\bibfnamefont {K.}~\bibnamefont {Chatziioannou}}, \bibinfo {author} {\bibfnamefont {E.}~\bibnamefont {Poisson}},\ and\ \bibinfo {author} {\bibfnamefont {N.}~\bibnamefont {Yunes}},\ }\bibfield  {title} {\bibinfo {title} {{Improved next-to-leading order tidal heating and torquing of a Kerr black hole}},\ }\href {https://doi.org/10.1103/PhysRevD.94.084043} {\bibfield  {journal} {\bibinfo  {journal} {Phys. Rev. D}\ }\textbf {\bibinfo {volume} {94}},\ \bibinfo {pages} {084043} (\bibinfo {year} {2016})},\ \Eprint {https://arxiv.org/abs/1608.02899} {arXiv:1608.02899 [gr-qc]} \BibitemShut {NoStop}%
\bibitem [{\citenamefont {Poisson}(2005)}]{Poisson:2005pi}%
  \BibitemOpen
  \bibfield  {author} {\bibinfo {author} {\bibfnamefont {E.}~\bibnamefont {Poisson}},\ }\bibfield  {title} {\bibinfo {title} {{Metric of a tidally distorted, nonrotating black hole}},\ }\href {https://doi.org/10.1103/PhysRevLett.94.161103} {\bibfield  {journal} {\bibinfo  {journal} {Phys. Rev. Lett.}\ }\textbf {\bibinfo {volume} {94}},\ \bibinfo {pages} {161103} (\bibinfo {year} {2005})},\ \Eprint {https://arxiv.org/abs/gr-qc/0501032} {arXiv:gr-qc/0501032} \BibitemShut {NoStop}%
\bibitem [{\citenamefont {Chia}\ \emph {et~al.}(2024)\citenamefont {Chia}, \citenamefont {Zhou},\ and\ \citenamefont {Ivanov}}]{Chia:2024bwc}%
  \BibitemOpen
  \bibfield  {author} {\bibinfo {author} {\bibfnamefont {H.~S.}\ \bibnamefont {Chia}}, \bibinfo {author} {\bibfnamefont {Z.}~\bibnamefont {Zhou}},\ and\ \bibinfo {author} {\bibfnamefont {M.~M.}\ \bibnamefont {Ivanov}},\ }\bibfield  {title} {\bibinfo {title} {{Bring the Heat: Tidal Heating Constraints for Black Holes and Exotic Compact Objects from the LIGO-Virgo-KAGRA Data}},\ }\href@noop {} {\  (\bibinfo {year} {2024})},\ \Eprint {https://arxiv.org/abs/2404.14641} {arXiv:2404.14641 [gr-qc]} \BibitemShut {NoStop}%
\bibitem [{\citenamefont {Jones}(2001)}]{Jones2001}%
  \BibitemOpen
  \bibfield  {author} {\bibinfo {author} {\bibfnamefont {P.~B.}\ \bibnamefont {Jones}},\ }\bibfield  {title} {\bibinfo {title} {Bulk viscosity of neutron-star matter},\ }\href {https://doi.org/10.1103/PhysRevD.64.084003} {\bibfield  {journal} {\bibinfo  {journal} {Phys. Rev. D}\ }\textbf {\bibinfo {volume} {64}},\ \bibinfo {pages} {084003} (\bibinfo {year} {2001})}\BibitemShut {NoStop}%
\bibitem [{\citenamefont {Lindblom}\ and\ \citenamefont {Owen}(2002)}]{Lindblom2002}%
  \BibitemOpen
  \bibfield  {author} {\bibinfo {author} {\bibfnamefont {L.}~\bibnamefont {Lindblom}}\ and\ \bibinfo {author} {\bibfnamefont {B.~J.}\ \bibnamefont {Owen}},\ }\bibfield  {title} {\bibinfo {title} {Effect of hyperon bulk viscosity on neutron-star r-modes},\ }\href {https://doi.org/10.1103/PhysRevD.65.063006} {\bibfield  {journal} {\bibinfo  {journal} {Phys. Rev. D}\ }\textbf {\bibinfo {volume} {65}},\ \bibinfo {pages} {063006} (\bibinfo {year} {2002})}\BibitemShut {NoStop}%
\bibitem [{\citenamefont {Passamonti}\ and\ \citenamefont {Glampedakis}(2012)}]{Passamonti_2012}%
  \BibitemOpen
  \bibfield  {author} {\bibinfo {author} {\bibfnamefont {A.}~\bibnamefont {Passamonti}}\ and\ \bibinfo {author} {\bibfnamefont {K.}~\bibnamefont {Glampedakis}},\ }\bibfield  {title} {\bibinfo {title} {Non-linear viscous damping and gravitational wave detectability of the f-mode instability in neutron stars: The f-mode instability in neutron stars},\ }\href {https://doi.org/10.1111/j.1365-2966.2012.20849.x} {\bibfield  {journal} {\bibinfo  {journal} {Monthly Notices of the Royal Astronomical Society}\ }\textbf {\bibinfo {volume} {422}},\ \bibinfo {pages} {3327–3338} (\bibinfo {year} {2012})}\BibitemShut {NoStop}%
\bibitem [{\citenamefont {Alford}\ \emph {et~al.}(2011)\citenamefont {Alford}, \citenamefont {Mahmoodifar},\ and\ \citenamefont {Schwenzer}}]{Alford_2011}%
  \BibitemOpen
  \bibfield  {author} {\bibinfo {author} {\bibfnamefont {M.}~\bibnamefont {Alford}}, \bibinfo {author} {\bibfnamefont {S.}~\bibnamefont {Mahmoodifar}},\ and\ \bibinfo {author} {\bibfnamefont {K.}~\bibnamefont {Schwenzer}},\ }\bibfield  {title} {\bibinfo {title} {Non-linear viscous saturation of r-modes},\ }in\ \href {https://doi.org/10.1063/1.3575100} {\emph {\bibinfo {booktitle} {AIP Conference Proceedings}}}\ (\bibinfo  {publisher} {AIP},\ \bibinfo {year} {2011})\BibitemShut {NoStop}%
\bibitem [{\citenamefont {Hernandez}\ \emph {et~al.}(2024)\citenamefont {Hernandez}, \citenamefont {Manuel},\ and\ \citenamefont {Tolos}}]{hernandez2024dampingdensityoscillationsbulk}%
  \BibitemOpen
  \bibfield  {author} {\bibinfo {author} {\bibfnamefont {J.~L.}\ \bibnamefont {Hernandez}}, \bibinfo {author} {\bibfnamefont {C.}~\bibnamefont {Manuel}},\ and\ \bibinfo {author} {\bibfnamefont {L.}~\bibnamefont {Tolos}},\ }\href {https://arxiv.org/abs/2402.06595} {\bibinfo {title} {Damping of density oscillations from bulk viscosity in quark matter}} (\bibinfo {year} {2024}),\ \Eprint {https://arxiv.org/abs/2402.06595} {arXiv:2402.06595 [hep-ph]} \BibitemShut {NoStop}%
\bibitem [{\citenamefont {Alford}\ \emph {et~al.}(2018)\citenamefont {Alford}, \citenamefont {Bovard}, \citenamefont {Hanauske}, \citenamefont {Rezzolla},\ and\ \citenamefont {Schwenzer}}]{Alford_2017PRL}%
  \BibitemOpen
  \bibfield  {author} {\bibinfo {author} {\bibfnamefont {M.~G.}\ \bibnamefont {Alford}}, \bibinfo {author} {\bibfnamefont {L.}~\bibnamefont {Bovard}}, \bibinfo {author} {\bibfnamefont {M.}~\bibnamefont {Hanauske}}, \bibinfo {author} {\bibfnamefont {L.}~\bibnamefont {Rezzolla}},\ and\ \bibinfo {author} {\bibfnamefont {K.}~\bibnamefont {Schwenzer}},\ }\bibfield  {title} {\bibinfo {title} {Viscous dissipation and heat conduction in binary neutron-star mergers},\ }\href {https://doi.org/10.1103/PhysRevLett.120.041101} {\bibfield  {journal} {\bibinfo  {journal} {Phys. Rev. Lett.}\ }\textbf {\bibinfo {volume} {120}},\ \bibinfo {pages} {041101} (\bibinfo {year} {2018})}\BibitemShut {NoStop}%
\bibitem [{\citenamefont {Most}\ \emph {et~al.}(2021)\citenamefont {Most}, \citenamefont {Harris}, \citenamefont {Plumberg}, \citenamefont {Alford}, \citenamefont {Noronha}, \citenamefont {Noronha-Hostler}, \citenamefont {Pretorius}, \citenamefont {Witek},\ and\ \citenamefont {Yunes}}]{Most_2021}%
  \BibitemOpen
  \bibfield  {author} {\bibinfo {author} {\bibfnamefont {E.~R.}\ \bibnamefont {Most}}, \bibinfo {author} {\bibfnamefont {S.~P.}\ \bibnamefont {Harris}}, \bibinfo {author} {\bibfnamefont {C.}~\bibnamefont {Plumberg}}, \bibinfo {author} {\bibfnamefont {M.~G.}\ \bibnamefont {Alford}}, \bibinfo {author} {\bibfnamefont {J.}~\bibnamefont {Noronha}}, \bibinfo {author} {\bibfnamefont {J.}~\bibnamefont {Noronha-Hostler}}, \bibinfo {author} {\bibfnamefont {F.}~\bibnamefont {Pretorius}}, \bibinfo {author} {\bibfnamefont {H.}~\bibnamefont {Witek}},\ and\ \bibinfo {author} {\bibfnamefont {N.}~\bibnamefont {Yunes}},\ }\bibfield  {title} {\bibinfo {title} {Projecting the likely importance of weak-interaction-driven bulk viscosity in neutron star mergers},\ }\href {https://doi.org/10.1093/mnras/stab2793} {\bibfield  {journal} {\bibinfo  {journal} {Monthly Notices of the Royal Astronomical Society}\ }\textbf {\bibinfo {volume} {509}},\ \bibinfo {pages} {1096–1108} (\bibinfo {year} {2021})}\BibitemShut {NoStop}%
\bibitem [{\citenamefont {Celora}\ \emph {et~al.}(2022)\citenamefont {Celora}, \citenamefont {Hawke}, \citenamefont {Hammond}, \citenamefont {Andersson},\ and\ \citenamefont {Comer}}]{Celora_2022}%
  \BibitemOpen
  \bibfield  {author} {\bibinfo {author} {\bibfnamefont {T.}~\bibnamefont {Celora}}, \bibinfo {author} {\bibfnamefont {I.}~\bibnamefont {Hawke}}, \bibinfo {author} {\bibfnamefont {P.~C.}\ \bibnamefont {Hammond}}, \bibinfo {author} {\bibfnamefont {N.}~\bibnamefont {Andersson}},\ and\ \bibinfo {author} {\bibfnamefont {G.~L.}\ \bibnamefont {Comer}},\ }\bibfield  {title} {\bibinfo {title} {Formulating bulk viscosity for neutron star simulations},\ }\bibfield  {journal} {\bibinfo  {journal} {Physical Review D}\ }\textbf {\bibinfo {volume} {105}},\ \href {https://doi.org/10.1103/physrevd.105.103016} {10.1103/physrevd.105.103016} (\bibinfo {year} {2022})\BibitemShut {NoStop}%
\bibitem [{\citenamefont {Most}\ \emph {et~al.}(2024)\citenamefont {Most}, \citenamefont {Haber}, \citenamefont {Harris}, \citenamefont {Zhang}, \citenamefont {Alford},\ and\ \citenamefont {Noronha}}]{Most_2024}%
  \BibitemOpen
  \bibfield  {author} {\bibinfo {author} {\bibfnamefont {E.~R.}\ \bibnamefont {Most}}, \bibinfo {author} {\bibfnamefont {A.}~\bibnamefont {Haber}}, \bibinfo {author} {\bibfnamefont {S.~P.}\ \bibnamefont {Harris}}, \bibinfo {author} {\bibfnamefont {Z.}~\bibnamefont {Zhang}}, \bibinfo {author} {\bibfnamefont {M.~G.}\ \bibnamefont {Alford}},\ and\ \bibinfo {author} {\bibfnamefont {J.}~\bibnamefont {Noronha}},\ }\bibfield  {title} {\bibinfo {title} {Emergence of microphysical bulk viscosity in binary neutron star postmerger dynamics},\ }\href {https://doi.org/10.3847/2041-8213/ad454f} {\bibfield  {journal} {\bibinfo  {journal} {The Astrophysical Journal Letters}\ }\textbf {\bibinfo {volume} {967}},\ \bibinfo {pages} {L14} (\bibinfo {year} {2024})}\BibitemShut {NoStop}%
\bibitem [{\citenamefont {Chabanov}\ and\ \citenamefont {Rezzolla}(2023)}]{chabanov2023impactbulkviscositypostmerger}%
  \BibitemOpen
  \bibfield  {author} {\bibinfo {author} {\bibfnamefont {M.}~\bibnamefont {Chabanov}}\ and\ \bibinfo {author} {\bibfnamefont {L.}~\bibnamefont {Rezzolla}},\ }\href {https://arxiv.org/abs/2307.10464} {\bibinfo {title} {Impact of bulk viscosity on the post-merger gravitational-wave signal from merging neutron stars}} (\bibinfo {year} {2023}),\ \Eprint {https://arxiv.org/abs/2307.10464} {arXiv:2307.10464 [gr-qc]} \BibitemShut {NoStop}%
\bibitem [{\citenamefont {{Meszaros}}\ and\ \citenamefont {{Rees}}(1992)}]{Rees1992}%
  \BibitemOpen
  \bibfield  {author} {\bibinfo {author} {\bibfnamefont {P.}~\bibnamefont {{Meszaros}}}\ and\ \bibinfo {author} {\bibfnamefont {M.~J.}\ \bibnamefont {{Rees}}},\ }\bibfield  {title} {\bibinfo {title} {{Tidal Heating and Mass Loss in Neutron Star Binaries: Implications for Gamma-Ray Burst Models}},\ }\href {https://doi.org/10.1086/171813} {\bibfield  {journal} {\bibinfo  {journal} {The Astrophysical Journal}\ }\textbf {\bibinfo {volume} {397}},\ \bibinfo {pages} {570} (\bibinfo {year} {1992})}\BibitemShut {NoStop}%
\bibitem [{\citenamefont {{Kochanek}}(1992)}]{Kochanek1992}%
  \BibitemOpen
  \bibfield  {author} {\bibinfo {author} {\bibfnamefont {C.~S.}\ \bibnamefont {{Kochanek}}},\ }\bibfield  {title} {\bibinfo {title} {{Coalescing Binary Neutron Stars}},\ }\href {https://doi.org/10.1086/171851} {\bibfield  {journal} {\bibinfo  {journal} {The Astrophysical Journal}\ }\textbf {\bibinfo {volume} {398}},\ \bibinfo {pages} {234} (\bibinfo {year} {1992})}\BibitemShut {NoStop}%
\bibitem [{\citenamefont {{Bildsten}}\ and\ \citenamefont {{Cutler}}(1992)}]{Bildsten1992}%
  \BibitemOpen
  \bibfield  {author} {\bibinfo {author} {\bibfnamefont {L.}~\bibnamefont {{Bildsten}}}\ and\ \bibinfo {author} {\bibfnamefont {C.}~\bibnamefont {{Cutler}}},\ }\bibfield  {title} {\bibinfo {title} {{Tidal Interactions of Inspiraling Compact Binaries}},\ }\href {https://doi.org/10.1086/171983} {\bibfield  {journal} {\bibinfo  {journal} {The Astrophysical Journal}\ }\textbf {\bibinfo {volume} {400}},\ \bibinfo {pages} {175} (\bibinfo {year} {1992})}\BibitemShut {NoStop}%
\bibitem [{\citenamefont {Lai}(1994)}]{Lai:1993di}%
  \BibitemOpen
  \bibfield  {author} {\bibinfo {author} {\bibfnamefont {D.}~\bibnamefont {Lai}},\ }\bibfield  {title} {\bibinfo {title} {{Resonant oscillations and tidal heating in coalescing binary neutron stars}},\ }\href {https://doi.org/10.1093/mnras/270.3.611} {\bibfield  {journal} {\bibinfo  {journal} {Mon. Not. Roy. Astron. Soc.}\ }\textbf {\bibinfo {volume} {270}},\ \bibinfo {pages} {611} (\bibinfo {year} {1994})},\ \Eprint {https://arxiv.org/abs/astro-ph/9404062} {arXiv:astro-ph/9404062} \BibitemShut {NoStop}%
\bibitem [{\citenamefont {Arras}\ and\ \citenamefont {Weinberg}(2019)}]{Arras:2018fxj}%
  \BibitemOpen
  \bibfield  {author} {\bibinfo {author} {\bibfnamefont {P.}~\bibnamefont {Arras}}\ and\ \bibinfo {author} {\bibfnamefont {N.~N.}\ \bibnamefont {Weinberg}},\ }\bibfield  {title} {\bibinfo {title} {{Urca reactions during neutron star inspiral}},\ }\href {https://doi.org/10.1093/mnras/stz880} {\bibfield  {journal} {\bibinfo  {journal} {Mon. Not. Roy. Astron. Soc.}\ }\textbf {\bibinfo {volume} {486}},\ \bibinfo {pages} {1424} (\bibinfo {year} {2019})},\ \Eprint {https://arxiv.org/abs/1806.04163} {arXiv:1806.04163 [astro-ph.HE]} \BibitemShut {NoStop}%
\bibitem [{\citenamefont {Ripley}\ \emph {et~al.}(2023{\natexlab{a}})\citenamefont {Ripley}, \citenamefont {Hegade K.~R.},\ and\ \citenamefont {Yunes}}]{Ripley:2023qxo}%
  \BibitemOpen
  \bibfield  {author} {\bibinfo {author} {\bibfnamefont {J.~L.}\ \bibnamefont {Ripley}}, \bibinfo {author} {\bibfnamefont {A.}~\bibnamefont {Hegade K.~R.}},\ and\ \bibinfo {author} {\bibfnamefont {N.}~\bibnamefont {Yunes}},\ }\bibfield  {title} {\bibinfo {title} {{Probing internal dissipative processes of neutron stars with gravitational waves during the inspiral of neutron star binaries}},\ }\href {https://doi.org/10.1103/PhysRevD.108.103037} {\bibfield  {journal} {\bibinfo  {journal} {Phys. Rev. D}\ }\textbf {\bibinfo {volume} {108}},\ \bibinfo {pages} {103037} (\bibinfo {year} {2023}{\natexlab{a}})},\ \Eprint {https://arxiv.org/abs/2306.15633} {arXiv:2306.15633 [gr-qc]} \BibitemShut {NoStop}%
\bibitem [{\citenamefont {Ripley}\ \emph {et~al.}(2023{\natexlab{b}})\citenamefont {Ripley}, \citenamefont {Hegade K.~R.}, \citenamefont {Chandramouli},\ and\ \citenamefont {Yunes}}]{Ripley:2023lsq}%
  \BibitemOpen
  \bibfield  {author} {\bibinfo {author} {\bibfnamefont {J.~L.}\ \bibnamefont {Ripley}}, \bibinfo {author} {\bibfnamefont {A.}~\bibnamefont {Hegade K.~R.}}, \bibinfo {author} {\bibfnamefont {R.~S.}\ \bibnamefont {Chandramouli}},\ and\ \bibinfo {author} {\bibfnamefont {N.}~\bibnamefont {Yunes}},\ }\bibfield  {title} {\bibinfo {title} {{First constraint on the dissipative tidal deformability of neutron stars}},\ }\href@noop {} {\  (\bibinfo {year} {2023}{\natexlab{b}})},\ \Eprint {https://arxiv.org/abs/2312.11659} {arXiv:2312.11659 [gr-qc]} \BibitemShut {NoStop}%
\bibitem [{\citenamefont {Hegade K.~R.}\ \emph {et~al.}(2024{\natexlab{a}})\citenamefont {Hegade K.~R.}, \citenamefont {Ripley},\ and\ \citenamefont {Yunes}}]{HegadeKR:2024slr}%
  \BibitemOpen
  \bibfield  {author} {\bibinfo {author} {\bibfnamefont {A.}~\bibnamefont {Hegade K.~R.}}, \bibinfo {author} {\bibfnamefont {J.~L.}\ \bibnamefont {Ripley}},\ and\ \bibinfo {author} {\bibfnamefont {N.}~\bibnamefont {Yunes}},\ }\bibfield  {title} {\bibinfo {title} {{Dissipative tidal effects to next-to-leading order and constraints on the dissipative tidal deformability using gravitational wave data}},\ }\href@noop {} {\  (\bibinfo {year} {2024}{\natexlab{a}})},\ \Eprint {https://arxiv.org/abs/2407.02584} {arXiv:2407.02584 [gr-qc]} \BibitemShut {NoStop}%
\bibitem [{\citenamefont {Pitre}\ and\ \citenamefont {Poisson}(2024)}]{Pitre:2023xsr}%
  \BibitemOpen
  \bibfield  {author} {\bibinfo {author} {\bibfnamefont {T.}~\bibnamefont {Pitre}}\ and\ \bibinfo {author} {\bibfnamefont {E.}~\bibnamefont {Poisson}},\ }\bibfield  {title} {\bibinfo {title} {{General relativistic dynamical tides in binary inspirals without modes}},\ }\href {https://doi.org/10.1103/PhysRevD.109.064004} {\bibfield  {journal} {\bibinfo  {journal} {Phys. Rev. D}\ }\textbf {\bibinfo {volume} {109}},\ \bibinfo {pages} {064004} (\bibinfo {year} {2024})},\ \Eprint {https://arxiv.org/abs/2311.04075} {arXiv:2311.04075 [gr-qc]} \BibitemShut {NoStop}%
\bibitem [{\citenamefont {Hegade K.~R.}\ \emph {et~al.}(2024{\natexlab{b}})\citenamefont {Hegade K.~R.}, \citenamefont {Ripley},\ and\ \citenamefont {Yunes}}]{HegadeKR:2024agt}%
  \BibitemOpen
  \bibfield  {author} {\bibinfo {author} {\bibfnamefont {A.}~\bibnamefont {Hegade K.~R.}}, \bibinfo {author} {\bibfnamefont {J.~L.}\ \bibnamefont {Ripley}},\ and\ \bibinfo {author} {\bibfnamefont {N.}~\bibnamefont {Yunes}},\ }\bibfield  {title} {\bibinfo {title} {{Dynamical tidal response of non-rotating relativistic stars}},\ }\href@noop {} {\  (\bibinfo {year} {2024}{\natexlab{b}})},\ \Eprint {https://arxiv.org/abs/2403.03254} {arXiv:2403.03254 [gr-qc]} \BibitemShut {NoStop}%
\bibitem [{\citenamefont {Chakrabarti}\ \emph {et~al.}(2013)\citenamefont {Chakrabarti}, \citenamefont {Delsate},\ and\ \citenamefont {Steinhoff}}]{Chakrabarti:2013lua}%
  \BibitemOpen
  \bibfield  {author} {\bibinfo {author} {\bibfnamefont {S.}~\bibnamefont {Chakrabarti}}, \bibinfo {author} {\bibfnamefont {T.}~\bibnamefont {Delsate}},\ and\ \bibinfo {author} {\bibfnamefont {J.}~\bibnamefont {Steinhoff}},\ }\bibfield  {title} {\bibinfo {title} {{New perspectives on neutron star and black hole spectroscopy and dynamic tides}},\ }\href@noop {} {\  (\bibinfo {year} {2013})},\ \Eprint {https://arxiv.org/abs/1304.2228} {arXiv:1304.2228 [gr-qc]} \BibitemShut {NoStop}%
\bibitem [{\citenamefont {Ivanov}\ \emph {et~al.}(2024)\citenamefont {Ivanov}, \citenamefont {Li}, \citenamefont {Parra-Martinez},\ and\ \citenamefont {Zhou}}]{Ivanov:2024sds}%
  \BibitemOpen
  \bibfield  {author} {\bibinfo {author} {\bibfnamefont {M.~M.}\ \bibnamefont {Ivanov}}, \bibinfo {author} {\bibfnamefont {Y.-Z.}\ \bibnamefont {Li}}, \bibinfo {author} {\bibfnamefont {J.}~\bibnamefont {Parra-Martinez}},\ and\ \bibinfo {author} {\bibfnamefont {Z.}~\bibnamefont {Zhou}},\ }\bibfield  {title} {\bibinfo {title} {{Gravitational Raman Scattering in Effective Field Theory: A Scalar Tidal Matching at O(G3)}},\ }\href {https://doi.org/10.1103/PhysRevLett.132.131401} {\bibfield  {journal} {\bibinfo  {journal} {Phys. Rev. Lett.}\ }\textbf {\bibinfo {volume} {132}},\ \bibinfo {pages} {131401} (\bibinfo {year} {2024})},\ \Eprint {https://arxiv.org/abs/2401.08752} {arXiv:2401.08752 [hep-th]} \BibitemShut {NoStop}%
\bibitem [{\citenamefont {Saketh}\ and\ \citenamefont {Vines}(2022)}]{Saketh:2022wap}%
  \BibitemOpen
  \bibfield  {author} {\bibinfo {author} {\bibfnamefont {M.~V.~S.}\ \bibnamefont {Saketh}}\ and\ \bibinfo {author} {\bibfnamefont {J.}~\bibnamefont {Vines}},\ }\bibfield  {title} {\bibinfo {title} {{Scattering of gravitational waves off spinning compact objects with an effective worldline theory}},\ }\href {https://doi.org/10.1103/PhysRevD.106.124026} {\bibfield  {journal} {\bibinfo  {journal} {Phys. Rev. D}\ }\textbf {\bibinfo {volume} {106}},\ \bibinfo {pages} {124026} (\bibinfo {year} {2022})},\ \Eprint {https://arxiv.org/abs/2208.03170} {arXiv:2208.03170 [gr-qc]} \BibitemShut {NoStop}%
\bibitem [{\citenamefont {Bautista}\ \emph {et~al.}(2023{\natexlab{a}})\citenamefont {Bautista}, \citenamefont {Guevara}, \citenamefont {Kavanagh},\ and\ \citenamefont {Vines}}]{Bautista:2021wfy}%
  \BibitemOpen
  \bibfield  {author} {\bibinfo {author} {\bibfnamefont {Y.~F.}\ \bibnamefont {Bautista}}, \bibinfo {author} {\bibfnamefont {A.}~\bibnamefont {Guevara}}, \bibinfo {author} {\bibfnamefont {C.}~\bibnamefont {Kavanagh}},\ and\ \bibinfo {author} {\bibfnamefont {J.}~\bibnamefont {Vines}},\ }\bibfield  {title} {\bibinfo {title} {{Scattering in black hole backgrounds and higher-spin amplitudes. Part I}},\ }\href {https://doi.org/10.1007/JHEP03(2023)136} {\bibfield  {journal} {\bibinfo  {journal} {JHEP}\ }\textbf {\bibinfo {volume} {03}},\ \bibinfo {pages} {136}},\ \Eprint {https://arxiv.org/abs/2107.10179} {arXiv:2107.10179 [hep-th]} \BibitemShut {NoStop}%
\bibitem [{\citenamefont {Scheopner}\ and\ \citenamefont {Vines}(2023)}]{Scheopner:2023rzp}%
  \BibitemOpen
  \bibfield  {author} {\bibinfo {author} {\bibfnamefont {T.}~\bibnamefont {Scheopner}}\ and\ \bibinfo {author} {\bibfnamefont {J.}~\bibnamefont {Vines}},\ }\bibfield  {title} {\bibinfo {title} {{Dynamical Implications of the Kerr Multipole Moments for Spinning Black Holes}},\ }\href@noop {} {\  (\bibinfo {year} {2023})},\ \Eprint {https://arxiv.org/abs/2311.18421} {arXiv:2311.18421 [gr-qc]} \BibitemShut {NoStop}%
\bibitem [{\citenamefont {Bautista}\ \emph {et~al.}(2024)\citenamefont {Bautista}, \citenamefont {Bonelli}, \citenamefont {Iossa}, \citenamefont {Tanzini},\ and\ \citenamefont {Zhou}}]{Bautista:2023sdf}%
  \BibitemOpen
  \bibfield  {author} {\bibinfo {author} {\bibfnamefont {Y.~F.}\ \bibnamefont {Bautista}}, \bibinfo {author} {\bibfnamefont {G.}~\bibnamefont {Bonelli}}, \bibinfo {author} {\bibfnamefont {C.}~\bibnamefont {Iossa}}, \bibinfo {author} {\bibfnamefont {A.}~\bibnamefont {Tanzini}},\ and\ \bibinfo {author} {\bibfnamefont {Z.}~\bibnamefont {Zhou}},\ }\bibfield  {title} {\bibinfo {title} {{Black hole perturbation theory meets CFT2: Kerr-Compton amplitudes from Nekrasov-Shatashvili functions}},\ }\href {https://doi.org/10.1103/PhysRevD.109.084071} {\bibfield  {journal} {\bibinfo  {journal} {Phys. Rev. D}\ }\textbf {\bibinfo {volume} {109}},\ \bibinfo {pages} {084071} (\bibinfo {year} {2024})},\ \Eprint {https://arxiv.org/abs/2312.05965} {arXiv:2312.05965 [hep-th]} \BibitemShut {NoStop}%
\bibitem [{\citenamefont {Bautista}\ \emph {et~al.}(2023{\natexlab{b}})\citenamefont {Bautista}, \citenamefont {Guevara}, \citenamefont {Kavanagh},\ and\ \citenamefont {Vines}}]{Bautista:2022wjf}%
  \BibitemOpen
  \bibfield  {author} {\bibinfo {author} {\bibfnamefont {Y.~F.}\ \bibnamefont {Bautista}}, \bibinfo {author} {\bibfnamefont {A.}~\bibnamefont {Guevara}}, \bibinfo {author} {\bibfnamefont {C.}~\bibnamefont {Kavanagh}},\ and\ \bibinfo {author} {\bibfnamefont {J.}~\bibnamefont {Vines}},\ }\bibfield  {title} {\bibinfo {title} {{Scattering in black hole backgrounds and higher-spin amplitudes. Part II}},\ }\href {https://doi.org/10.1007/JHEP05(2023)211} {\bibfield  {journal} {\bibinfo  {journal} {JHEP}\ }\textbf {\bibinfo {volume} {05}},\ \bibinfo {pages} {211}},\ \Eprint {https://arxiv.org/abs/2212.07965} {arXiv:2212.07965 [hep-th]} \BibitemShut {NoStop}%
\bibitem [{\citenamefont {Andersson}\ and\ \citenamefont {Pnigouras}(2020)}]{Andersson:2019ahb}%
  \BibitemOpen
  \bibfield  {author} {\bibinfo {author} {\bibfnamefont {N.}~\bibnamefont {Andersson}}\ and\ \bibinfo {author} {\bibfnamefont {P.}~\bibnamefont {Pnigouras}},\ }\bibfield  {title} {\bibinfo {title} {{Exploring the effective tidal deformability of neutron stars}},\ }\href {https://doi.org/10.1103/PhysRevD.101.083001} {\bibfield  {journal} {\bibinfo  {journal} {Phys. Rev. D}\ }\textbf {\bibinfo {volume} {101}},\ \bibinfo {pages} {083001} (\bibinfo {year} {2020})},\ \Eprint {https://arxiv.org/abs/1906.08982} {arXiv:1906.08982 [astro-ph.HE]} \BibitemShut {NoStop}%
\bibitem [{\citenamefont {Ranea-Sandoval}\ \emph {et~al.}(2018)\citenamefont {Ranea-Sandoval}, \citenamefont {Guilera}, \citenamefont {Mariani},\ and\ \citenamefont {Orsaria}}]{Ranea_Sandoval_2018}%
  \BibitemOpen
  \bibfield  {author} {\bibinfo {author} {\bibfnamefont {I.~F.}\ \bibnamefont {Ranea-Sandoval}}, \bibinfo {author} {\bibfnamefont {O.~M.}\ \bibnamefont {Guilera}}, \bibinfo {author} {\bibfnamefont {M.}~\bibnamefont {Mariani}},\ and\ \bibinfo {author} {\bibfnamefont {M.~G.}\ \bibnamefont {Orsaria}},\ }\bibfield  {title} {\bibinfo {title} {Oscillation modes of hybrid stars within the relativistic cowling approximation},\ }\href {https://doi.org/10.1088/1475-7516/2018/12/031} {\bibfield  {journal} {\bibinfo  {journal} {Journal of Cosmology and Astroparticle Physics}\ }\textbf {\bibinfo {volume} {2018}}\bibinfo  {number} { (12)},\ \bibinfo {pages} {031–031}}\BibitemShut {NoStop}%
\bibitem [{\citenamefont {Tran}\ \emph {et~al.}(2023)\citenamefont {Tran}, \citenamefont {Ghosh}, \citenamefont {Lozano}, \citenamefont {Chatterjee},\ and\ \citenamefont {Jaikumar}}]{Tran_2023}%
  \BibitemOpen
\bibfield  {number} {  }\bibfield  {author} {\bibinfo {author} {\bibfnamefont {V.}~\bibnamefont {Tran}}, \bibinfo {author} {\bibfnamefont {S.}~\bibnamefont {Ghosh}}, \bibinfo {author} {\bibfnamefont {N.}~\bibnamefont {Lozano}}, \bibinfo {author} {\bibfnamefont {D.}~\bibnamefont {Chatterjee}},\ and\ \bibinfo {author} {\bibfnamefont {P.}~\bibnamefont {Jaikumar}},\ }\bibfield  {title} {\bibinfo {title} {g-mode oscillations in neutron stars with hyperons},\ }\bibfield  {journal} {\bibinfo  {journal} {Physical Review C}\ }\textbf {\bibinfo {volume} {108}},\ \href {https://doi.org/10.1103/physrevc.108.015803} {10.1103/physrevc.108.015803} (\bibinfo {year} {2023})\BibitemShut {NoStop}%
\bibitem [{\citenamefont {Casals}\ and\ \citenamefont {Ottewill}(2015)}]{Casals:2015nja}%
  \BibitemOpen
  \bibfield  {author} {\bibinfo {author} {\bibfnamefont {M.}~\bibnamefont {Casals}}\ and\ \bibinfo {author} {\bibfnamefont {A.~C.}\ \bibnamefont {Ottewill}},\ }\bibfield  {title} {\bibinfo {title} {{High-order tail in Schwarzschild spacetime}},\ }\href {https://doi.org/10.1103/PhysRevD.92.124055} {\bibfield  {journal} {\bibinfo  {journal} {Phys. Rev. D}\ }\textbf {\bibinfo {volume} {92}},\ \bibinfo {pages} {124055} (\bibinfo {year} {2015})},\ \Eprint {https://arxiv.org/abs/1509.04702} {arXiv:1509.04702 [gr-qc]} \BibitemShut {NoStop}%
\bibitem [{\citenamefont {Savi\'c}\ and\ \citenamefont {Vernizzi}(2026)}]{Vernizzi2026}%
  \BibitemOpen
  \bibfield  {author} {\bibinfo {author} {\bibfnamefont {N.}~\bibnamefont {Savi\'c}}\ and\ \bibinfo {author} {\bibfnamefont {F.}~\bibnamefont {Vernizzi}},\ }\bibfield  {title} {\bibinfo {title} {{Dissipative Tidal Love Numbers of Neutron Stars from Viscous Heating}},\ }\href@noop {} {\  (\bibinfo {year} {2026})},\ \Eprint {https://arxiv.org/abs/2611.30466} {arXiv:2611.30466 [gr-qc]} \BibitemShut {NoStop}%
\bibitem [{\citenamefont {{Cutler}}\ and\ \citenamefont {{Lindblom}}(1987)}]{Cutler_1987}%
  \BibitemOpen
  \bibfield  {author} {\bibinfo {author} {\bibfnamefont {C.}~\bibnamefont {{Cutler}}}\ and\ \bibinfo {author} {\bibfnamefont {L.}~\bibnamefont {{Lindblom}}},\ }\bibfield  {title} {\bibinfo {title} {{The Effect of Viscosity on Neutron Star Oscillations}},\ }\href {https://doi.org/10.1086/165052} {\bibfield  {journal} {\bibinfo  {journal} {\apj}\ }\textbf {\bibinfo {volume} {314}},\ \bibinfo {pages} {234} (\bibinfo {year} {1987})}\BibitemShut {NoStop}%
\bibitem [{\citenamefont {Buonanno}\ \emph {et~al.}(2009)\citenamefont {Buonanno}, \citenamefont {Iyer}, \citenamefont {Ochsner}, \citenamefont {Pan},\ and\ \citenamefont {Sathyaprakash}}]{Buonanno:2009zt}%
  \BibitemOpen
  \bibfield  {author} {\bibinfo {author} {\bibfnamefont {A.}~\bibnamefont {Buonanno}}, \bibinfo {author} {\bibfnamefont {B.}~\bibnamefont {Iyer}}, \bibinfo {author} {\bibfnamefont {E.}~\bibnamefont {Ochsner}}, \bibinfo {author} {\bibfnamefont {Y.}~\bibnamefont {Pan}},\ and\ \bibinfo {author} {\bibfnamefont {B.~S.}\ \bibnamefont {Sathyaprakash}},\ }\bibfield  {title} {\bibinfo {title} {{Comparison of post-Newtonian templates for compact binary inspiral signals in gravitational-wave detectors}},\ }\href {https://doi.org/10.1103/PhysRevD.80.084043} {\bibfield  {journal} {\bibinfo  {journal} {Phys. Rev. D}\ }\textbf {\bibinfo {volume} {80}},\ \bibinfo {pages} {084043} (\bibinfo {year} {2009})},\ \Eprint {https://arxiv.org/abs/0907.0700} {arXiv:0907.0700 [gr-qc]} \BibitemShut {NoStop}%
\bibitem [{\citenamefont {Arun}\ \emph {et~al.}(2009)\citenamefont {Arun}, \citenamefont {Buonanno}, \citenamefont {Faye},\ and\ \citenamefont {Ochsner}}]{Arun:2008kb}%
  \BibitemOpen
  \bibfield  {author} {\bibinfo {author} {\bibfnamefont {K.~G.}\ \bibnamefont {Arun}}, \bibinfo {author} {\bibfnamefont {A.}~\bibnamefont {Buonanno}}, \bibinfo {author} {\bibfnamefont {G.}~\bibnamefont {Faye}},\ and\ \bibinfo {author} {\bibfnamefont {E.}~\bibnamefont {Ochsner}},\ }\bibfield  {title} {\bibinfo {title} {{Higher-order spin effects in the amplitude and phase of gravitational waveforms emitted by inspiraling compact binaries: Ready-to-use gravitational waveforms}},\ }\href {https://doi.org/10.1103/PhysRevD.79.104023} {\bibfield  {journal} {\bibinfo  {journal} {Phys. Rev. D}\ }\textbf {\bibinfo {volume} {79}},\ \bibinfo {pages} {104023} (\bibinfo {year} {2009})},\ \bibinfo {note} {[Erratum: Phys.Rev.D 84, 049901 (2011)]},\ \Eprint {https://arxiv.org/abs/0810.5336} {arXiv:0810.5336 [gr-qc]} \BibitemShut {NoStop}%
\bibitem [{\citenamefont {Datta}\ \emph {et~al.}(2021)\citenamefont {Datta}, \citenamefont {Phukon},\ and\ \citenamefont {Bose}}]{Datta:2020gem}%
  \BibitemOpen
  \bibfield  {author} {\bibinfo {author} {\bibfnamefont {S.}~\bibnamefont {Datta}}, \bibinfo {author} {\bibfnamefont {K.~S.}\ \bibnamefont {Phukon}},\ and\ \bibinfo {author} {\bibfnamefont {S.}~\bibnamefont {Bose}},\ }\bibfield  {title} {\bibinfo {title} {{Recognizing black holes in gravitational-wave observations: Challenges in telling apart impostors in mass-gap binaries}},\ }\href {https://doi.org/10.1103/PhysRevD.104.084006} {\bibfield  {journal} {\bibinfo  {journal} {Phys. Rev. D}\ }\textbf {\bibinfo {volume} {104}},\ \bibinfo {pages} {084006} (\bibinfo {year} {2021})},\ \Eprint {https://arxiv.org/abs/2004.05974} {arXiv:2004.05974 [gr-qc]} \BibitemShut {NoStop}%
\bibitem [{\citenamefont {Blanchet}\ \emph {et~al.}(2023{\natexlab{a}})\citenamefont {Blanchet}, \citenamefont {Faye}, \citenamefont {Henry}, \citenamefont {Larrouturou},\ and\ \citenamefont {Trestini}}]{Blanchet:2023soy}%
  \BibitemOpen
  \bibfield  {author} {\bibinfo {author} {\bibfnamefont {L.}~\bibnamefont {Blanchet}}, \bibinfo {author} {\bibfnamefont {G.}~\bibnamefont {Faye}}, \bibinfo {author} {\bibfnamefont {Q.}~\bibnamefont {Henry}}, \bibinfo {author} {\bibfnamefont {F.}~\bibnamefont {Larrouturou}},\ and\ \bibinfo {author} {\bibfnamefont {D.}~\bibnamefont {Trestini}},\ }\bibfield  {title} {\bibinfo {title} {{Gravitational waves from compact binaries to the fourth post-Newtonian order}},\ }in\ \href@noop {} {\emph {\bibinfo {booktitle} {{57th Rencontres de Moriond on Gravitation}}}}\ (\bibinfo {year} {2023})\ \Eprint {https://arxiv.org/abs/2304.13647} {arXiv:2304.13647 [gr-qc]} \BibitemShut {NoStop}%
\bibitem [{\citenamefont {Chen}\ and\ \citenamefont {Piekarewicz}(2014)}]{Chen2014}%
  \BibitemOpen
  \bibfield  {author} {\bibinfo {author} {\bibfnamefont {W.-C.}\ \bibnamefont {Chen}}\ and\ \bibinfo {author} {\bibfnamefont {J.}~\bibnamefont {Piekarewicz}},\ }\bibfield  {title} {\bibinfo {title} {Building relativistic mean field models for finite nuclei and neutron stars},\ }\href {https://doi.org/10.1103/PhysRevC.90.044305} {\bibfield  {journal} {\bibinfo  {journal} {Phys. Rev. C}\ }\textbf {\bibinfo {volume} {90}},\ \bibinfo {pages} {044305} (\bibinfo {year} {2014})}\BibitemShut {NoStop}%
\bibitem [{\citenamefont {Glendenning}\ and\ \citenamefont {Moszkowski}(1991)}]{GM}%
  \BibitemOpen
  \bibfield  {author} {\bibinfo {author} {\bibfnamefont {N.~K.}\ \bibnamefont {Glendenning}}\ and\ \bibinfo {author} {\bibfnamefont {S.~A.}\ \bibnamefont {Moszkowski}},\ }\bibfield  {title} {\bibinfo {title} {Reconciliation of neutron-star masses and binding of the $\lambda$ in hypernuclei},\ }\href {https://doi.org/10.1103/PhysRevLett.67.2414} {\bibfield  {journal} {\bibinfo  {journal} {Phys. Rev. Lett.}\ }\textbf {\bibinfo {volume} {67}},\ \bibinfo {pages} {2414} (\bibinfo {year} {1991})}\BibitemShut {NoStop}%
\bibitem [{\citenamefont {Hornick}\ \emph {et~al.}(2018)\citenamefont {Hornick}, \citenamefont {Tolos}, \citenamefont {Zacchi}, \citenamefont {Christian},\ and\ \citenamefont {Schaffner-Bielich}}]{HZTCS}%
  \BibitemOpen
  \bibfield  {author} {\bibinfo {author} {\bibfnamefont {N.}~\bibnamefont {Hornick}}, \bibinfo {author} {\bibfnamefont {L.}~\bibnamefont {Tolos}}, \bibinfo {author} {\bibfnamefont {A.}~\bibnamefont {Zacchi}}, \bibinfo {author} {\bibfnamefont {J.-E.}\ \bibnamefont {Christian}},\ and\ \bibinfo {author} {\bibfnamefont {J.}~\bibnamefont {Schaffner-Bielich}},\ }\bibfield  {title} {\bibinfo {title} {Relativistic parameterizations of neutron matter and implications for neutron stars},\ }\href {https://doi.org/10.1103/PhysRevC.98.065804} {\bibfield  {journal} {\bibinfo  {journal} {Phys. Rev. C}\ }\textbf {\bibinfo {volume} {98}},\ \bibinfo {pages} {065804} (\bibinfo {year} {2018})}\BibitemShut {NoStop}%
\bibitem [{\citenamefont {Blanchet}\ \emph {et~al.}(2023{\natexlab{b}})\citenamefont {Blanchet}, \citenamefont {Faye}, \citenamefont {Henry}, \citenamefont {Larrouturou},\ and\ \citenamefont {Trestini}}]{Blanchet:2023bwj}%
  \BibitemOpen
  \bibfield  {author} {\bibinfo {author} {\bibfnamefont {L.}~\bibnamefont {Blanchet}}, \bibinfo {author} {\bibfnamefont {G.}~\bibnamefont {Faye}}, \bibinfo {author} {\bibfnamefont {Q.}~\bibnamefont {Henry}}, \bibinfo {author} {\bibfnamefont {F.}~\bibnamefont {Larrouturou}},\ and\ \bibinfo {author} {\bibfnamefont {D.}~\bibnamefont {Trestini}},\ }\bibfield  {title} {\bibinfo {title} {{Gravitational-Wave Phasing of Quasicircular Compact Binary Systems to the Fourth-and-a-Half Post-Newtonian Order}},\ }\href {https://doi.org/10.1103/PhysRevLett.131.121402} {\bibfield  {journal} {\bibinfo  {journal} {Phys. Rev. Lett.}\ }\textbf {\bibinfo {volume} {131}},\ \bibinfo {pages} {121402} (\bibinfo {year} {2023}{\natexlab{b}})},\ \Eprint {https://arxiv.org/abs/2304.11185} {arXiv:2304.11185 [gr-qc]} \BibitemShut {NoStop}%
\bibitem [{\citenamefont {Porto}(2008)}]{Porto:2007qi}%
  \BibitemOpen
  \bibfield  {author} {\bibinfo {author} {\bibfnamefont {R.~A.}\ \bibnamefont {Porto}},\ }\bibfield  {title} {\bibinfo {title} {{Absorption effects due to spin in the worldline approach to black hole dynamics}},\ }\href {https://doi.org/10.1103/PhysRevD.77.064026} {\bibfield  {journal} {\bibinfo  {journal} {Phys. Rev. D}\ }\textbf {\bibinfo {volume} {77}},\ \bibinfo {pages} {064026} (\bibinfo {year} {2008})},\ \Eprint {https://arxiv.org/abs/0710.5150} {arXiv:0710.5150 [hep-th]} \BibitemShut {NoStop}%
\bibitem [{\citenamefont {Goldberger}(2022{\natexlab{a}})}]{Goldberger:2022ebt}%
  \BibitemOpen
  \bibfield  {author} {\bibinfo {author} {\bibfnamefont {W.~D.}\ \bibnamefont {Goldberger}},\ }\bibfield  {title} {\bibinfo {title} {{Effective field theories of gravity and compact binary dynamics: A Snowmass 2021 whitepaper}},\ }in\ \href@noop {} {\emph {\bibinfo {booktitle} {{Snowmass 2021}}}}\ (\bibinfo {year} {2022})\ \Eprint {https://arxiv.org/abs/2206.14249} {arXiv:2206.14249 [hep-th]} \BibitemShut {NoStop}%
\bibitem [{\citenamefont {Goldberger}(2022{\natexlab{b}})}]{Goldberger:2022rqf}%
  \BibitemOpen
  \bibfield  {author} {\bibinfo {author} {\bibfnamefont {W.~D.}\ \bibnamefont {Goldberger}},\ }\bibfield  {title} {\bibinfo {title} {{Effective Field Theory for Compact Binary Dynamics}},\ }\href@noop {} {\  (\bibinfo {year} {2022}{\natexlab{b}})},\ \Eprint {https://arxiv.org/abs/2212.06677} {arXiv:2212.06677 [hep-th]} \BibitemShut {NoStop}%
\bibitem [{\citenamefont {Poisson}\ and\ \citenamefont {Will}(2014)}]{poisson2014gravity}%
  \BibitemOpen
  \bibfield  {author} {\bibinfo {author} {\bibfnamefont {E.}~\bibnamefont {Poisson}}\ and\ \bibinfo {author} {\bibfnamefont {C.~M.}\ \bibnamefont {Will}},\ }\bibfield  {title} {\bibinfo {title} {Gravity},\ }\href@noop {} {\bibfield  {journal} {\bibinfo  {journal} {Gravity}\ } (\bibinfo {year} {2014})}\BibitemShut {NoStop}%
\bibitem [{\citenamefont {Kojima}(1992)}]{Kojima:1992ie}%
  \BibitemOpen
  \bibfield  {author} {\bibinfo {author} {\bibfnamefont {Y.}~\bibnamefont {Kojima}},\ }\bibfield  {title} {\bibinfo {title} {{Equations governing the nonradial oscillations of a slowly rotating relativistic star}},\ }\href {https://doi.org/10.1103/PhysRevD.46.4289} {\bibfield  {journal} {\bibinfo  {journal} {Phys. Rev. D}\ }\textbf {\bibinfo {volume} {46}},\ \bibinfo {pages} {4289} (\bibinfo {year} {1992})}\BibitemShut {NoStop}%
\bibitem [{\citenamefont {Tolman}(1939)}]{Tolman:1939jz}%
  \BibitemOpen
  \bibfield  {author} {\bibinfo {author} {\bibfnamefont {R.~C.}\ \bibnamefont {Tolman}},\ }\bibfield  {title} {\bibinfo {title} {{Static solutions of Einstein's field equations for spheres of fluid}},\ }\href {https://doi.org/10.1103/PhysRev.55.364} {\bibfield  {journal} {\bibinfo  {journal} {Phys. Rev.}\ }\textbf {\bibinfo {volume} {55}},\ \bibinfo {pages} {364} (\bibinfo {year} {1939})}\BibitemShut {NoStop}%
\bibitem [{\citenamefont {Oppenheimer}\ and\ \citenamefont {Volkoff}(1939)}]{Oppenheimer:1939ne}%
  \BibitemOpen
  \bibfield  {author} {\bibinfo {author} {\bibfnamefont {J.~R.}\ \bibnamefont {Oppenheimer}}\ and\ \bibinfo {author} {\bibfnamefont {G.~M.}\ \bibnamefont {Volkoff}},\ }\bibfield  {title} {\bibinfo {title} {{On Massive neutron cores}},\ }\href {https://doi.org/10.1103/PhysRev.55.374} {\bibfield  {journal} {\bibinfo  {journal} {Phys. Rev.}\ }\textbf {\bibinfo {volume} {55}},\ \bibinfo {pages} {374} (\bibinfo {year} {1939})}\BibitemShut {NoStop}%
\bibitem [{\citenamefont {Fiorella~Burgio}\ and\ \citenamefont {Fantina}(2018)}]{EoSreview}%
  \BibitemOpen
  \bibfield  {author} {\bibinfo {author} {\bibfnamefont {G.}~\bibnamefont {Fiorella~Burgio}}\ and\ \bibinfo {author} {\bibfnamefont {A.~F.}\ \bibnamefont {Fantina}},\ }\bibfield  {title} {\bibinfo {title} {Nuclear equation of state for compact stars and supernovae},\ }\href {https://doi.org/10.1007/978-3-319-97616-7_6} {\bibfield  {journal} {\bibinfo  {journal} {Astrophysics and Space Science Library}\ ,\ \bibinfo {pages} {255–335}} (\bibinfo {year} {2018})}\BibitemShut {NoStop}%
\bibitem [{\citenamefont {Drischler}\ \emph {et~al.}(2020)\citenamefont {Drischler}, \citenamefont {Furnstahl}, \citenamefont {Melendez},\ and\ \citenamefont {Phillips}}]{Drischler2020}%
  \BibitemOpen
  \bibfield  {author} {\bibinfo {author} {\bibfnamefont {C.}~\bibnamefont {Drischler}}, \bibinfo {author} {\bibfnamefont {R.~J.}\ \bibnamefont {Furnstahl}}, \bibinfo {author} {\bibfnamefont {J.~A.}\ \bibnamefont {Melendez}},\ and\ \bibinfo {author} {\bibfnamefont {D.~R.}\ \bibnamefont {Phillips}},\ }\bibfield  {title} {\bibinfo {title} {How well do we know the neutron-matter equation of state at the densities inside neutron stars? a bayesian approach with correlated uncertainties},\ }\href {https://doi.org/10.1103/PhysRevLett.125.202702} {\bibfield  {journal} {\bibinfo  {journal} {Phys. Rev. Lett.}\ }\textbf {\bibinfo {volume} {125}},\ \bibinfo {pages} {202702} (\bibinfo {year} {2020})}\BibitemShut {NoStop}%
\bibitem [{\citenamefont {Ghosh}\ \emph {et~al.}(2022{\natexlab{b}})\citenamefont {Ghosh}, \citenamefont {Pradhan}, \citenamefont {Chatterjee},\ and\ \citenamefont {Schaffner-Bielich}}]{Ghosh2022_b}%
  \BibitemOpen
  \bibfield  {author} {\bibinfo {author} {\bibfnamefont {S.}~\bibnamefont {Ghosh}}, \bibinfo {author} {\bibfnamefont {B.~K.}\ \bibnamefont {Pradhan}}, \bibinfo {author} {\bibfnamefont {D.}~\bibnamefont {Chatterjee}},\ and\ \bibinfo {author} {\bibfnamefont {J.}~\bibnamefont {Schaffner-Bielich}},\ }\bibfield  {title} {\bibinfo {title} {Multi-physics constraints at different densities to probe nuclear symmetry energy in hyperonic neutron stars},\ }\bibfield  {journal} {\bibinfo  {journal} {Frontiers in Astronomy and Space Sciences}\ }\textbf {\bibinfo {volume} {9}},\ \href {https://doi.org/10.3389/fspas.2022.864294} {10.3389/fspas.2022.864294} (\bibinfo {year} {2022}{\natexlab{b}})\BibitemShut {NoStop}%
\bibitem [{\citenamefont {Traversi}\ \emph {et~al.}(2020)\citenamefont {Traversi}, \citenamefont {Char},\ and\ \citenamefont {Pagliara}}]{Traversi_2020}%
  \BibitemOpen
  \bibfield  {author} {\bibinfo {author} {\bibfnamefont {S.}~\bibnamefont {Traversi}}, \bibinfo {author} {\bibfnamefont {P.}~\bibnamefont {Char}},\ and\ \bibinfo {author} {\bibfnamefont {G.}~\bibnamefont {Pagliara}},\ }\bibfield  {title} {\bibinfo {title} {Bayesian inference of dense matter equation of state within relativistic mean field models using astrophysical measurements},\ }\href {https://doi.org/10.3847/1538-4357/ab99c1} {\bibfield  {journal} {\bibinfo  {journal} {The Astrophysical Journal}\ }\textbf {\bibinfo {volume} {897}},\ \bibinfo {pages} {165} (\bibinfo {year} {2020})}\BibitemShut {NoStop}%
\bibitem [{\citenamefont {Typel}\ \emph {et~al.}(2010)\citenamefont {Typel}, \citenamefont {R\"opke}, \citenamefont {Kl\"ahn}, \citenamefont {Blaschke},\ and\ \citenamefont {Wolter}}]{Crust}%
  \BibitemOpen
  \bibfield  {author} {\bibinfo {author} {\bibfnamefont {S.}~\bibnamefont {Typel}}, \bibinfo {author} {\bibfnamefont {G.}~\bibnamefont {R\"opke}}, \bibinfo {author} {\bibfnamefont {T.}~\bibnamefont {Kl\"ahn}}, \bibinfo {author} {\bibfnamefont {D.}~\bibnamefont {Blaschke}},\ and\ \bibinfo {author} {\bibfnamefont {H.~H.}\ \bibnamefont {Wolter}},\ }\bibfield  {title} {\bibinfo {title} {Composition and thermodynamics of nuclear matter with light clusters},\ }\href {https://doi.org/10.1103/PhysRevC.81.015803} {\bibfield  {journal} {\bibinfo  {journal} {Phys. Rev. C}\ }\textbf {\bibinfo {volume} {81}},\ \bibinfo {pages} {015803} (\bibinfo {year} {2010})}\BibitemShut {NoStop}%
\bibitem [{\citenamefont {Oertel}\ \emph {et~al.}(2017)\citenamefont {Oertel}, \citenamefont {Hempel}, \citenamefont {Klaehn},\ and\ \citenamefont {Typel}}]{Compose}%
  \BibitemOpen
  \bibfield  {author} {\bibinfo {author} {\bibfnamefont {M.}~\bibnamefont {Oertel}}, \bibinfo {author} {\bibfnamefont {M.}~\bibnamefont {Hempel}}, \bibinfo {author} {\bibfnamefont {T.}~\bibnamefont {Klaehn}},\ and\ \bibinfo {author} {\bibfnamefont {S.}~\bibnamefont {Typel}},\ }\href@noop {} {}\bibinfo {howpublished} {\url{(https://compose.obspm.fr/)}} (\bibinfo {year} {2017})\BibitemShut {NoStop}%
\bibitem [{\citenamefont {Yakovlev}\ \emph {et~al.}(2005)\citenamefont {Yakovlev}, \citenamefont {Gnedin}, \citenamefont {Gusakov}, \citenamefont {Kaminker}, \citenamefont {Levenfish},\ and\ \citenamefont {Potekhin}}]{Yakovlev_2005}%
  \BibitemOpen
  \bibfield  {author} {\bibinfo {author} {\bibfnamefont {D.}~\bibnamefont {Yakovlev}}, \bibinfo {author} {\bibfnamefont {O.}~\bibnamefont {Gnedin}}, \bibinfo {author} {\bibfnamefont {M.}~\bibnamefont {Gusakov}}, \bibinfo {author} {\bibfnamefont {A.}~\bibnamefont {Kaminker}}, \bibinfo {author} {\bibfnamefont {K.}~\bibnamefont {Levenfish}},\ and\ \bibinfo {author} {\bibfnamefont {A.}~\bibnamefont {Potekhin}},\ }\bibfield  {title} {\bibinfo {title} {Neutron star cooling},\ }\href {https://doi.org/10.1016/j.nuclphysa.2005.02.061} {\bibfield  {journal} {\bibinfo  {journal} {Nuclear Physics A}\ }\textbf {\bibinfo {volume} {752}},\ \bibinfo {pages} {590–599} (\bibinfo {year} {2005})}\BibitemShut {NoStop}%
\bibitem [{\citenamefont {Haskell}\ \emph {et~al.}(2012)\citenamefont {Haskell}, \citenamefont {Degenaar},\ and\ \citenamefont {Ho}}]{Haskell:2012vg}%
  \BibitemOpen
  \bibfield  {author} {\bibinfo {author} {\bibfnamefont {B.}~\bibnamefont {Haskell}}, \bibinfo {author} {\bibfnamefont {N.}~\bibnamefont {Degenaar}},\ and\ \bibinfo {author} {\bibfnamefont {W.~C.~G.}\ \bibnamefont {Ho}},\ }\bibfield  {title} {\bibinfo {title} {{Constraining the physics of the r-mode instability in neutron stars with X-ray and UV observations}},\ }\href {https://doi.org/10.1111/j.1365-2966.2012.21171.x} {\bibfield  {journal} {\bibinfo  {journal} {Mon. Not. Roy. Astron. Soc.}\ }\textbf {\bibinfo {volume} {424}},\ \bibinfo {pages} {93} (\bibinfo {year} {2012})},\ \Eprint {https://arxiv.org/abs/1201.2101} {arXiv:1201.2101 [astro-ph.SR]} \BibitemShut {NoStop}%
\bibitem [{\citenamefont {{Flowers}}\ and\ \citenamefont {{Itoh}}(1976)}]{Flower_1976}%
  \BibitemOpen
  \bibfield  {author} {\bibinfo {author} {\bibfnamefont {E.}~\bibnamefont {{Flowers}}}\ and\ \bibinfo {author} {\bibfnamefont {N.}~\bibnamefont {{Itoh}}},\ }\bibfield  {title} {\bibinfo {title} {{Transport properties of dense matter.}},\ }\href {https://doi.org/10.1086/154375} {\bibfield  {journal} {\bibinfo  {journal} {\apj}\ }\textbf {\bibinfo {volume} {206}},\ \bibinfo {pages} {218} (\bibinfo {year} {1976})}\BibitemShut {NoStop}%
\bibitem [{\citenamefont {Shternin}\ and\ \citenamefont {Yakovlev}(2008)}]{Shternin_2008}%
  \BibitemOpen
  \bibfield  {author} {\bibinfo {author} {\bibfnamefont {P.~S.}\ \bibnamefont {Shternin}}\ and\ \bibinfo {author} {\bibfnamefont {D.~G.}\ \bibnamefont {Yakovlev}},\ }\bibfield  {title} {\bibinfo {title} {Shear viscosity in neutron star cores},\ }\href {https://doi.org/10.1103/PhysRevD.78.063006} {\bibfield  {journal} {\bibinfo  {journal} {Phys. Rev. D}\ }\textbf {\bibinfo {volume} {78}},\ \bibinfo {pages} {063006} (\bibinfo {year} {2008})}\BibitemShut {NoStop}%
\bibitem [{\citenamefont {Sawyer}(1989)}]{Sawyer1989}%
  \BibitemOpen
  \bibfield  {author} {\bibinfo {author} {\bibfnamefont {R.~F.}\ \bibnamefont {Sawyer}},\ }\bibfield  {title} {\bibinfo {title} {Bulk viscosity of hot neutron-star matter and the maximum rotation rates of neutron stars},\ }\href {https://doi.org/10.1103/PhysRevD.39.3804} {\bibfield  {journal} {\bibinfo  {journal} {Phys. Rev. D}\ }\textbf {\bibinfo {volume} {39}},\ \bibinfo {pages} {3804} (\bibinfo {year} {1989})}\BibitemShut {NoStop}%
\bibitem [{\citenamefont {Chatterjee}\ and\ \citenamefont {Bandyopadhyay}(2006)}]{Chatterjee2006}%
  \BibitemOpen
  \bibfield  {author} {\bibinfo {author} {\bibfnamefont {D.}~\bibnamefont {Chatterjee}}\ and\ \bibinfo {author} {\bibfnamefont {D.}~\bibnamefont {Bandyopadhyay}},\ }\bibfield  {title} {\bibinfo {title} {Effect of hyperon-hyperon interaction on bulk viscosity and $r$-mode instability in neutron stars},\ }\href {https://doi.org/10.1103/PhysRevD.74.023003} {\bibfield  {journal} {\bibinfo  {journal} {Phys. Rev. D}\ }\textbf {\bibinfo {volume} {74}},\ \bibinfo {pages} {023003} (\bibinfo {year} {2006})}\BibitemShut {NoStop}%
\bibitem [{\citenamefont {Ofengeim}\ and\ \citenamefont {Yakovlev}(2015)}]{Ofengeim_2015}%
  \BibitemOpen
  \bibfield  {author} {\bibinfo {author} {\bibfnamefont {D.~D.}\ \bibnamefont {Ofengeim}}\ and\ \bibinfo {author} {\bibfnamefont {D.~G.}\ \bibnamefont {Yakovlev}},\ }\bibfield  {title} {\bibinfo {title} {Shear viscosity in magnetized neutron star crust},\ }\href {https://doi.org/10.1209/0295-5075/112/59001} {\bibfield  {journal} {\bibinfo  {journal} {EPL (Europhysics Letters)}\ }\textbf {\bibinfo {volume} {112}},\ \bibinfo {pages} {59001} (\bibinfo {year} {2015})}\BibitemShut {NoStop}%
\bibitem [{\citenamefont {{Chugunov}}\ and\ \citenamefont {{Yakovlev}}(2005)}]{Yakolev_2005}%
  \BibitemOpen
  \bibfield  {author} {\bibinfo {author} {\bibfnamefont {A.~I.}\ \bibnamefont {{Chugunov}}}\ and\ \bibinfo {author} {\bibfnamefont {D.~G.}\ \bibnamefont {{Yakovlev}}},\ }\bibfield  {title} {\bibinfo {title} {{Shear Viscosity and Oscillations of Neutron Star Crust}},\ }\href {https://doi.org/10.1134/1.2045323} {\bibfield  {journal} {\bibinfo  {journal} {Astronomy Reports}\ }\textbf {\bibinfo {volume} {49}},\ \bibinfo {pages} {724} (\bibinfo {year} {2005})},\ \Eprint {https://arxiv.org/abs/astro-ph/0511300} {arXiv:astro-ph/0511300 [astro-ph]} \BibitemShut {NoStop}%
\bibitem [{\citenamefont {Yakovlev}\ \emph {et~al.}(2018)\citenamefont {Yakovlev}, \citenamefont {Gusakov},\ and\ \citenamefont {Haensel}}]{Yakovlev:2018jia}%
  \BibitemOpen
  \bibfield  {author} {\bibinfo {author} {\bibfnamefont {D.~G.}\ \bibnamefont {Yakovlev}}, \bibinfo {author} {\bibfnamefont {M.~E.}\ \bibnamefont {Gusakov}},\ and\ \bibinfo {author} {\bibfnamefont {P.}~\bibnamefont {Haensel}},\ }\bibfield  {title} {\bibinfo {title} {{Bulk viscosity in a neutron star mantle}},\ }\href {https://doi.org/10.1093/mnras/sty2639} {\bibfield  {journal} {\bibinfo  {journal} {Mon. Not. Roy. Astron. Soc.}\ }\textbf {\bibinfo {volume} {481}},\ \bibinfo {pages} {4924} (\bibinfo {year} {2018})},\ \Eprint {https://arxiv.org/abs/1809.08609} {arXiv:1809.08609 [astro-ph.HE]} \BibitemShut {NoStop}%
\bibitem [{\citenamefont {Lindblom}\ and\ \citenamefont {Detweiler}(1983)}]{Lindblom:1983ps}%
  \BibitemOpen
  \bibfield  {author} {\bibinfo {author} {\bibfnamefont {L.}~\bibnamefont {Lindblom}}\ and\ \bibinfo {author} {\bibfnamefont {S.~L.}\ \bibnamefont {Detweiler}},\ }\bibfield  {title} {\bibinfo {title} {{The quadrupole oscillations of neutron stars}},\ }\href {https://doi.org/10.1086/190884} {\bibfield  {journal} {\bibinfo  {journal} {Astrophys. J. Suppl.}\ }\textbf {\bibinfo {volume} {53}},\ \bibinfo {pages} {73} (\bibinfo {year} {1983})}\BibitemShut {NoStop}%
\bibitem [{\citenamefont {Ghosh}\ \emph {et~al.}(2016)\citenamefont {Ghosh} \emph {et~al.}}]{Ghosh:2016qgn}%
  \BibitemOpen
  \bibfield  {author} {\bibinfo {author} {\bibfnamefont {A.}~\bibnamefont {Ghosh}} \emph {et~al.},\ }\bibfield  {title} {\bibinfo {title} {{Testing general relativity using golden black-hole binaries}},\ }\href {https://doi.org/10.1103/PhysRevD.94.021101} {\bibfield  {journal} {\bibinfo  {journal} {Phys. Rev. D}\ }\textbf {\bibinfo {volume} {94}},\ \bibinfo {pages} {021101} (\bibinfo {year} {2016})},\ \Eprint {https://arxiv.org/abs/1602.02453} {arXiv:1602.02453 [gr-qc]} \BibitemShut {NoStop}%
\bibitem [{\citenamefont {Raithel}\ \emph {et~al.}(2019)\citenamefont {Raithel}, \citenamefont {Özel},\ and\ \citenamefont {Psaltis}}]{Raithel_2019}%
  \BibitemOpen
  \bibfield  {author} {\bibinfo {author} {\bibfnamefont {C.~A.}\ \bibnamefont {Raithel}}, \bibinfo {author} {\bibfnamefont {F.}~\bibnamefont {Özel}},\ and\ \bibinfo {author} {\bibfnamefont {D.}~\bibnamefont {Psaltis}},\ }\bibfield  {title} {\bibinfo {title} {Finite-temperature extension for cold neutron star equations of state},\ }\href {https://doi.org/10.3847/1538-4357/ab08ea} {\bibfield  {journal} {\bibinfo  {journal} {The Astrophysical Journal}\ }\textbf {\bibinfo {volume} {875}},\ \bibinfo {pages} {12} (\bibinfo {year} {2019})}\BibitemShut {NoStop}%
\bibitem [{\citenamefont {Raduta}\ \emph {et~al.}(2021)\citenamefont {Raduta}, \citenamefont {Nacu},\ and\ \citenamefont {Oertel}}]{Raduta:2021coc}%
  \BibitemOpen
  \bibfield  {author} {\bibinfo {author} {\bibfnamefont {A.~R.}\ \bibnamefont {Raduta}}, \bibinfo {author} {\bibfnamefont {F.}~\bibnamefont {Nacu}},\ and\ \bibinfo {author} {\bibfnamefont {M.}~\bibnamefont {Oertel}},\ }\bibfield  {title} {\bibinfo {title} {{Equations of state for hot neutron stars}},\ }\href {https://doi.org/10.1140/epja/s10050-021-00628-z} {\bibfield  {journal} {\bibinfo  {journal} {Eur. Phys. J. A}\ }\textbf {\bibinfo {volume} {57}},\ \bibinfo {pages} {329} (\bibinfo {year} {2021})},\ \Eprint {https://arxiv.org/abs/2109.00251} {arXiv:2109.00251 [nucl-th]} \BibitemShut {NoStop}%
\bibitem [{\citenamefont {Counsell}\ \emph {et~al.}(2024)\citenamefont {Counsell}, \citenamefont {Gittins},\ and\ \citenamefont {Andersson}}]{Counsell:2023pqp}%
  \BibitemOpen
  \bibfield  {author} {\bibinfo {author} {\bibfnamefont {A.~R.}\ \bibnamefont {Counsell}}, \bibinfo {author} {\bibfnamefont {F.}~\bibnamefont {Gittins}},\ and\ \bibinfo {author} {\bibfnamefont {N.}~\bibnamefont {Andersson}},\ }\bibfield  {title} {\bibinfo {title} {{The impact of nuclear reactions on the neutron-star g-mode spectrum}},\ }\href {https://doi.org/10.1093/mnras/stae1242} {\bibfield  {journal} {\bibinfo  {journal} {Mon. Not. Roy. Astron. Soc.}\ }\textbf {\bibinfo {volume} {531}},\ \bibinfo {pages} {1721} (\bibinfo {year} {2024})},\ \Eprint {https://arxiv.org/abs/2310.13586} {arXiv:2310.13586 [astro-ph.HE]} \BibitemShut {NoStop}%
\bibitem [{\citenamefont {Andersson}\ and\ \citenamefont {Pnigouras}(2019)}]{Andersson:2019mxp}%
  \BibitemOpen
  \bibfield  {author} {\bibinfo {author} {\bibfnamefont {N.}~\bibnamefont {Andersson}}\ and\ \bibinfo {author} {\bibfnamefont {P.}~\bibnamefont {Pnigouras}},\ }\bibfield  {title} {\bibinfo {title} {{The g-mode spectrum of reactive neutron star cores}},\ }\href {https://doi.org/10.1093/mnras/stz2449} {\bibfield  {journal} {\bibinfo  {journal} {Mon. Not. Roy. Astron. Soc.}\ }\textbf {\bibinfo {volume} {489}},\ \bibinfo {pages} {4043} (\bibinfo {year} {2019})},\ \Eprint {https://arxiv.org/abs/1905.00010} {arXiv:1905.00010 [gr-qc]} \BibitemShut {NoStop}%
\bibitem [{\citenamefont {Cowling}(1941)}]{Cowling:1941nqk}%
  \BibitemOpen
  \bibfield  {author} {\bibinfo {author} {\bibfnamefont {T.~G.}\ \bibnamefont {Cowling}},\ }\bibfield  {title} {\bibinfo {title} {{The Non-radial Oscillations of Polytropic Stars}},\ }\href {https://doi.org/10.1093/mnras/101.8.367} {\bibfield  {journal} {\bibinfo  {journal} {Mon. Not. Roy. Astron. Soc.}\ }\textbf {\bibinfo {volume} {101}},\ \bibinfo {pages} {367} (\bibinfo {year} {1941})}\BibitemShut {NoStop}%
\bibitem [{\citenamefont {Ipser}\ and\ \citenamefont {Price}(1991)}]{PhysRevD.43.1768}%
  \BibitemOpen
  \bibfield  {author} {\bibinfo {author} {\bibfnamefont {J.~R.}\ \bibnamefont {Ipser}}\ and\ \bibinfo {author} {\bibfnamefont {R.~H.}\ \bibnamefont {Price}},\ }\bibfield  {title} {\bibinfo {title} {Nonradial pulsations of stellar models in general relativity},\ }\href {https://doi.org/10.1103/PhysRevD.43.1768} {\bibfield  {journal} {\bibinfo  {journal} {Phys. Rev. D}\ }\textbf {\bibinfo {volume} {43}},\ \bibinfo {pages} {1768} (\bibinfo {year} {1991})}\BibitemShut {NoStop}%
\bibitem [{\citenamefont {Ipser}\ and\ \citenamefont {Managan}(1985)}]{ipser1985eulerian}%
  \BibitemOpen
  \bibfield  {author} {\bibinfo {author} {\bibfnamefont {J.~R.}\ \bibnamefont {Ipser}}\ and\ \bibinfo {author} {\bibfnamefont {R.~A.}\ \bibnamefont {Managan}},\ }\bibfield  {title} {\bibinfo {title} {An eulerian variational principle and a criterion for the occurrence of nonaxisymmetric neutral modes along rotating axisymmetric sequences},\ }\href@noop {} {\bibfield  {journal} {\bibinfo  {journal} {Astrophysical Journal, Part 1 (ISSN 0004-637X), vol. 292, May 15, 1985, p. 517-521.}\ }\textbf {\bibinfo {volume} {292}},\ \bibinfo {pages} {517} (\bibinfo {year} {1985})}\BibitemShut {NoStop}%
\bibitem [{\citenamefont {{Reisenegger}}\ and\ \citenamefont {{Goldreich}}(1992)}]{1992ApJ...395..240R}%
  \BibitemOpen
  \bibfield  {author} {\bibinfo {author} {\bibfnamefont {A.}~\bibnamefont {{Reisenegger}}}\ and\ \bibinfo {author} {\bibfnamefont {P.}~\bibnamefont {{Goldreich}}},\ }\bibfield  {title} {\bibinfo {title} {{A New Class of g-Modes in Neutron Stars}},\ }\href {https://doi.org/10.1086/171645} {\bibfield  {journal} {\bibinfo  {journal} {\apj}\ }\textbf {\bibinfo {volume} {395}},\ \bibinfo {pages} {240} (\bibinfo {year} {1992})}\BibitemShut {NoStop}%
\bibitem [{\citenamefont {Camelio}\ \emph {et~al.}(2023)\citenamefont {Camelio}, \citenamefont {Gavassino}, \citenamefont {Antonelli}, \citenamefont {Bernuzzi},\ and\ \citenamefont {Haskell}}]{Giovanni_2023}%
  \BibitemOpen
  \bibfield  {author} {\bibinfo {author} {\bibfnamefont {G.}~\bibnamefont {Camelio}}, \bibinfo {author} {\bibfnamefont {L.}~\bibnamefont {Gavassino}}, \bibinfo {author} {\bibfnamefont {M.}~\bibnamefont {Antonelli}}, \bibinfo {author} {\bibfnamefont {S.}~\bibnamefont {Bernuzzi}},\ and\ \bibinfo {author} {\bibfnamefont {B.}~\bibnamefont {Haskell}},\ }\bibfield  {title} {\bibinfo {title} {Simulating bulk viscosity in neutron stars. i. formalism},\ }\href {https://doi.org/10.1103/PhysRevD.107.103031} {\bibfield  {journal} {\bibinfo  {journal} {Phys. Rev. D}\ }\textbf {\bibinfo {volume} {107}},\ \bibinfo {pages} {103031} (\bibinfo {year} {2023})}\BibitemShut {NoStop}%
\bibitem [{\citenamefont {Mano}\ \emph {et~al.}(1996)\citenamefont {Mano}, \citenamefont {Suzuki},\ and\ \citenamefont {Takasugi}}]{Mano:1996mf}%
  \BibitemOpen
  \bibfield  {author} {\bibinfo {author} {\bibfnamefont {S.}~\bibnamefont {Mano}}, \bibinfo {author} {\bibfnamefont {H.}~\bibnamefont {Suzuki}},\ and\ \bibinfo {author} {\bibfnamefont {E.}~\bibnamefont {Takasugi}},\ }\bibfield  {title} {\bibinfo {title} {{Analytic solutions of the Regge-Wheeler equation and the postMinkowskian expansion}},\ }\href {https://doi.org/10.1143/PTP.96.549} {\bibfield  {journal} {\bibinfo  {journal} {Prog. Theor. Phys.}\ }\textbf {\bibinfo {volume} {96}},\ \bibinfo {pages} {549} (\bibinfo {year} {1996})},\ \Eprint {https://arxiv.org/abs/gr-qc/9605057} {arXiv:gr-qc/9605057} \BibitemShut {NoStop}%
\bibitem [{\citenamefont {Blanchet}\ \emph {et~al.}(2023{\natexlab{c}})\citenamefont {Blanchet}, \citenamefont {Faye}, \citenamefont {Henry}, \citenamefont {Larrouturou},\ and\ \citenamefont {Trestini}}]{Blanchet:2023sbv}%
  \BibitemOpen
  \bibfield  {author} {\bibinfo {author} {\bibfnamefont {L.}~\bibnamefont {Blanchet}}, \bibinfo {author} {\bibfnamefont {G.}~\bibnamefont {Faye}}, \bibinfo {author} {\bibfnamefont {Q.}~\bibnamefont {Henry}}, \bibinfo {author} {\bibfnamefont {F.}~\bibnamefont {Larrouturou}},\ and\ \bibinfo {author} {\bibfnamefont {D.}~\bibnamefont {Trestini}},\ }\bibfield  {title} {\bibinfo {title} {{Gravitational-wave flux and quadrupole modes from quasicircular nonspinning compact binaries to the fourth post-Newtonian order}},\ }\href {https://doi.org/10.1103/PhysRevD.108.064041} {\bibfield  {journal} {\bibinfo  {journal} {Phys. Rev. D}\ }\textbf {\bibinfo {volume} {108}},\ \bibinfo {pages} {064041} (\bibinfo {year} {2023}{\natexlab{c}})},\ \Eprint {https://arxiv.org/abs/2304.11186} {arXiv:2304.11186 [gr-qc]} \BibitemShut {NoStop}%
\bibitem [{\citenamefont {Owen}\ \emph {et~al.}(2023)\citenamefont {Owen}, \citenamefont {Haster}, \citenamefont {Perkins}, \citenamefont {Cornish},\ and\ \citenamefont {Yunes}}]{Owen:2023mid}%
  \BibitemOpen
  \bibfield  {author} {\bibinfo {author} {\bibfnamefont {C.~B.}\ \bibnamefont {Owen}}, \bibinfo {author} {\bibfnamefont {C.-J.}\ \bibnamefont {Haster}}, \bibinfo {author} {\bibfnamefont {S.}~\bibnamefont {Perkins}}, \bibinfo {author} {\bibfnamefont {N.~J.}\ \bibnamefont {Cornish}},\ and\ \bibinfo {author} {\bibfnamefont {N.}~\bibnamefont {Yunes}},\ }\bibfield  {title} {\bibinfo {title} {{Waveform accuracy and systematic uncertainties in current gravitational wave observations}},\ }\href {https://doi.org/10.1103/PhysRevD.108.044018} {\bibfield  {journal} {\bibinfo  {journal} {Phys. Rev. D}\ }\textbf {\bibinfo {volume} {108}},\ \bibinfo {pages} {044018} (\bibinfo {year} {2023})},\ \Eprint {https://arxiv.org/abs/2301.11941} {arXiv:2301.11941 [gr-qc]} \BibitemShut {NoStop}%
\bibitem [{\citenamefont {P\"urrer}\ and\ \citenamefont {Haster}(2020)}]{Purrer:2019jcp}%
  \BibitemOpen
  \bibfield  {author} {\bibinfo {author} {\bibfnamefont {M.}~\bibnamefont {P\"urrer}}\ and\ \bibinfo {author} {\bibfnamefont {C.-J.}\ \bibnamefont {Haster}},\ }\bibfield  {title} {\bibinfo {title} {{Gravitational waveform accuracy requirements for future ground-based detectors}},\ }\href {https://doi.org/10.1103/PhysRevResearch.2.023151} {\bibfield  {journal} {\bibinfo  {journal} {Phys. Rev. Res.}\ }\textbf {\bibinfo {volume} {2}},\ \bibinfo {pages} {023151} (\bibinfo {year} {2020})},\ \Eprint {https://arxiv.org/abs/1912.10055} {arXiv:1912.10055 [gr-qc]} \BibitemShut {NoStop}%
\bibitem [{\citenamefont {Hu}\ and\ \citenamefont {Veitch}(2022)}]{Hu:2022rjq}%
  \BibitemOpen
  \bibfield  {author} {\bibinfo {author} {\bibfnamefont {Q.}~\bibnamefont {Hu}}\ and\ \bibinfo {author} {\bibfnamefont {J.}~\bibnamefont {Veitch}},\ }\bibfield  {title} {\bibinfo {title} {{Assessing the model waveform accuracy of gravitational waves}},\ }\href {https://doi.org/10.1103/PhysRevD.106.044042} {\bibfield  {journal} {\bibinfo  {journal} {Phys. Rev. D}\ }\textbf {\bibinfo {volume} {106}},\ \bibinfo {pages} {044042} (\bibinfo {year} {2022})},\ \Eprint {https://arxiv.org/abs/2205.08448} {arXiv:2205.08448 [gr-qc]} \BibitemShut {NoStop}%
\bibitem [{\citenamefont {Ghosh}\ \emph {et~al.}(2026)\citenamefont {Ghosh}, \citenamefont {Haber}, \citenamefont {Andersson},\ and\ \citenamefont {Counsell}}]{Ghosh:2026hyp}%
  \BibitemOpen
  \bibfield  {author} {\bibinfo {author} {\bibfnamefont {S.}~\bibnamefont {Ghosh}}, \bibinfo {author} {\bibfnamefont {A.}~\bibnamefont {Haber}}, \bibinfo {author} {\bibfnamefont {N.}~\bibnamefont {Andersson}},\ and\ \bibinfo {author} {\bibfnamefont {A.~R.}\ \bibnamefont {Counsell}},\ }\bibfield  {title} {\bibinfo {title} {{Oscillations of dissipative neutron stars: The impact of hyperonic reaction rates}},\ }\href {https://doi.org/10.1103/n24c-v54v} {\bibfield  {journal} {\bibinfo  {journal} {Phys. Rev. D (accepted)}\ } (\bibinfo {year} {2026})},\ \Eprint {https://arxiv.org/abs/2608.07311} {arXiv:2608.07311 [gr-qc]} \BibitemShut {NoStop}%
\bibitem [{\citenamefont {Ghosh}\ \emph {et~al.}(2025)\citenamefont {Ghosh}, \citenamefont {Mukherjee}, \citenamefont {Bose},\ and\ \citenamefont {Chatterjee}}]{Ghosh:2025tid}%
  \BibitemOpen
  \bibfield  {author} {\bibinfo {author} {\bibfnamefont {S.}~\bibnamefont {Ghosh}}, \bibinfo {author} {\bibfnamefont {S.}~\bibnamefont {Mukherjee}}, \bibinfo {author} {\bibfnamefont {S.}~\bibnamefont {Bose}},\ and\ \bibinfo {author} {\bibfnamefont {D.}~\bibnamefont {Chatterjee}},\ }\bibfield  {title} {\bibinfo {title} {{Tidal dissipation in binary neutron star inspiral: Bias study and modeling of frequency domain phase}},\ }\href@noop {} {\  (\bibinfo {year} {2025})},\ \Eprint {https://arxiv.org/abs/2503.14606} {arXiv:2503.14606 [gr-qc]} \BibitemShut {NoStop}%
\bibitem [{\citenamefont {Saketh}\ \emph {et~al.}(2026)\citenamefont {Saketh}, \citenamefont {Ghosh},\ and\ \citenamefont {Andersson}}]{Saketh:2026dyn}%
  \BibitemOpen
  \bibfield  {author} {\bibinfo {author} {\bibfnamefont {M.~V.~S.}\ \bibnamefont {Saketh}}, \bibinfo {author} {\bibfnamefont {S.}~\bibnamefont {Ghosh}},\ and\ \bibinfo {author} {\bibfnamefont {N.}~\bibnamefont {Andersson}},\ }\bibfield  {title} {\bibinfo {title} {{Dynamical tidal response of neutron stars via scattering amplitudes}},\ }\href@noop {} {\  (\bibinfo {year} {2026})},\ \Eprint {https://arxiv.org/abs/2606.14405} {arXiv:2606.14405 [gr-qc]} \BibitemShut {NoStop}%
\bibitem [{\citenamefont {Flanagan}\ and\ \citenamefont {Racine}(2007)}]{Flanagan:2006sb}%
  \BibitemOpen
  \bibfield  {author} {\bibinfo {author} {\bibfnamefont {E.~E.}\ \bibnamefont {Flanagan}}\ and\ \bibinfo {author} {\bibfnamefont {E.}~\bibnamefont {Racine}},\ }\bibfield  {title} {\bibinfo {title} {{Gravitomagnetic resonant excitation of Rossby modes in coalescing neutron star binaries}},\ }\href {https://doi.org/10.1103/PhysRevD.75.044001} {\bibfield  {journal} {\bibinfo  {journal} {Phys. Rev. D}\ }\textbf {\bibinfo {volume} {75}},\ \bibinfo {pages} {044001} (\bibinfo {year} {2007})},\ \Eprint {https://arxiv.org/abs/gr-qc/0601029} {arXiv:gr-qc/0601029} \BibitemShut {NoStop}%
\bibitem [{\citenamefont {Poisson}(2020)}]{Poisson:2020mdi}%
  \BibitemOpen
  \bibfield  {author} {\bibinfo {author} {\bibfnamefont {E.}~\bibnamefont {Poisson}},\ }\bibfield  {title} {\bibinfo {title} {{Gravitomagnetic Love tensor of a slowly rotating body: post-Newtonian theory}},\ }\href {https://doi.org/10.1103/PhysRevD.102.064059} {\bibfield  {journal} {\bibinfo  {journal} {Phys. Rev. D}\ }\textbf {\bibinfo {volume} {102}},\ \bibinfo {pages} {064059} (\bibinfo {year} {2020})},\ \Eprint {https://arxiv.org/abs/2007.01678} {arXiv:2007.01678 [gr-qc]} \BibitemShut {NoStop}%
\bibitem [{\citenamefont {Ma}\ \emph {et~al.}(2021)\citenamefont {Ma}, \citenamefont {Yu},\ and\ \citenamefont {Chen}}]{Ma:2020oni}%
  \BibitemOpen
  \bibfield  {author} {\bibinfo {author} {\bibfnamefont {S.}~\bibnamefont {Ma}}, \bibinfo {author} {\bibfnamefont {H.}~\bibnamefont {Yu}},\ and\ \bibinfo {author} {\bibfnamefont {Y.}~\bibnamefont {Chen}},\ }\bibfield  {title} {\bibinfo {title} {{Detecting resonant tidal excitations of Rossby modes in coalescing neutron-star binaries with third-generation gravitational-wave detectors}},\ }\href {https://doi.org/10.1103/PhysRevD.103.063020} {\bibfield  {journal} {\bibinfo  {journal} {Phys. Rev. D}\ }\textbf {\bibinfo {volume} {103}},\ \bibinfo {pages} {063020} (\bibinfo {year} {2021})},\ \Eprint {https://arxiv.org/abs/2010.03066} {arXiv:2010.03066 [gr-qc]} \BibitemShut {NoStop}%
\bibitem [{\citenamefont {Gupta}\ \emph {et~al.}(2021)\citenamefont {Gupta}, \citenamefont {Steinhoff},\ and\ \citenamefont {Hinderer}}]{Gupta:2020lnv}%
  \BibitemOpen
  \bibfield  {author} {\bibinfo {author} {\bibfnamefont {P.~K.}\ \bibnamefont {Gupta}}, \bibinfo {author} {\bibfnamefont {J.}~\bibnamefont {Steinhoff}},\ and\ \bibinfo {author} {\bibfnamefont {T.}~\bibnamefont {Hinderer}},\ }\bibfield  {title} {\bibinfo {title} {{Relativistic effective action of dynamical gravitomagnetic tides for slowly rotating neutron stars}},\ }\href {https://doi.org/10.1103/PhysRevResearch.3.013147} {\bibfield  {journal} {\bibinfo  {journal} {Phys. Rev. Res.}\ }\textbf {\bibinfo {volume} {3}},\ \bibinfo {pages} {013147} (\bibinfo {year} {2021})},\ \Eprint {https://arxiv.org/abs/2011.03508} {arXiv:2011.03508 [gr-qc]} \BibitemShut {NoStop}%
\bibitem [{\citenamefont {Yu}\ \emph {et~al.}(2024)\citenamefont {Yu}, \citenamefont {Arras},\ and\ \citenamefont {Weinberg}}]{Yu:2024uxt}%
  \BibitemOpen
  \bibfield  {author} {\bibinfo {author} {\bibfnamefont {H.}~\bibnamefont {Yu}}, \bibinfo {author} {\bibfnamefont {P.}~\bibnamefont {Arras}},\ and\ \bibinfo {author} {\bibfnamefont {N.~N.}\ \bibnamefont {Weinberg}},\ }\bibfield  {title} {\bibinfo {title} {{Dynamical tides during the inspiral of rapidly spinning neutron stars: Solutions beyond mode resonance}},\ }\href@noop {} {\  (\bibinfo {year} {2024})},\ \Eprint {https://arxiv.org/abs/2404.00147} {arXiv:2404.00147 [gr-qc]} \BibitemShut {NoStop}%
\bibitem [{\citenamefont {Ma}\ \emph {et~al.}(2020)\citenamefont {Ma}, \citenamefont {Yu},\ and\ \citenamefont {Chen}}]{Ma:2020rak}%
  \BibitemOpen
  \bibfield  {author} {\bibinfo {author} {\bibfnamefont {S.}~\bibnamefont {Ma}}, \bibinfo {author} {\bibfnamefont {H.}~\bibnamefont {Yu}},\ and\ \bibinfo {author} {\bibfnamefont {Y.}~\bibnamefont {Chen}},\ }\bibfield  {title} {\bibinfo {title} {{Excitation of f-modes during mergers of spinning binary neutron star}},\ }\href {https://doi.org/10.1103/PhysRevD.101.123020} {\bibfield  {journal} {\bibinfo  {journal} {Phys. Rev. D}\ }\textbf {\bibinfo {volume} {101}},\ \bibinfo {pages} {123020} (\bibinfo {year} {2020})},\ \Eprint {https://arxiv.org/abs/2003.02373} {arXiv:2003.02373 [gr-qc]} \BibitemShut {NoStop}%
\bibitem [{\citenamefont {Steinhoff}\ \emph {et~al.}(2021)\citenamefont {Steinhoff}, \citenamefont {Hinderer}, \citenamefont {Dietrich},\ and\ \citenamefont {Foucart}}]{Steinhoff:2021dsn}%
  \BibitemOpen
  \bibfield  {author} {\bibinfo {author} {\bibfnamefont {J.}~\bibnamefont {Steinhoff}}, \bibinfo {author} {\bibfnamefont {T.}~\bibnamefont {Hinderer}}, \bibinfo {author} {\bibfnamefont {T.}~\bibnamefont {Dietrich}},\ and\ \bibinfo {author} {\bibfnamefont {F.}~\bibnamefont {Foucart}},\ }\bibfield  {title} {\bibinfo {title} {{Spin effects on neutron star fundamental-mode dynamical tides: Phenomenology and comparison to numerical simulations}},\ }\href {https://doi.org/10.1103/PhysRevResearch.3.033129} {\bibfield  {journal} {\bibinfo  {journal} {Phys. Rev. Res.}\ }\textbf {\bibinfo {volume} {3}},\ \bibinfo {pages} {033129} (\bibinfo {year} {2021})},\ \Eprint {https://arxiv.org/abs/2103.06100} {arXiv:2103.06100 [gr-qc]} \BibitemShut {NoStop}%
\bibitem [{\citenamefont {Thorne}(1969)}]{Thorne:1969rba}%
  \BibitemOpen
  \bibfield  {author} {\bibinfo {author} {\bibfnamefont {K.~S.}\ \bibnamefont {Thorne}},\ }\bibfield  {title} {\bibinfo {title} {{Nonradial Pulsation of General-Relativistic Stellar Models.IV. The Weakfield Limit}},\ }\href {https://doi.org/10.1086/150259} {\bibfield  {journal} {\bibinfo  {journal} {Astrophys. J.}\ }\textbf {\bibinfo {volume} {158}},\ \bibinfo {pages} {997} (\bibinfo {year} {1969})}\BibitemShut {NoStop}%
\bibitem [{\citenamefont {Smeyers}\ and\ \citenamefont {Van~Hoolst}(2011)}]{smeyers2011linear}%
  \BibitemOpen
  \bibfield  {author} {\bibinfo {author} {\bibfnamefont {P.}~\bibnamefont {Smeyers}}\ and\ \bibinfo {author} {\bibfnamefont {T.}~\bibnamefont {Van~Hoolst}},\ }\href@noop {} {\emph {\bibinfo {title} {Linear Isentropic Oscillations of Stars: Theoretical Foundations}}},\ Vol.\ \bibinfo {volume} {371}\ (\bibinfo  {publisher} {Springer Science \& Business Media},\ \bibinfo {year} {2011})\BibitemShut {NoStop}%
\bibitem [{\citenamefont {Chandrasekhar}(1960)}]{chandrasekhar1960general}%
  \BibitemOpen
  \bibfield  {author} {\bibinfo {author} {\bibfnamefont {S.}~\bibnamefont {Chandrasekhar}},\ }\bibfield  {title} {\bibinfo {title} {A general variational principle governing the radial and the non-radial oscillations of gaseous masses},\ }\href@noop {} {\bibfield  {journal} {\bibinfo  {journal} {VI. Ellipsoidal Figures of Equilibrium}\ }\textbf {\bibinfo {volume} {1}} (\bibinfo {year} {1960})}\BibitemShut {NoStop}%
\end{thebibliography}%
\end{document}